# Frustrated Interlayer Coupling Creates a Higher-Order Helical Superstructure in a Twist-Bend Smectic Phase

Abigail Pearson[1], Damian Pociecha[2], Magdalena M. Majewska[2], Nataša Vaupotič[3,4,*], Rebecca Walker[1,*], Ewa Gorecka[2,*]

---

[1] School of Natural and Computing Sciences, University of Aberdeen, Aberdeen AB24 3UE, Great Britain

[2] Faculty of Chemistry, University of Warsaw, ul. Pasteura 1, 02-093 Warsaw, Poland

[3] Faculty of Natural Sciences and Mathematics, University of Maribor Maribor 2000, Slovenia

[4] Jozef Stefan Institute, Ljubljana 1000, Slovenia

[*] natasa.vaupotic@um.si, rebecca.walker@abdn.ac.uk, gorecka@chem.uw.edu.pl

**Significance statement (50 to max 120 words)**:

Hierarchical self-organization allows simple molecular building blocks to generate structures with increasing levels of complexity. We report a new twist-bend smectic phase in which an established four-layer helical unit undergoes an additional periodic modulation, producing a higher-order helical superstructure. Resonant soft X-ray scattering reveals a distinctive four-peak splitting pattern that directly reflects this hierarchy. The results demonstrate how competing intermolecular packing tendencies can be resolved through spontaneous formation of multilevel structures in soft matter.

## Author contribution

A. Pearson: investigation, writing – review and editing
D. Pociecha: conceptualization, investigation, writing – original draft, writing – review and editing, data curation, methodology
M. M. Majewska: investigation, writing – review and editing, project administration, funding acquisition
N. Vaupotič: investigation, writing – original draft, writing – review and editing
R. Walker: conceptualization, investigation, writing – review and editing, project administration, supervision
E. Gorecka: conceptualization, investigation, funding acquisition, writing – original draft, writing – review and editing, supervision.

## Competing Interest Statement

The authors declare no competing interests.

# Frustrated Interlayer Coupling Creates a Higher-Order Helical Superstructure in a Twist-Bend Smectic Phase

Abigail Pearson[1], Damian Pociecha[2], Magdalena M. Majewska[2], Nataša Vaupotič[3,4,*], Rebecca Walker[1,*], Ewa Gorecka[2,*]

[1] School of Natural and Computing Sciences, University of Aberdeen, Aberdeen AB24 3UE, Great Britain

[2] Faculty of Chemistry, University of Warsaw, ul. Pasteura 1, 02-093 Warsaw, Poland

[3] Faculty of Natural Sciences and Mathematics, University of Maribor Maribor 2000, Slovenia

[4] Jozef Stefan Institute, Ljubljana 1000, Slovenia

[*] natasa.vaupotic@um.si, rebecca.walker@abdn.ac.uk, gorecka@chem.uw.edu.pl

**Abstract:** Bent liquid-crystalline dimers provide a model system for studying the spontaneous emergence of chirality from achiral molecules. Depending on terminal-chain length, these materials exhibit a competition between heliconical, synclinic, and anticlinic packing tendencies in tilted smectic phases. In short homologues, the tendency to form twist-bend phases dominates, leading to two distinct helical phases: the $SmC_{TB\text{-}SH}$ phase, with a helical periodicity of four molecular lengths, and at lower temperatures, a double helical structure - the $SmC_{TB\text{-}DH}$ phase, with progressive rotation of the four-layer unit. With increasing terminal chain length, this twist-bend tendency is gradually suppressed, favoring phases with simple synclinic and anticlinic tilt arrangement in successive layers. Remarkably, in the regime where these tendencies become comparable, a new modulated smectic phase, $SmC_{TB\text{-}mod}$ develops; a double helical structure in which the internal four-layer unit undergoes a periodic modulation. This additional level of structural organization provides a means of accommodating competing heliconical and synclinic/anticlinic tendencies and produces a characteristic splitting of the resonant x-ray diffraction peaks into four signals symmetrically located around a signal of four-layer periodicity. These results reveal how bent dimers resolve local-packing-mode frustration through complex spontaneous hierarchical structural organization.

Hierarchical self-organization, whereby molecular building blocks assemble into increasingly complex structures across multiple length scales, is a pervasive principle underlying both natural and synthetic materials [1]. In biological systems, hierarchical organization is essential for functionality, enabling the emergence of sophisticated mechanical, optical, and transport properties from comparatively simple molecular constituents. Inspired by such examples, the design and control of such architectures have become central goals in materials science, because organization at multiple levels often gives rise to emergent phenomena that cannot be inferred from the properties of individual molecules alone. Hierarchical structures are therefore of fundamental interest for understanding self-assembly and symmetry breaking, while also offering opportunities for applications in photonics, sensing, energy conversion, information storage, and responsive materials.

Soft matter systems provide an especially rich platform for investigating hierarchical self-organization due to the delicate balance between the energy of intermolecular interactions and that of thermal fluctuations. Among them, liquid crystals are unique as they combine fluidity with long-range orientational and, in some cases, limited positional order [2]. The structures of liquid crystal mesophases are extremely sensitive to molecular architecture, making them the ideal model system for exploring how nanoscale molecular design translates into mesoscale and macroscopic organization. One of the most effective approaches for establishing structure-property relationships in liquid crystals is the study of homologous series, in which a single feature of molecular structure is systematically varied. In particular, the length of terminal alkyl chains strongly influences molecular packing through the competition between chain microsegregation, steric constraints, and orientational order. As a result, seemingly subtle molecular modifications can generate profound changes in self-assembled structures and lead to the emergence of new levels of hierarchical organization [3].

A typical phase behavior of rod-like liquid-crystalline molecules as a function of terminal alkyl-chain length is well established [4]: short-chain homologues generally exhibit only a nematic phase, whereas elongation of the terminal chains promotes the appearance of smectic phases. The formation of layers, characteristic of smectic ordering, is driven by microsegregation between chemically-incompatible flexible alkyl chains and rigid aromatic mesogenic cores. The phase behavior of liquid-crystalline dimers is considerably more complex because it is strongly influenced by the length and parity of flexible spacer linking two rigid mesogenic units [5,6]. Dimers might form smectic phases with layer spacing equal to the molecular length, half of the molecular length or, in the case of asymmetric dimers, also double the molecular length [7]. Moreover, dimers containing odd-membered spacers often exhibit modulated structures as a consequence of their on-average bent shape. The orientation of such dimers is therefore described not only by a molecular long axis connecting the molecular termini, but also by a secondary axis associated with the bend direction. Thus, orientational ordering may involve both alignment of the molecular long axes and correlation of molecular bend orientation, making their packing substantially more complex than that of conventional calamitic liquid crystals [8].

The bent molecular geometry of odd-membered dimers also promotes spontaneous elastic deformation of the director field [9], giving rise to the twist-bend nematic, $N_{TB}$ [10,11] and smectic phases [12]. These phases exhibit increasingly complex hierarchical organization, including structures characterized by short-pitch, few-molecular-length periodicity ($N_{TB}$, $SmC_{TB\text{-}SH}$) and the recently discovered $SmC_{TB\text{-}DH}$ phase, in which a long-wavelength helical modulation is superimposed on the primary four-layer helix to generate a double-helical architecture [13,14]. These phases represent striking examples of multilevel self-organization, where chirality spontaneously emerges sequentially on distinct length scales in systems made of achiral molecules.

Here, we investigate four homologous series of asymmetric liquid-crystalline dimers containing odd-membered spacers and substantially extended terminal alkyl chains, reaching up to eighteen carbon atoms (Fig. 1). The selected compounds span a broad family of closely related dimeric mesogens that differ in their terminal-chain length and mesogenic core architecture, while preserving the key structural features responsible for twist-bend-type molecular arrangement: namely, an odd-membered spacer endowing an overall bent molecular shape. However, the influence of this geometry on the resulting phase behaviour is expected to be screened by elongated terminal chains. Indeed, across all series, a remarkably consistent phase behavior emerges, revealing distinct regions of the phase diagram governed by different intermolecular interactions and thus packing mechanisms. As the terminal-chain length increases, the balance between molecular bend and alkyl-chain microsegregation evolves systematically, producing a rich sequence of hierarchical structures. The resulting phase diagrams, summarized in Fig. 1, provide a unified framework for understanding this structural evolution and reveal the emergence of a previously unrecognized modulated twist-bend smectic phase, $SmC_{TB\text{-}mod}$. In this phase, the established four-layer heliconical organization develops an additional long-period modulation, giving rise to a higher-order helical superstructure. Resonant soft X-ray scattering and theoretical analysis show that this hierarchy emerges from frustration between competing interlayer packing tendencies, providing a new example of how complex multilevel organizations can spontaneously arise in soft-matter systems.

## Results

Liquid crystalline phases formed by the studied mesogens were identified by a combination of experimental techniques including polarized-light optical microscopy and x-ray diffraction, both conventional and resonant. The resonant soft x-ray scattering (RSoXS), with beam energy tuned to the carbon absorption edge, is uniquely sensitive to orientational modulations of molecular ordering and therefore particularly suited for studies of heliconical and other complex liquid-crystalline structures [15-18]. By probing variations in molecular orientation, rather than electron density alone, RSoXS provides direct access to the periodicity of the structures defined solely by orientation of molecules.

All four of the studied homologous series exhibit the same fundamental trends; therefore we will focus on the CT4O2Me.*m* series as a representative model system, which captures the essential features observed throughout the entire family, while the remaining series are discussed in terms of their similarities and deviations from this general behavior. On cooling, the shortest homologues in the CT4O2Me.*m* series exhibit the phase sequence N → $N_{TB}$ → $SmC_{TB\text{-}C}$, where the $SmC_{TB\text{-}C}$ phase is a tilted intercalated smectic with layer spacing equal half the molecular length (Fig. S1), each sublayer made of mesogenic cores is separated by alkyl sublayer made of terminal and spacer chains. The thermal stability of this intercalated smectic phase reaches a maximum for the homologue with $m = 4$. This can be attributed to the close correspondence between the lengths of the terminal alkyl chains and the flexible spacer, which optimizes molecular packing within the intercalated arrangement [19]. A locally intercalated structure is observed also for the nematic and twist-bend nematic phases of short homologues (Fig. S1). The helical pitch in the $N_{TB}$ phase determined by RSoXS has a typical temperature dependence, the pitch is close to 4 molecular lengths at low temperatures but critically unwinds on approaching the $N_{TB}$-N phase transition (Fig. 2a,d) [20,21].

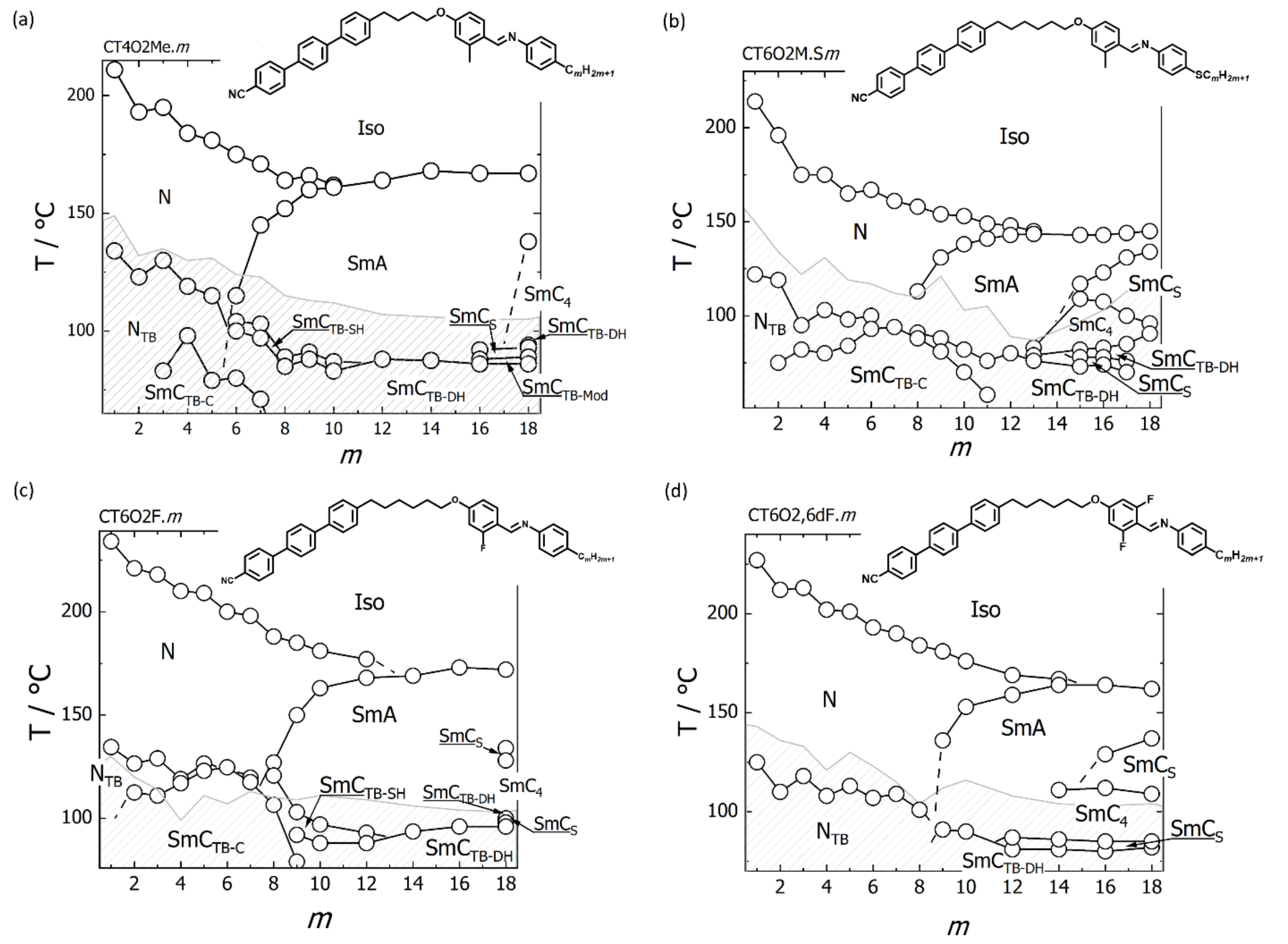

**Figure 1**. **Phase behavior of the four studied homologous series**. Phase diagrams plotted as a function of the number of carbon atoms in the molecular terminal chain, *m*, for dimeric molecules of (a) CT4O2Me.*m*, (b) CT602Me.S*m*, (c) CT6O2F.*m* and (d) CT6O2,6dF.*m* homologue series. Molecular structures are given above each panel. The $SmC_{TB}$-type phases exhibit helical structures, the $SmC_S$ phase is synclinic, whereas the $SmC_4$ phase is characterized by alternating synclinic and anticlinic interlayer arrangements. Phase transition temperatures were determined mainly from DSC and optical birefringence measurements, supported by observation of optical textures, however for some of the longest homologues the precise temperature ranges and structural characteristics of the phases formed could not always be determined unambiguously, due to their short temperature ranges and similarity of optical textures. The $SmC_{TB\text{-}Mod}$ phase has been indicated only for these compounds for which RSoXS experiments proved its presence unumbigously.

A distinct change in phase behavior, from an intercalated packing regime to a bilayer regime, occurs for homologues with terminal chains $m = 6 - 14$. Within this regime, the nematic phase is gradually replaced by the smectic A phase and the twist-bend nematic phase is replaced by a sequence of heliconical smectic phases, yielding the typical phase sequence SmA → $SmC_{TB\text{-}SH}$ → $SmC_{TB\text{-}DH}$.. Smectic phases, studied by conventional x-ray diffraction, show electron density modulation with periodicity close to twice the molecular length, indicative of antiparallel pairing of neighboring dimers in successive molecular layers; the repeating unit is composed of two molecular layers (Fig. S2). In RSoXS experiments the $SmC_{TB\text{-}SH}$ phase exhibits a characteristic resonant x-ray reflection corresponding to a four-molecular-layer periodicity, at $q_4 = 2\pi/(4d_0)$, where $d_0$ is the molecular layer thickness (Fig. 2). Throughout this work, resonant reflections are labeled according to their positions in reciprocal space; we therefore refer to this signal as the $\mathbf{q_4}$ reflection [13]. The observation of the $\mathbf{q_4}$ reflection, together with its biaxial optical character – evidenced by a schlieren texture observed when light propagates along the layer normal – and a decrease of optical birefringence compared to the preceding SmA phase when the light propagates perpendicular to layer normal (Fig. S3), demonstrates that the four-layer heliconical unit in $SmC_{TB-SH}$ phase is distorted from an ideal clock-structure. In the $SmC_{TB-DH}$ phase, the $\mathbf{q_4}$ refection splits symmetrically into two reflections located at $\mathbf{q_4}+\mathbf{q_m}$ and $\mathbf{q_4}-\mathbf{q_m}$ (Fig. 2b,e), revealing the emergence of an additional long-wavelength rotation of the four-layer structural unit [13] - the phase is thus an analogue of chiral $SmC_{FI2}$ phase [22], but we must emphasize that our present system is built from achiral molecules. Upon cooling, the separation between the two resonant peaks increases, indicating a progressive shortening of the pitch ($p = \frac{2\pi}{q_m}$) associated with this secondary modulation. The $SmC_{TB-DH}$ phase is optically uniaxial because the superimposed helix produces an azimuthal rotation of the molecular tilt direction, averaging the in-plane optical anisotropy. As the terminal alkyl chain becomes longer, the temperature range of the $SmC_{TB-SH}$ phase gradually narrows. Consequently, for the CT4O2Me.$m$ homologues with $m = 12$ and 14, this phase disappears and a direct SmA → $SmC_{TB-DH}$ transition takes place (Fig. 2c,f).

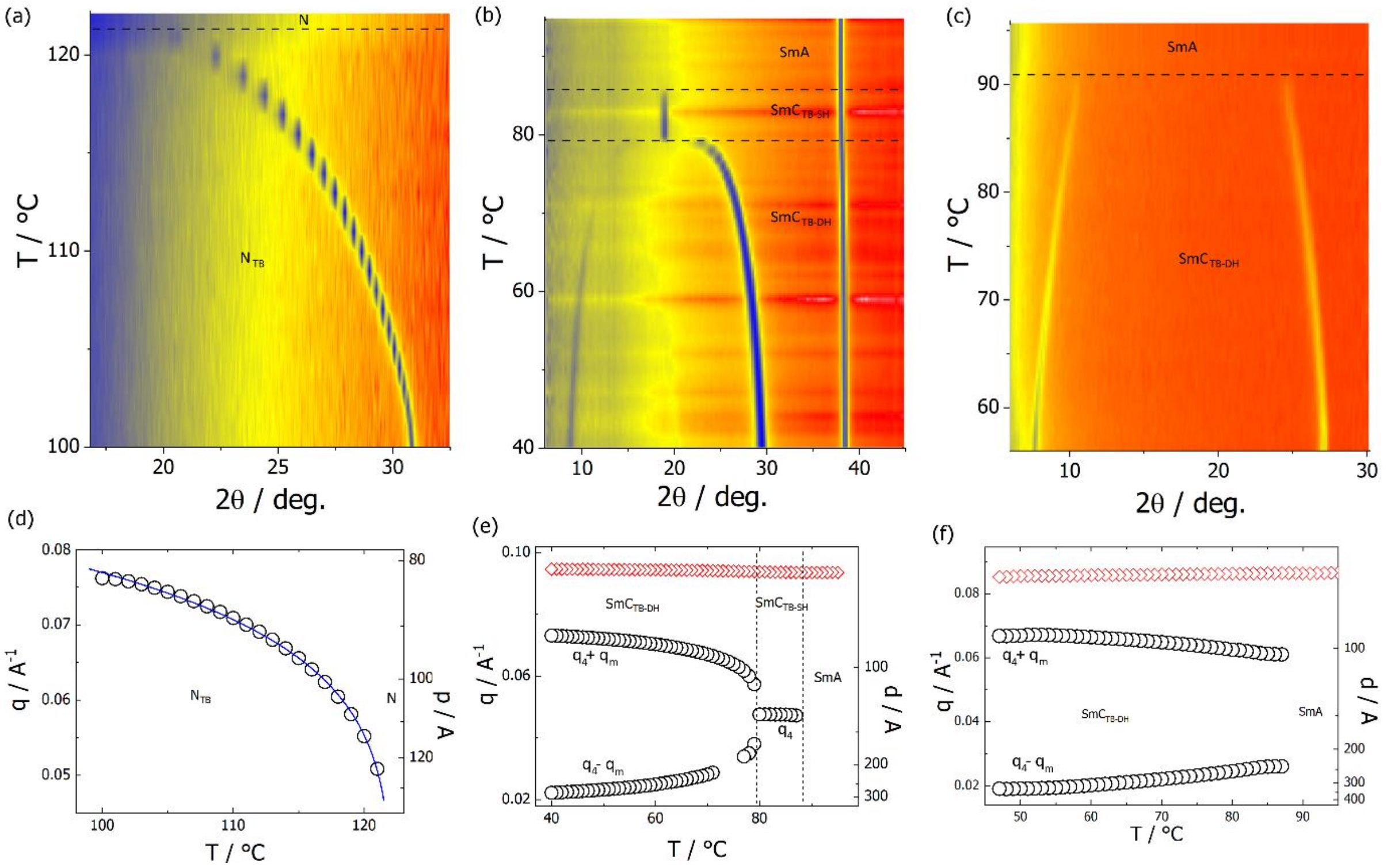

**Figure 2. RSoXS data.** Temperature evolution of the resonant x-ray diffraction patterns for (a) CT4O2Me.4 in the $N_{TB}$ phase and (b) CT4O2Me.10 across the SmA-$SmC_{TB-SH}$ and $SmC_{TB-DH}$ phase sequence. The signal observed at diffraction angle of approximately 38 deg. (nearly insensitive to the temperature) is non-resonant and corresponds to the electron density modulation associated with the bilayer periodicity of the smectic phases and (c) CT4O2Me.14 showing a SmA - $SmC_{TB-DH}$ phase sequence. Panels (d)-(e) show the positions of the diffraction signals recalculated to the scattering vector magnitude $q$, and the corresponding real-space periodicity, $d$. The $\mathbf{q_4}$ reflection corresponds to a four-layer periodicity; the increase in the splitting of the resonant signals $\mathbf{q_4}+\mathbf{q_m}$ and $\mathbf{q_4}-\mathbf{q_m}$ in the $SmC_{TB-DH}$ phase indicates a decreasing pitch of the secondary helix.

The longest homologues ($m = 16$ and 18) define the third region in the phase diagram, in which alkyl chain segregation becomes the dominant factor governing self-assembly. X-ray diffraction measurements indicate a bilayer organization throughout all smectic phases in this regime (Fig. S4). Detailed RSoXS studies, performed for the homologue with $m = 18$, identified a sequence of 5 tilted smectic phases below the SmA phase (Fig. 3a,b), with the lowest temperature phase being the $SmC_{TB-DH}$ phase, as in the shorter homologues.

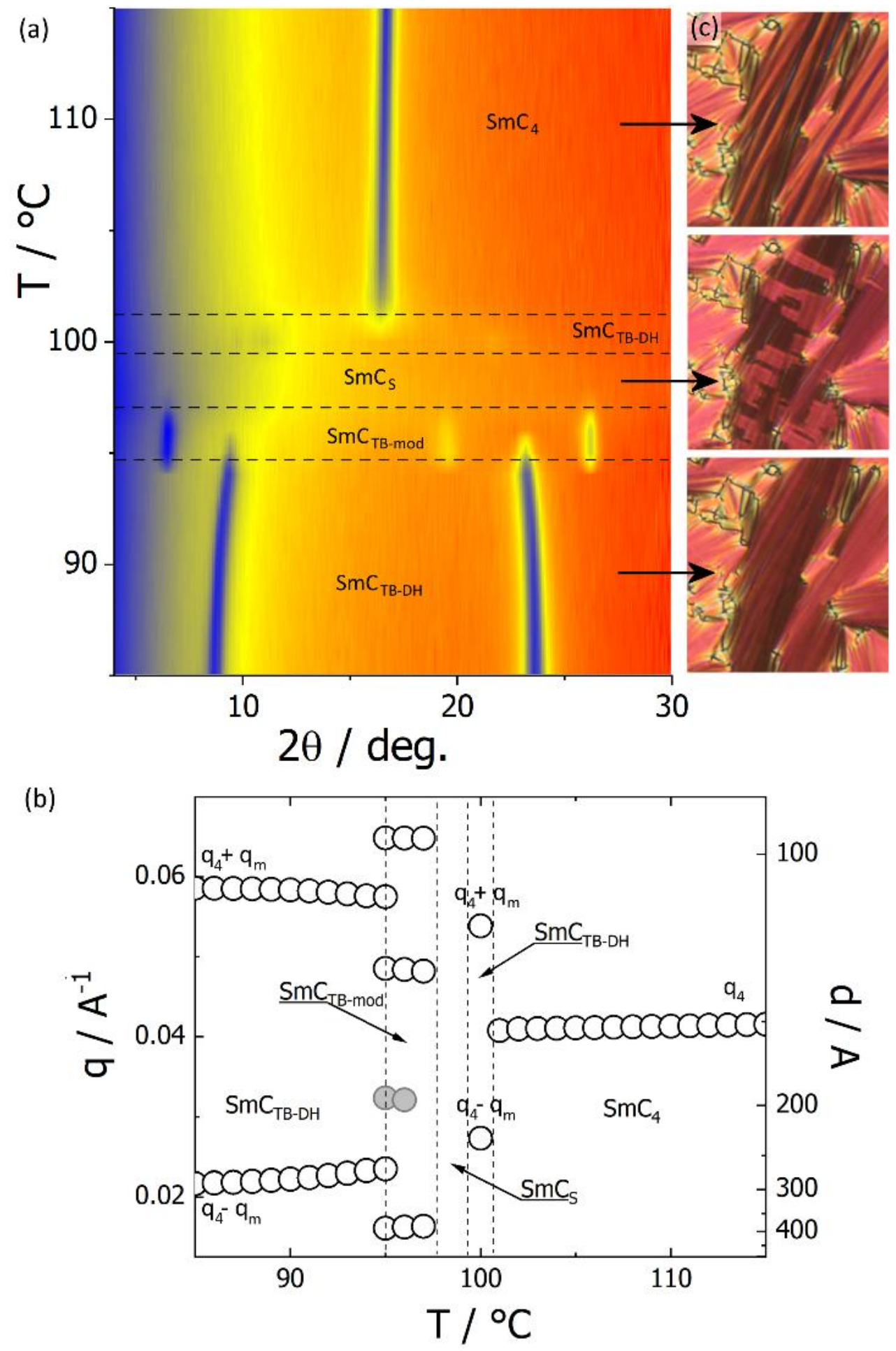

**Figure 3. Phase sequence for CT4O2Me.18.** Temperature evolution of (a) the RSoXS pattern and (b) corresponding diffraction peak positions. In the $SmC_4$ phase, the resonant diffraction signal corresponds to a four-layer periodicity. Upon transition to the $SmC_{TB-DH}$ phase, the resonant signal splits into two reflections $\mathbf{q_4}+\mathbf{q_m}$ and $\mathbf{q_4}-\mathbf{q_m}$. The resonant signals disappear in the $SmC_S$ phase. In the $SmC_{TB-Mod}$ phase, four resonant peaks emerge, positioned symmetrically around the $\mathbf{q_4}+\mathbf{q_m}$ and $\mathbf{q_4}-\mathbf{q_m}$ reflections. Upon entering the lowest temperature $SmC_{TB-DH}$ phase, the signals $\mathbf{q_4}+\mathbf{q_m}$ and $\mathbf{q_4}-\mathbf{q_m}$ reappear. (c) Optical textures observed for selected smectic phases. The temperature range over which the resonant signal disappears completely corresponds to the optical texture in which tilt domains can be clearly distinguished, the $SmC_4$ and $SmC_{TB-DH}$ phases are optically uniaxial with optical axis oriented along layer normal.

In the phase formed directly below the SmA phase, resonant scattering measurements revealed the signal related to a four-layer periodicity. However, the onset temperature of this phase, which we have termed here $SmC_4$, is substantially higher than that of the $SmC_{TB-SH}$ phase found in shorter homologues (Fig. 1), indicating that the underlying mechanisms responsible for four-layer periodicity in the $SmC_4$ and $SmC_{TB-SH}$ phases are distinct. While the $SmC_{TB-SH}$ phase can transform continuously into the double-helical $SmC_{TB-DH}$ phase [23], as observed for the shorter homologues, the evolution from the $SmC_4$ phase to the stable $SmC_{TB-DH}$ phase in homologue $m = 18$ proceeds through several intermediate states. Although immediately below the $SmC_4$ phase the resonant peak associated with the four-layer periodicity splits into a pair of satellites, signaling the narrow temperature range (~1K) of $SmC_{TB-DH}$, upon further cooling the resonant scattering signals disappear. In this temperature range, optical texture with tilt domains is formed in planar cells, characteristic for synclinic $SmC_S$ phase (Fig. 3c). At further cooling, in the phase named $SmC_{TB\text{-}mod}$, resonant x-ray diffraction signals reappear, but with a fundamentally different pattern. Two pairs of reflections are observed, one located around $\mathbf{q_4}+\mathbf{q_m}$ and the other around $\mathbf{q_4}-\mathbf{q_m}$, producing a total of four resonant peaks symmetrically distributed around the original $\boldsymbol{q}_4$ peak of the $SmC_4$ phase. The outer satellites exhibit substantially larger intensity than the inner ones. Finally, this complex diffraction pattern disappears, and the characteristic scattering signature of the $SmC_{TB\text{-}DH}$ phase is recovered, changing only weakly on further cooling. The unusual resonant diffraction pattern with four peaks observed in the intermediate $SmC_{TB\text{-}mod}$ phase cannot be explained within the framework of the previously established double-helical structure.[13] The diffraction pattern therefore reveals a new level of hierarchical organization superimposed on the already existing double-helical structure. The $SmC_{TB\text{-}mod}$ phase has been also observed in homologue $m = 16$, however in this homologue the $SmC_4$ phase is missing, the observed sequence is SmA-$SmC_S$-$SmC_{TB\text{-}mod}$-$SmC_{TB\text{-}DH}$ (Fig. S5).

A similar overall evolution of phase behavior with increasing *m* has been observed for several related homologous series of bent liquid-crystalline dimers (Fig. 1), suggesting that the sequence of structural transformations reflects a general consequence of extending terminal chain length. In all of these systems, the phase diagrams can be divided into three regions characterized by different balance between interlayer coupling, molecular tilt organization, and nanosegregation. A notable difference, however, is that whereas in the CT4O2Me.*m* series the $SmC_4$ phase emerges directly from the orthogonal SmA phase, in materials of other studied homologue series it is preceded by a synclinic $SmC_S$ phase, characterized by well-developed tilt domains in optical textures observed in planar-aligned cells (Fig. 4a).

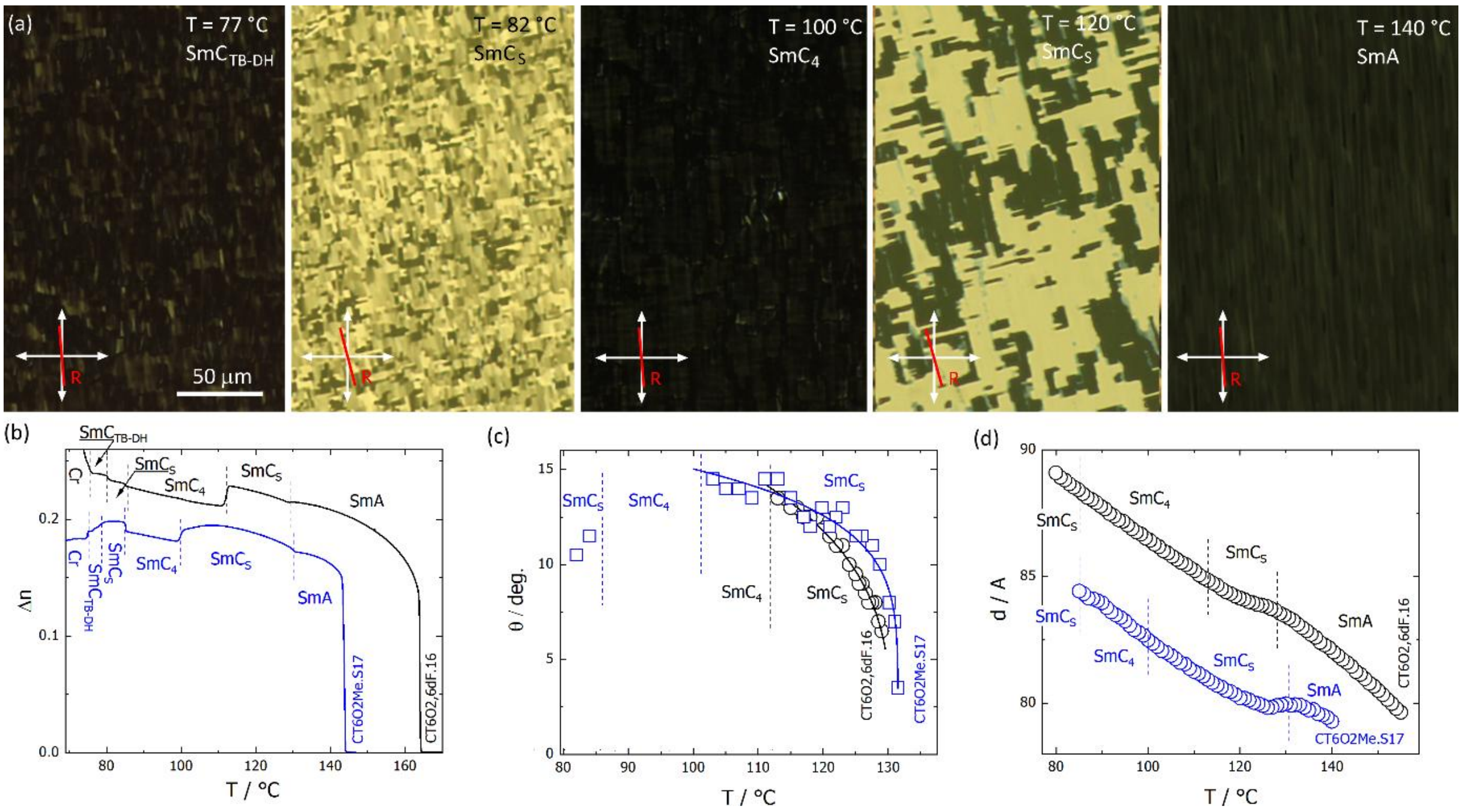

**Figure 4. Phase sequence and physical properties of tilted phases.** (a) Optical textures observed between crossed polarizers for a sequence of smectic phases: SmA, SmCs, $SmC_4$, reentrant $SmC_S$ and $SmC_{TB\text{-}DH}$ for the CT6O2Me.S17 homologue. In the $SmC_S$ phases the sample was rotated to obtain light extinction in one of the tilted domains, whereas in the $SmC_4$ and $SmC_{TB\text{-}DH}$ phases extinction occurs when the rubbing direction (layer

normal) is aligned with the polarizer axis. Temperature dependence of (b) birefringence, (c) tilt, and (d) layer spacing for the CT602Me.S17 (blue) and CT6O2,6dF.16 (black) compounds.

In all studied series the lowest temperature smectic phase formed by long homologues is $SmC_{TB\text{-}DH}$, and the evolution from the $SmC_4$ phase again involves several intermediate structures, however it was not possible to unambiguously determine whether the exact path is the same as observed for CT4O2Me.*m* series. Nevertheless, it should be noted that the tilt angle magnitude, as well as smectic layer thickness evolves smoothly upon cooling through the sequence of phases (Fig. 4c,d), from SmA to $SmC_{TB\text{-}DH}$, suggesting that all the phase transitions are driven by re-arrangement of the azimuthal tilt directions in consecutive molecular layers. Additional structural information about intermediate phases can be revealed from optical studies. $SmC_{TB\text{-}DH}$ and $SmC_{TB\text{-}mod}$ are optically uniaxial and thus give homeotropic textures under the polarized optical microscope, while in the $SmC_{TB\text{-}SH}$, $SmC_S$ and $SmC_4$ phases, schlieren textures evidence optical biaxiality. In wedge cells with homeotropic anchoring condition, the $SmC_S$ phase forms a regular array of stripes originating from the discrete variation in the number of smectic layers across the continuously varying cell thickness (Fig. 5a).

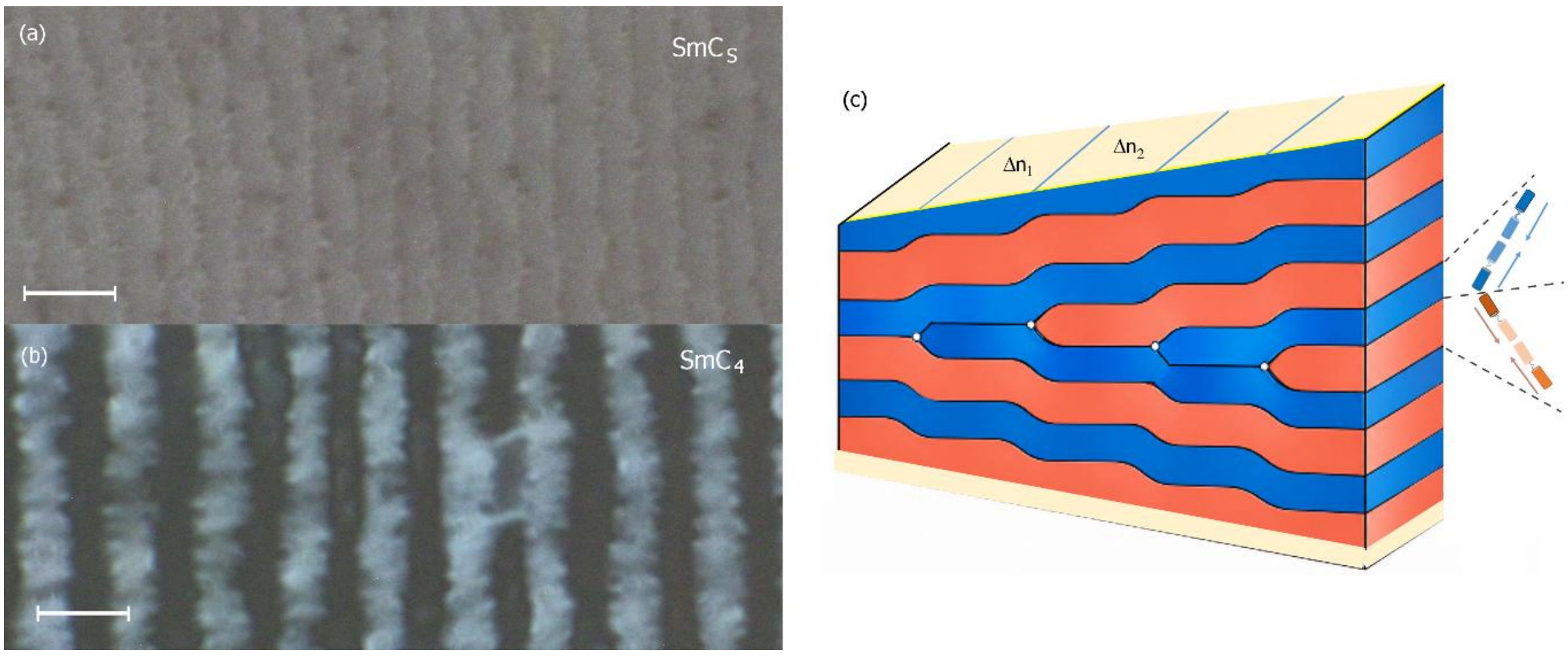

**Figure 5. Optical textures observed in a wedge cell.** Optical textures observed between crossed polarizers in a wedge cell ($\tan\alpha = 0.001$) with homeotropic anchoring for compound CT6O2,6dF.16 in (a) $SmC_S$ and (b) $SmC_4$ phases. Scale bars correspond to 20 μm. The dark lines correspond to edge dislocations. In both phases the spacing between neighboring dislocations corresponds to a cell-thickness change of approximately 9 nm, equal to a bilayer period. In SmCs phase, neighboring terraces exhibit essentially identical birefringence, whereas in the $SmC_4$ phase alternation of high- and low-birefringence terraces is observed. (c) Schematic illustration of the wedge-cell geometry. In both the $SmC_4$ and $SmC_s$ phases, each edge dislocation corresponds to the insertion of one molecular bilayer. In the $SmC_4$ phase, with a four-layer basic unit, successive bilayers are characterized by opposite tilt orientations, indicated by the blue and orange colors. Regions containing an integer number of four-layer units can adopt the thermodynamically preferred structure (for example, an *all-in-one-plane-like* arrangement), whereas regions that do not contain an integer number of four-layer repeats may experience structural frustration, and consequently exhibit different birefringence, $\Delta n$.

Thickness accommodation occurs through the formation of edge dislocations [24,25], here associated with the molecular bilayer, such that neighboring terraces differ by an integer number of bilayer units. All terraces exhibit essentially identical birefringence and a nearly uniform azimuthal orientation of the molecular tilt. Stripe textures in wedge cells are also observed for both the $SmC_4$ and $SmC_{TB\text{-}SH}$ phases; despite the basic four-layer periodicity in these phases the dislocations again correspond to the addition or removal of molecular bilayers. In contrast to the $SmC_S$ phase however, neighboring terraces exhibit alternating high and low birefringence (Fig. 5b). Depending on the sample preparation, alignment quality, and equilibration time, the birefringence in every second stripe may either decrease nearly to zero or remain only partially reduced. These observations indicate that neighboring terraces differing by only two molecular layers are not optically equivalent. Although the birefringence modulation observed in wedge cells is clearly induced by confinement, its microscopic origin remains unresolved. While the underlying dislocation structure is defined by the bilayer periodicity, the optical response remains sensitive to the four-layer structural repeating unit. Regions containing an integer number of four-layer units can adopt the thermodynamically-preferred structure (e.g. *all-in-one-plane-like* or *clock-like*), whereas regions that do not contain a complete number of four-layer repeats experience structural frustration. If molecular orientation remains continuous at the confining surfaces, terraces whose thickness is incompatible with an integer number of four-layer repeats cannot preserve the preferred interlayer registry throughout the sample. The resulting frustration may be relieved through a gradual redistribution of the azimuthal orientation between neighboring layers, causing the molecular arrangement near the dislocations to acquire a more clock-like character. Consequently, the two registries could generate distinct surface-induced distortions of the four-layer structure and therefore exhibit different optical birefringence.

**Discussion**

Our starting point in analyzing the RSoXS patterns of new phases is the structural framework developed previously for the interpretation of resonant diffraction in twist-bend smectic phases [13,14]. These phases are smectic analogues of the twist-bend nematic phase and originate from the tendency of bent molecular dimers to form heliconical orientational structures. In the simplest $SmC_{TB\text{-}SH}$ phase, the molecular tilt directions form a distorted four-layer clock arrangement (Fig. 6a,b), in which the azimuthal orientation of the tilt rotates from layer to layer. The interlayer azimuthal increments alternate between $\delta$ and $\pi - \delta$, so that an ideal four-layer clock structure corresponds to $\delta = \pi/2$, and an ideal all-in-one-plane structure to $\delta = 0$, deviations of $\delta$ describe distortions of the four-layer helical unit. In the $SmC_{TB\text{-}DH}$ phase, this four-layer motif is preserved, but an additional long-wavelength helical modulation develops, with an angle $\varepsilon$ describing the gradual rotation of successive four-layer units about the layer normal (Fig. 6c).The resulting structure can therefore be regarded as a superposition of two helical modulations: a short-pitch four-layer modulation governed by $\delta$ and a long-pitch modulation governed by $\varepsilon$. This hierarchical double-helical organization reproduces the characteristic pair of resonant x-ray diffraction peaks observed in a wide range of twist-bend smectic systems, and accounts for optical uniaxiality of the $SmC_{TB\text{-}DH}$ phase. The systematic evolution of the phase diagram with increasing terminal alkyl-chain length suggests that the interactions stabilizing the heliconical arrangement become progressively weaker as self-segregation between terminal chains and mesogenic cores becomes more dominant. Longer terminal chains increase the flexible, non-mesogenic fraction of the molecule and reduce the relative influence of the bent aromatic core on collective orientational ordering. Consequently, the tendency toward long-range heliconical organization competes increasingly with more conventional synclinic and anticlinic interlayer arrangements, and in the longest homologues the latter prevails, leading to newly observed $SmC_4$ and $SmC_{TB\text{-}mod}$ phases. We therefore propose that the $SmC_4$ phase has an *all-in-one-plane* structure comprising alternating synclinic and anticlinic interfaces between successive layers.

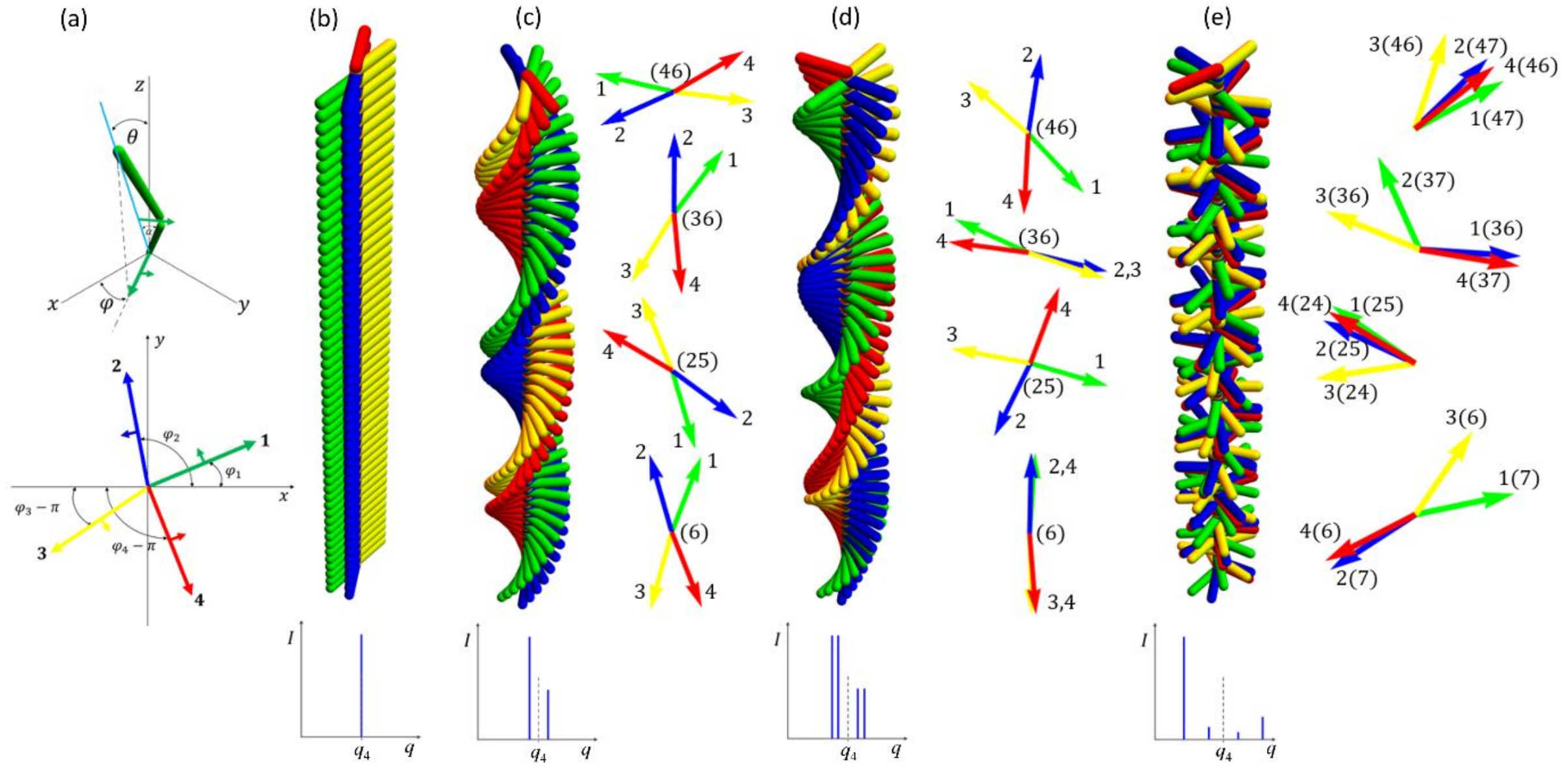

**Figure 6: Heliconical structure of $SmC_{TB\text{-}SH}$, $SmC_{TB\text{-}DH}$, and $SmC_{TB\text{-}mod}$.** (a) Description of molecular orientation in smectic layers. The axis connecting the termini of a dimer molecule is tilted by an angle $\theta$ with respect to the smectic layer normal (along $z$ axis). The direction of its projection vector onto the smectic plane $(xy)$ is defined by the angle $\varphi$. The apex angle of the dimer is $\alpha$. Orientation of the molecular tip is indicated by the short arrow perpendicular to the projection vector and may be identical in all layers (the homochiral case shown here) or alternate between adjacent layers; angles $\varphi_1$ to $\varphi_4$ describe tilt direction in consecutive layers and are defined in the text. (b-d) Projection vectors on the layer plane, colors denote the vector projections in consecutive layers of four layer unit cell. (b) $SmC_{TB\text{-}SH}$ phase with a 4-layer pitch ($C_2 = 0.6$, $C_3 = \pi$, $C_4 = \pi + 0.6$, $\varepsilon = \gamma = 0$). Only a single peak is observed in resonant XRD, at $q_4 = 2\pi/(4d_0)$, where $d_0$ is the thickness of one smectic layer. (c) $SmC_{TB\text{-}DH}$ phase, having a double-helical structure with a long, secondary helix superimposed on the shorter ($C_2 = 0.6$, $C_3 = \pi$, $C_4 = \pi + 0.6$, $\varepsilon = 0.05$, $\gamma = 0$), four-layer helix in consecutive are shown for the four-layer stacks with $j = 6, 25, 26, 46.$ In $SmC_{TB\text{-}DH}$ the basic four-layer unit rotates while preserving its distorted clock like structure. In this phase the $q_4$ peak splits into two signals whose relative intensities depend on the sign and magnitude of the apex angle and the tilt $\theta$ [13] (d) $SmC_{TB\text{-}mod}$ phase with structure analogous to the $SmC_{TB\text{-}DH}$ phase, but with the basic four-layer unit periodically modulated along the secondary helix. At a chosen set of parameters ($C_2 = 0.6$, $C_3 = \pi$, $C_4 = \pi + 0.6$, $\varepsilon = 0.05$, $\gamma = 0.05$), the four-layer unit structure varies between all-in-one plane and clock-like configurations. This additional modulation leads to four resonant reflections positioned symmetrically around $q_4$ reflection. (e) $SmC_{TB\text{-}mod}$ phase modeled with parameters $C_2 = C_4 = 0$ and $C_3 = \pi$ , $\varepsilon = 0.62$ and $\gamma = 1.14$ that reproduce main features of the experimental resonant XRD pattern, namely the much higher intensity of the lowest $q$-peak compared with the highest-$q$ peak and the reduced intensity of the inner peaks, for this set of parameter the four-layer unit structure varies between a nearly anticlinic and a nearly synclinic configuration along the secondary helix.

The $SmC_{TB\text{-}mod}$ phase, characterized by a complex RSoXS pattern, may then be viewed as an intermediate state between the heliconical $SmC_{TB\text{-}DH}$ phase and the *all-in-one-plane* $SmC_4$ structure. In this picture, competing ordering tendencies cause the four-layer unit itself to vary periodically along the layer-stacking direction: neighboring four-layer units are no longer identical but evolve continuously between configurations with different degrees of synclinic and anticlinic character. Such a periodically varying four-layer motif naturally introduces an additional hierarchical length scale and provides a natural explanation for the observed splitting of the already split $\mathbf{q_4+q_m}$ and $\mathbf{q_4-q_m}$ resonant X-ray reflections. To describe this behavior, we extend the existing $SmC_{TB-DH}$ model by introducing a modulation parameter, $\gamma$, in addition to the conventional double-helix parameter $\varepsilon$ (details of the modelling are in Supplementary Information). Whereas $\varepsilon$ determines the gradual rotation of successive four-layer units, $\gamma$ describes a periodic modulation of the internal azimuthal structure of the four-layer building block itself. The resulting orientational structure therefore contains three hierarchical levels: the four-layer distorted-clock arrangement described by $\delta$, and the two longer-scale modulations governed by $\varepsilon$ and $\gamma$. The azimuthal orientations of the molecular tilt directions in the $j$-th four-layer stack are therefore written as $\varphi_1 = 4j\epsilon + j\gamma$, $\varphi_2 = C_2 + \epsilon + 4j\epsilon - j\gamma$, $\varphi_3 = C_3 + 2\epsilon + 4j\epsilon + j\gamma$ and $\varphi_4 = C_4 + 3\epsilon + 4j\epsilon - j\gamma$, where $C_i$, $i = 2,3,4$ are constants; for example, with $C_2 = \delta$, $C_3 = \pi$, $C_4 = \pi + \delta$ and $j = 0$, one obtains a four-layer structure of $SmC_{TB-SH}$ phase, shown in Figure 6b. With $\epsilon \neq 0$ and $\gamma = 0$, we have a structure of $SmC_{TB-DH}$ phase with a double twist (Fig. 6c), which leads to splitting of the RSoXS $\mathbf{q_4}$-peak into pair: $\mathbf{q_4}$+$\mathbf{q_m}$ and $\mathbf{q_4}$-$\mathbf{q_m}$. With $\gamma \neq 0$, however, we add an additional modulation along the smectic layer normal. For the chosen signs in front of $j\gamma$, the angle between the first and third layer and between the second and the fourth layer is fixed, but the molecular tilt rotate on the cone by $\gamma$ more (for layers 1 and 3) or less (for layers 2 and 4) than $4\epsilon$, when moving to the next set of four layers (Fig. 6d,e). The proposed structure gives splitting of the $\mathbf{q_4}$ peak into 4 satellite peaks, two pairs on each side of the position of the $\mathbf{q_4}$ peak. These peaks are positioned at

$$q = \frac{q_0}{4}\left(1 \pm \frac{2\epsilon}{\pi} \pm \frac{\gamma}{2\pi}\right) \ .$$

Values of parameters $\epsilon$ and $\gamma$ can be obtained from resonant diffraction peak positions, and for the homologue CT4O2Me.18 are found to be $\epsilon = 0.67$ and $\gamma = 1.14$. The intensity of the peaks strongly depends on the values of parameters $C_i$ and to some extent also on the apex angle $\alpha$ and the homo- or heterochirality of the layers. For all the materials studied the outer peaks with $q = \frac{q_0}{4}\left(1 + \frac{2\epsilon}{\pi} + \frac{\gamma}{2\pi}\right)$ and $q = \frac{q_0}{4}\left(1 - \frac{2\epsilon}{\pi} - \frac{\gamma}{2\pi}\right)$ are much stronger than two inner peaks. This pronounced intensity asymmetry therefore provides an important additional constraint on the structural model beyond the peak positions alone. We find that such an intensity distribution can be reproduced only when layers 1 and 3 are coupled and rotate by $4\epsilon + \gamma$, whereas layers 2 and 4 are coupled and rotate by $4\epsilon - \gamma$ from one four-layer stack to another (as opposed to layers 1 and 2 being coupled).

## Conclusions

The structural evolution observed in bent liquid-crystalline dimers demonstrates that the phase sequence is governed by a reorganization of the azimuthal directions of molecular tilt. Across the phase sequence, the azimuthal orientation of the tilt cone exhibits a four-layer periodicity. This periodicity constitutes a common structural framework for the different phases; however, the organization of the tilt in this basic unit changes with the balance of intermolecular interactions. The $SmC_{TB-SH}$ phase can be described in terms of a distorted four-layer clock arrangement, whereas $SmC_{TB-DH}$ retains this short-period modulation and additionally develops a longer-wavelength helical modulation superimposed on the four-layer units. As shown previously, progressive shortening of the secondary pitch in the $SmC_{TB-DH}$ phase upon cooling ultimately drives the system toward an anticlinic arrangement of molecules in consecutive layers. Such continuous evolution appears to be a general behavior of a system with an intrinsic tendency toward bend deformation, whereas transitions between structures with different tilt organizations in rod-like mesogens typically occur as first-order phase transition.

However, with increasing terminal chain length, the tendency toward twist-bend organization progressively weakens, while conventional synclinic and anticlinic interlayer arrangements become increasingly important. In long chain homologues, this is reflected in the formation of either a phase with synclinic tilt between consecutive layers, $SmC_S$, or the $SmC_4$ phase, characterized by alternating synclinic and anticlinic interfaces. Within this framework, the newly identified $SmC_{TB\text{-}mod}$ phase can be understood as a direct consequence of the competition between these two ordering tendencies: the formation of a synclinic/anticlinic structure and the tendency toward twist-bend organization. As a result, the system adopts neither a fully developed heliconical arrangement nor a fixed synclinic/anticlinic registry. Instead, it retains the fundamental four-layer building block while periodically modifying its internal azimuthal organization. In other words, the four-layer unit itself becomes modulated along the layer-stacking direction. Neighboring four-layer units therefore are not required to be structurally identical, but can continuously evolve between configurations with different degrees of synclinic and anticlinic character. This provides a physical picture of $SmC_{TB\text{-}mod}$ as a frustrated intermediate structure, in which competing ordering tendencies are accommodated through modulation rather than by selecting a single interlayer arrangement. The additional modulation introduces a new hierarchical level of structural organization and an associated length scale and naturally accounts for the characteristic splitting of the resonant reflections into four satellite peaks symmetrically distributed around the $q_4$ reflection associated with the four-layer periodicity. The resulting structure thus combines the underlying four-layer periodicity with longer-scale modulations, providing a consistent description of the complex diffraction patterns observed experimentally.

## Materials and Methods

**Organic synthesis and analytical data** for the studied materials are given in Supplementary information.

**Optical studies.** Phase characterization was performed by polarized optical microscopy (POM) using a Zeiss AxioImager A2m microscope equipped with a Linkam T96 heating stage. Samples were sandwiched between glass plates to form films several μm thick. Selected samples were also prepared in commercial cells purchased from WAT, with a cell thickness ranging from 1.5 to 5 μm and treated for planar or homeotropic anchoring. Measurements of optical birefringence were performed with a set-up based on a photoelastic modulator (PEM-90, Hinds) working at a modulation frequency $f = 50\,\text{kHz}$. A halogen lamp (Hamamatsu LC8), equipped with a narrow bandpass filter (532 nm), was used as a light source. The intensity of transmitted light vs. time recorded with a photodiode (FLC Electronics PIN−20) was de-convoluted with a lock-in amplifier (EG&G 7265) into $1f$ and $2f$ components. The ratio of these components is proportional to the optical retardation.
The tilt angle was determined from the angular difference between azimuthal sample positions corresponding to light extinction in neighboring tilt domains observed in planar textures between crossed polarizers.

Wedge cells with homeotropic anchoring and defined inclination angle were purchased from WAT.

**Differential scanning calorimetry.** Studies were performed using Mettler-Toledo DSC1 or DSC3 instruments, fitted with an intracooler and calibrated using indium and zinc as standards. Heating and cooling rates were 10 K min−1, and all samples were measured under a nitrogen atmosphere. Transition temperatures and associated enthalpy changes were extracted from the second heating trace unless otherwise noted. For each sample, two aliquots were measured, and the data listed are the average of the two sets of data.

**X-ray diffraction.** 2D XRD patterns were obtained with the Bruker D8 GADDS system (CuKα radiation, Goebel mirror monochromator, 0.5 mm point beam collimator, VANTEC 2000 area detector), equipped with a modified Linkam heating stage. Samples were prepared as droplets on a heated surface. The temperature dependence of the smectic layer thickness was determined from the small-angle X-ray diffraction experiments performed with the Bruker D8 Discover system (CuKα line, Goebel mirror, Anton Paar DCS350 heating stage, scintillation counter) working in the reflection mode. Homeotropically aligned samples were prepared as a thin film on a silicon wafer. The small-angle X-ray diffraction (SAXS) patterns for powder samples were obtained with a Bruker Nanostar system using CuKα radiation, and patterns were collected with a VANTEC 2000 area detector. The temperature of the samples was controlled within ±0.1 K using an MRI heating stage.

**Resonant soft X-ray diffraction (RSoXS).** Studies were performed at the Advanced Light Source, Lawrence Berkeley National Laboratory on the soft X-ray beam line 11.0.1.2 and at the National Synchrotron Light Source II, Brookhaven National Laboratory on 7-ID-1 SST1 beam line. The energy of the incident beam was tuned to carbon K-edge absorption energy (283 eV). Samples with a thickness lower than 1 μm were placed on a transmission electron microscopy grid. The scattering intensity was recorded by 2D detectors.

**Acknowledgements**

The work was financed by the National Centre for Research and Development (Poland) through the project EIG CONCERT-JAPAN/9/89/FerroFluid/2023 (MM). The beamlines: 11.0.1.2 at the Advanced Light Source at the Lawrence Berkeley National Laboratory and 7-ID-1 SST1 at National Synchrotron Light Source II, Brookhaven National Laboratory are supported by the U.S. Department of Energy under Contract No. DE-AC02-05CH11231 and Contract No. DE-SC0012704, respectively.

**References**:

[1] G. M. Whitesides, B. Grzybowski, Self-assembly at all scales. *Science,* **295**, 2418-2421, (2002). https://doi.org/10.1126/science.1070821

[2] Y.-K. Kim, J. Noh, K. Nayani, N. L. Abbott, Soft matter from liquid crystals. *Soft Matter*, **15**, 6913–6929 (2019). https://doi.org/10.1039/C9SM01424A

[3] C. Tschierske, Microsegregation: From Basic Concepts to Complexity in Liquid Crystal Self-Assembly, *Isr. J. Chem.*, **52**, 935–959 (2012). https://doi.org/10.1002/ijch.201200053

[4] J. W. Goodby, E. J. Davis, R. J. Mandle, S. J. Cowling, Nano-Segregation and Directed Self-Assembly in the Formation of Functional Liquid Crystals. *Isr. J. Chem.* **52**, 863–880 (2012). https://doi.org/10.1002/ijch.201200059

[5] C. T. Imrie, P. A. Henderson, Liquid crystal dimers and higher oligomers: between monomers and polymers, *Chem. Soc. Rev*. **36**, 2096–2124 (2007). https://doi.org/10.1039/B714102E

[6] M. Cestari, E. Frezza, A. Ferrarini, G. R. Luckhurst, Crucial role of molecular curvature for the bend elastic and flexoelectric properties of liquid crystals: mesogenic dimers as a case study. *J. Mater. Chem*. **21**, 12303–12308 (2011). https://doi.org/0.1039/C1JM12233A

[7] C. T. Imrie, R. Walker, J. M. D. Storey, E. Górecka, D. Pociecha, Liquid Crystal Dimers and Smectic Phases from the Intercalated to the Twist-Bend. *Crystals*, **12**, 1245 (2022). https://doi.org/10.3390/cryst12091245

[8] Y. Takanishi, M. Toshimitsu, M. Nakata, N. Takada, T. Izumi, K. Ishikawa, H. Takezoe, J. Watanabe, Y. Takahashi, A. Iida, Frustrated smectic layer structures in bent-shaped dimer liquid crystals studied by x-ray microbeam diffraction. *Phys. Rev. E*, **74**, 051703 (2006). https://doi.org/10.1103/PhysRevE.74.051703

[9] I. Dozov, On the spontaneous symmetry breaking in the mesophases of achiral banana-shaped molecules. *Europhys. Lett*. **56**, 247–253 (2001). https://doi.org/10.1209/epl/i2001-00513-x

[10] V. Borshch, Y.-K. Kim, J. Xiang, M. Gao, A. Jákli, V. P. Panov, J. K. Vij, C. T. Imrie, M. G. Tamba, G. H. Mehl, O. D. Lavrentovich, Nematic twist-bend phase with nanoscale modulation of molecular orientation. *Nat. Commun*. **4**, 2635 (2013). https://doi.org/10.1038/ncomms3635

[11] D. Chen, J. H. Porada, J. B. Hooper, A. Klittnick, Y. Shen, M. R. Tuchband, E. Korblova, D. Bedrov, D. M. Walba, M. A. Glaser, J. E. Maclennan, N. A. Clark, Chiral heliconical ground state of nanoscale pitch in a nematic liquid crystal of achiral molecular dimers. *Proc. Natl. Acad. Sci. USA*, **110**, 15931–15936 (2013). https://doi.org/10.1073/pnas.1314654110

[12] J. P. Abberley, R. Killah, R. Walker, J. M. D. Storey, C. T. Imrie, M. Salamończyk, C. Zhu, E. Górecka, D. Pociecha, Heliconical smectic phases formed by achiral molecules, *Nat. Commun.* **9**, 228 (2018). https://doi.org/10.1038/s41467-017-02626-6

[13] M. Salamończyk, N. Vaupotič, D. Pociecha, R. Walker, J. M. D. Storey, C. T. Imrie, C. Wang, C. Zhu, E. Górecka. Multi-level chirality in liquid crystals formed by achiral molecules, *Nat. Commun.* **10**, 1922 (2019). https://doi.org/10.1038/s41467-019-09862-y

[14] A. Pearson, A. Alshammari, G. J. Strachan, M. Majewska, D. Pociecha, J. M. D. Storey, C. T. Imrie, N. Vaupotič, R. Walker, C. Zhu, M. Salamończyk, E. Górecka, Diverse helical structures made of achiral mesogenic dimers, *Nat. Commun.*, **17**, 6090 (2026). https://doi.org/10.1038/s41467-026-72565-8

[15] H. F. Gleeson, L. S. Hirst, Resonant X-ray Scattering: A Tool for Structure Elucidation in Liquid Crystals. *ChemPhysChem*, **7**, 321–328 (2006). https://doi.org/10.1002/cphc.200500426

[16] Y. Cao, M. Alaasar, A. Nallapaneni, M. Salamończyk, P. Marinko, E. Górecka, C. Tschierske, F. Liu, N. Vaupotič, C. Zhu, Molecular Packing in Double Gyroid Cubic Phases Revealed via Resonant Soft X-Ray Scattering, *Phys. Rev. Lett*. **125**, 027801 (2020). https://doi.org/10.1103/PhysRevLett.125.027801

[17] W. Lewandowski, N. Vaupotič, D. Pociecha, E. Górecka, L. M. Liz-Marzán, Chirality of Liquid Crystals Formed from Achiral Molecules Revealed by Resonant X-Ray Scattering, *Adv. Mater*. **32**, 1905591 (2020). https://doi.org/10.1002/adma.201905591

[18] Y. Cao, C. Feng, A. Jákli, C. Zhu, F. Liu, Deciphering chiral structures in soft materials via resonant soft and tender X-ray scattering, *Giant* **2**, 100018 (2020), https://doi.org/10.1016/j.giant.2020.100018

[19] A. F. Alshammari, D. Pociecha, R. Walker, J. M. D. Storey, E. Gorecka and C. T. Imrie, New patterns of twist-bend liquid crystal phase behaviour: the synthesis and characterisation of the 1-(4-cyanobiphenyl-4'-yl)-10-(4-alkylanilinebenzylidene-4'-oxy)decanes (CB10O.m), *Soft Matter,* **18,** 4679-4688 (2022). https://doi.org/10.1039/D2SM00162D

[20] C. Zhu, M. R. Tuchband, A. Young, M. Shuai, A. Scarbrough, D. M. Walba, J. E. Maclennan, C. Wang, A. Hexemer, N. A. Clark, Resonant carbon K-edge soft X-ray scattering from lattice-free heliconical molecular ordering: Soft dilative elasticity of the twist-bend liquid crystal phase, *Phys. Rev. Lett*. 116, 147803,(2016). https://doi.org/10.1103/PhysRevLett.116.147803

[21] E. Cruickshank, M. Salamończyk, D. Pociecha, G. J Strachan, J. M. D. Storey, C. Wang, J. Feng, C. Zhu, E. Gorecka, C. T. Imrie, Sulfur-linked cyanobiphenyl-based liquid crystal dimers and the twist-bend nematic phase, *Liq. Cryst*. **104**, 1595-1609, (2019). https://doi.org/10.1080/02678292.2019.1641638

[22] A. Cady, J. A. Pitney, R. Pindak, L. S. Matkin, S. J. Watson, H. F. Gleeson, P. Cluzeau, P. Barois, A. M. Levelut, W. Caliebe, J. W. Goodby, M. Hird, C. C. Huang, Orientational ordering in the chiral smectic-$C_{FI2}$ liquid crystal phase determined by resonant polarized x-ray diffraction, *Phys. Rev. E* **64**, 050702(R) (2001). https://doi.org/10.1103/PhysRevE.64.050702

[23] D. Pociecha, N. Vaupotič, M. Majewska, E. Cruickshank, R. Walker, J. M. D. Storey, C. T. Imrie, C. Wang, E. Gorecka, Photonic band gap in achiral liquid crystals – a twist on a twist, *Adv. Mater*. **33,** 2103288, (2021). https://doi.org/10.1002/adma.202103288

[24] C. Blanc, N. Zuodar, I. Lelidis, M. Kleman, J.-L. Martin, Defect dynamics in a smectic Grandjean-Cano wedge. *Phys. Rev. E,* **69**, 011705 (2004). https://doi.org/10.1103/PhysRevE.69.011705

[25] C. Zhang, A. M. Grubb, A. J. Seed, P. Sampson, A. Jákli, O. D. Lavrentovich, Nanostructure of Edge Dislocations in a Smectic-*C** Liquid Crystal, *Phys. Rev. Lett*. **115**, 087801 (2015). https://doi.org/10.1103/PhysRevLett.115.087801

Supporting Information:

# Frustrated Interlayer Coupling Creates a Higher-Order Helical Superstructure in a Twist-Bend Smectic Phase

Abigail Pearson[1], Damian Pociecha[2], Magdalena Majewska[2], Nataša Vaupotič[3,4], Rebecca Walker[1], Ewa Gorecka[2]

[1]School of Natural and Computing Sciences, University of Aberdeen, Aberdeen AB24 3UE, Great Britain

[2]Faculty of Chemistry, University of Warsaw, ul. Pasteura 1, 02-093 Warsaw, Poland

[3] Faculty of Natural Sciences and Mathematics. University of Maribor Maribor 2000, Slovenia

[4] Jozef Stefan Institute, Ljubljana 1000, Slovenia

## 1. Additional results

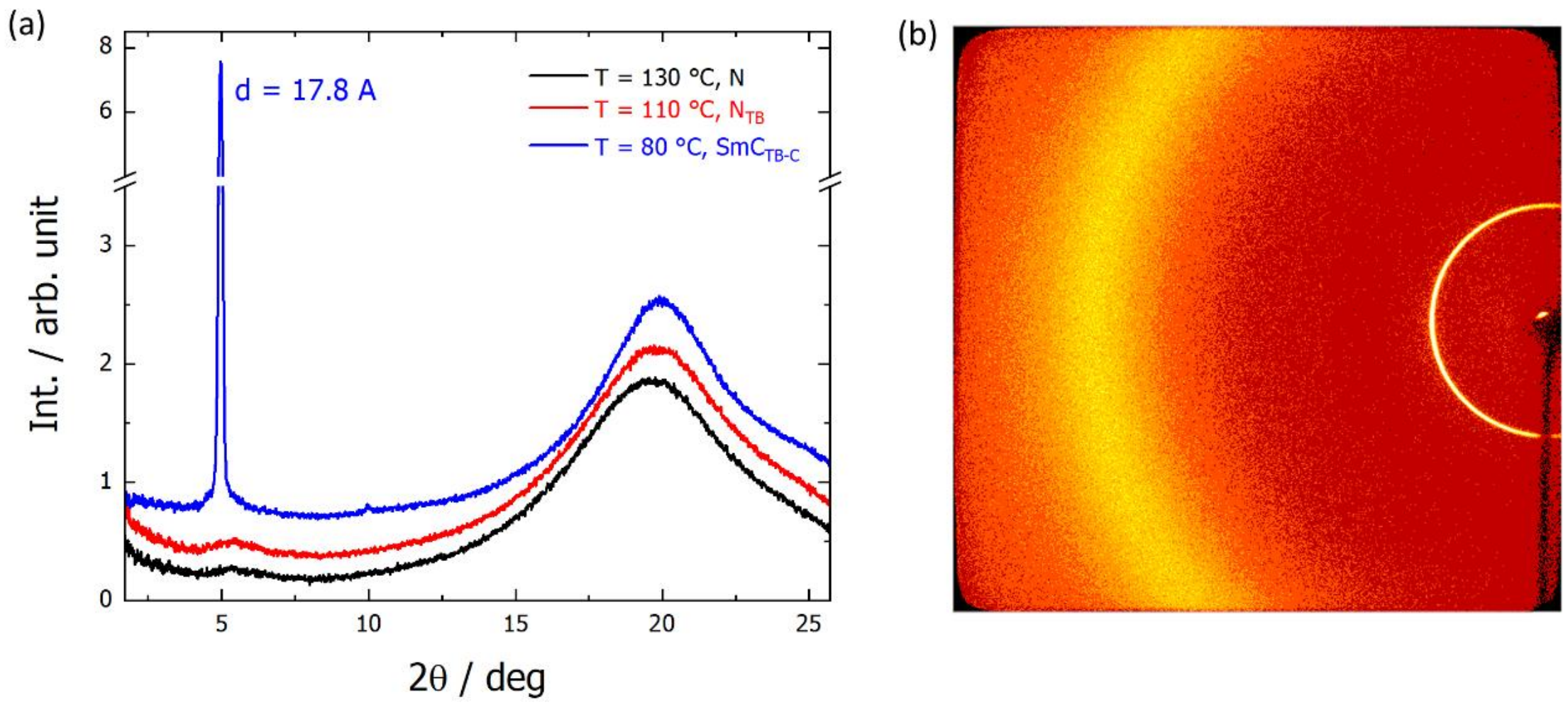

**Figure S1**: (a) X-ray diffractograms for compound CT4O2Me.5 in N, $N_{TB}$ and $SmC_{TB-C}$ phases. Layer spacing in $SmC_{TB-C}$ phase (d=17.8A) corresponds to half molecular length, evidencing intercalated lamellar structure. (b) 2D X-ray diffraction pattern recorded in $SmC_{TB-C}$ phase, diffused signal at high angle confirms liquid like order of molecules inside the layers.

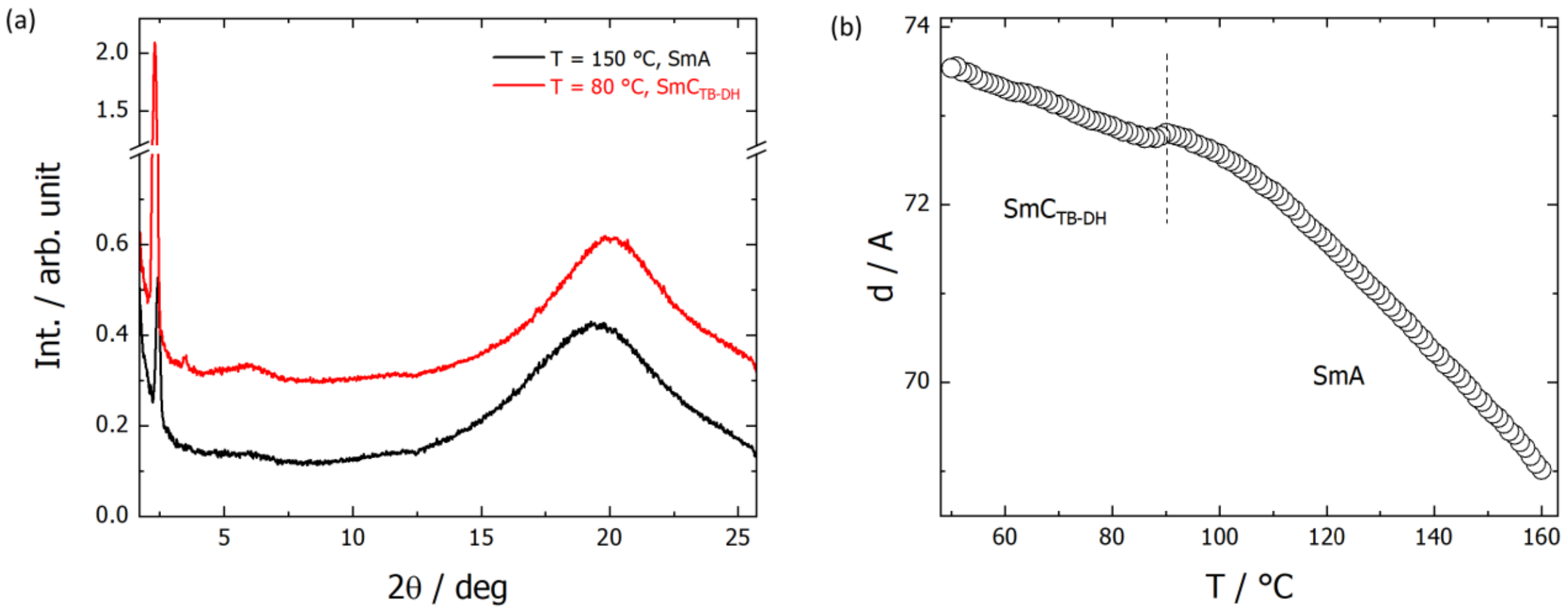

**Figure S2**: (a) X-ray diffractorams for compound CT4O2Me.14 in SmA, and SmC$_{TB-DH}$ phases. (b) Layer spacing in smectic phases measured as a function of temperature. In both phases layer signal corresponds to double molecular length, evidencing bilayer-type lamellar structure, which is realized due to antiparallel pairing of strongly polar cyanoterphenyl cores in successive molecular layers.

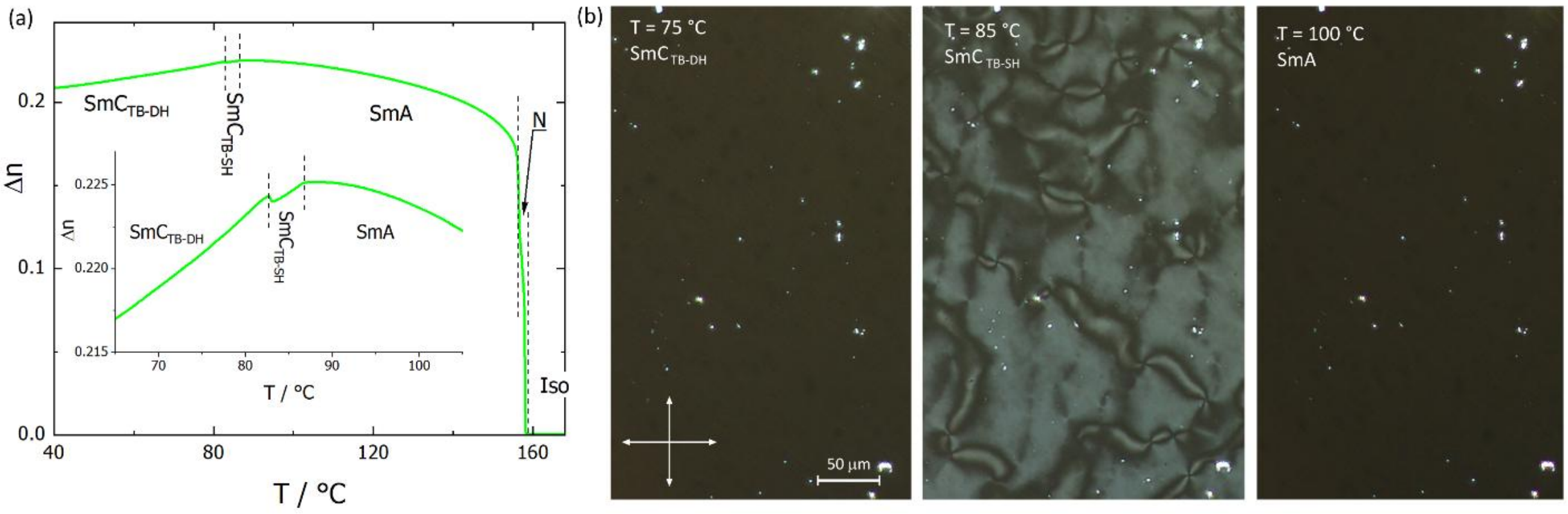

**Figure S3**: (a) Optical birefringence measured with green light (λ=532 nm) for compound CT4O2Me.10 with the phase sequence Iso - N - SmA - SmC$_{TB-SH}$ - SmC$_{TB-DH}$. In the inset engaged temperature range at SmA - SmC$_{TB-SH}$ - SmC$_{TB-DH}$ phase transitions (b) Optical textures recorded in one-free-surface sample of CT4O2Me.10, schlieren texture seen for SmC$_{TB-SH}$ phase evidences its biaxial character, all the other phases are optically uniaxial, thus exhibit homeotropic texture between crossed polarizers.

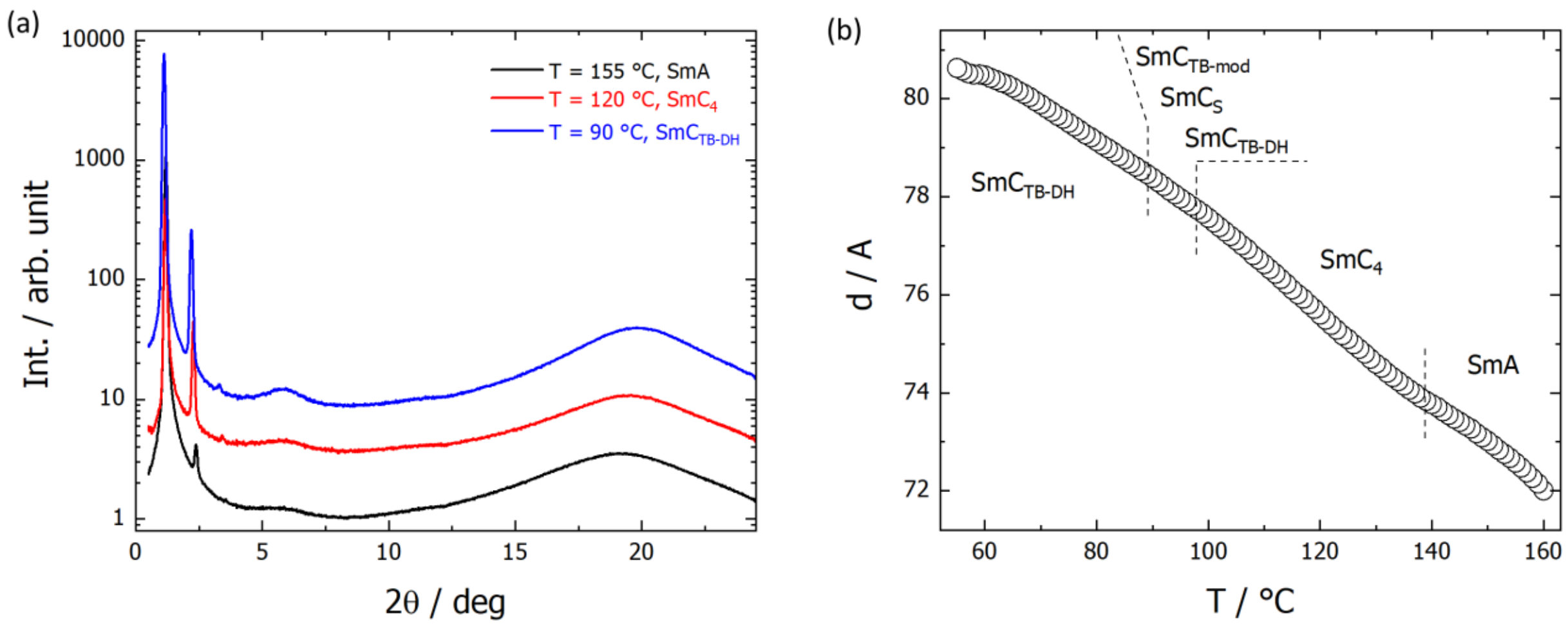

**Figure S4**: (a) X-ray diffractorams for compound CT4O2Me.18 in SmA, $SmC_4$ and $SmC_{TB\text{-}DH}$ phases. (b) Layer spacing in smectic phases measured as a function of temperature. In all the smectic phases bilayer-type lamellar structure is evidenced by layer periodicity being close to double molecular length. No changes in layer spacing at the transition between various tilted smectics are detected.

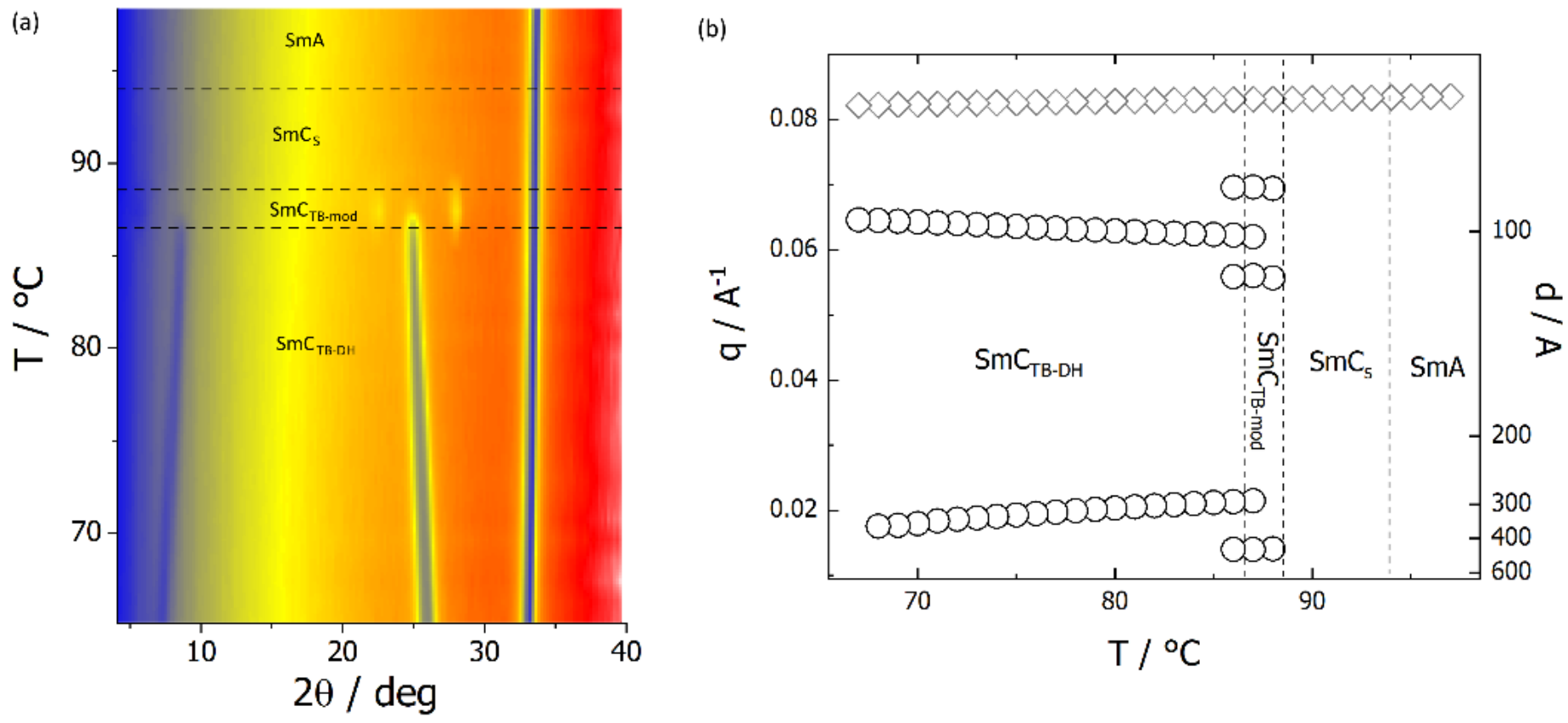

**Figure S5**: (a) Temperature evolution of the RSoXS pattern (a) and corresponding diffraction peak positions (b) for compound CT4O2Me.16. Signal positioned around 2θ angle at 33 deg, (shown by grey rhombs in panel (b)) is a non-resonant diffraction peak due to electron density modulations defining bi-layer type lamellar structure in all smectic phases of the compound. Other signal are resonant.

## 2. Theoretical modelling

Orientation of dimers within a smectic layer is characterized by the angles defined in Figure S6. The axis connecting the two ends of a dimer, whose direction is represented by a unit vector $\vec{n}$, is tilted by an angle $\theta$ with respect to the smectic layer normal, which is aligned with the $z$ axis. The projection of $\vec{n}$ onto the smectic plane ($xy$ plane) is along the unit vector $\vec{c}$, which forms an angle $\varphi$ with the $x$ axis. The apex angle of the dimer is $\pi - 2\alpha$. The four-layer unit is defined by the azimuthal angles of the $\vec{c}$ vectors in four successive layers, with

$$\varphi_1 = 4j\varepsilon + j\gamma \ ,$$

$$\varphi_2 = C_2 + \epsilon + 4j\varepsilon - j\gamma \ ,$$

$$\varphi_3 = C_3 + 2\varepsilon + 4j\varepsilon + j\gamma \ ,$$

$$\varphi_4 = C_4 + 3\varepsilon + 4j\varepsilon - j\gamma \ ,$$

where $C_i$ $(i = 2,3,4)$ are constants; $C_1$ is conveniently set to 0 and is therefore omitted from the expression for $\varphi_1$.

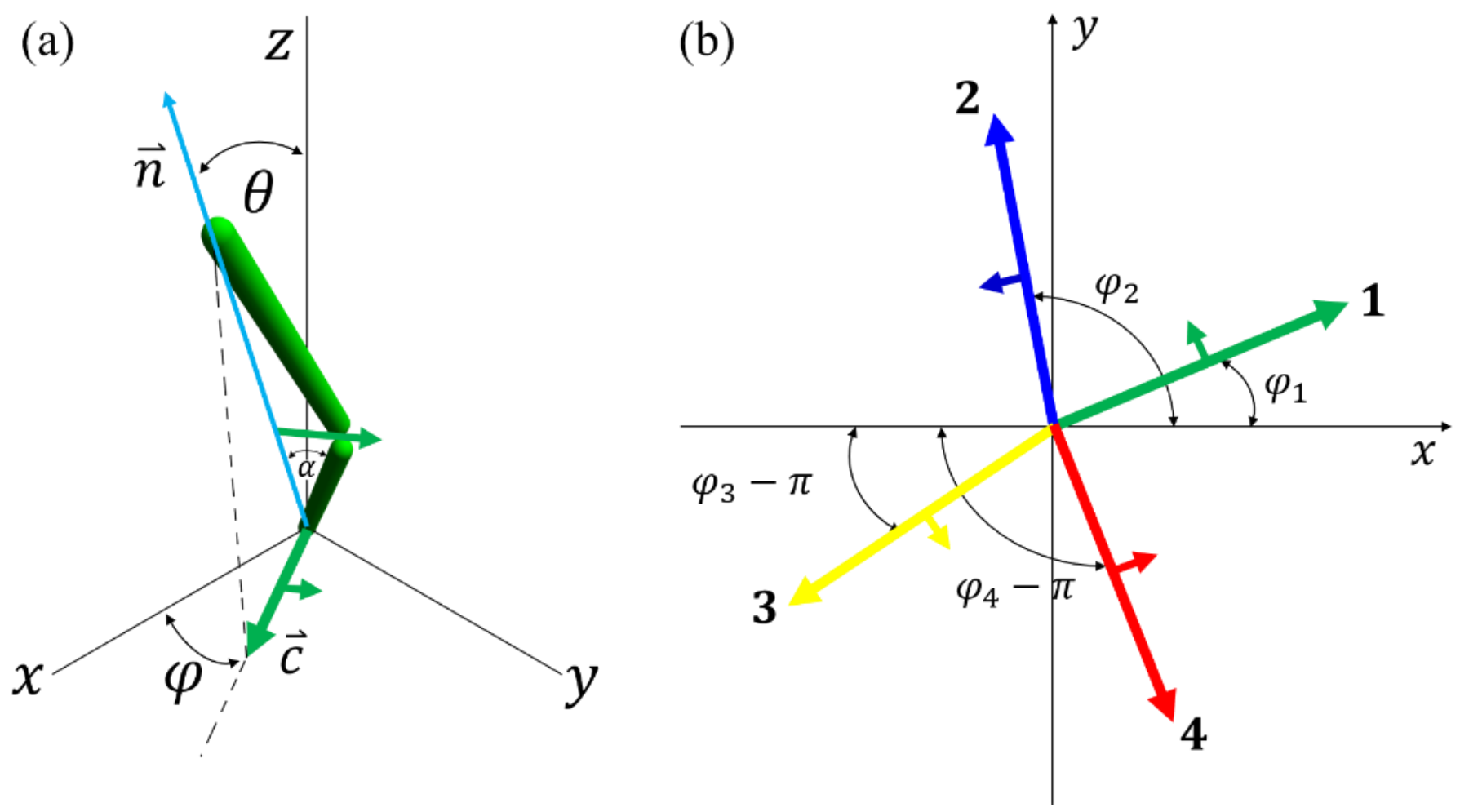

**Figure S6** (a) The orientation of the long dimer axis, represented by the unit vector $\vec{n}$, in the laboratory coordinate system is specified by the tilt angle θ and the azimuthal angle φ; $\vec{c}$ is a unit vector along the projection of $\vec{n}$ onto the $xy$-plane. The apex angle of the dimer is $\pi - 2\alpha$. The apex of the dimer lies in the direction perpendicular to the plane spanned by $\vec{n}$ and $\hat{z}$, and is therefore also perpendicular to $\vec{c}$. (b) In the four-layer unit, successive layers differ in the orientation of the vector $\vec{c}$. The figure shows the directions of $\vec{c}$ within a four-layer stack. A homochiral case with the dimer tips in all four layers pointing in the same direction is shown.

The resonant x-ray scattering response was calculated using the molecular tensor model introduced in Ref. S1. For completeness, the key elements of the model are summarized below. Each arm of the bent dimer is represented by a resonant dipole characterized by a uniaxial polarizability tensor

$$\mathbf{F}_{ei} = f_0 \begin{pmatrix} 1 & 0 & 0 \\ 0 & 1 & 0 \\ 0 & 0 & -2 \end{pmatrix}, \tag{1}$$

where the local $z$-axis is aligned along the arm of the dimer. And $f_0$ determines the overall scattering amplitude. The tensor form factor in the laboratory coordinate system ($\mathbf{F}$) is obtained by successive rotations about the $x$-, $y$-, and $z$-axes through angles $\alpha$, $\theta$, and $\varphi$, respectively,

$$\mathbf{F} = \mathbf{R}_\varphi \mathbf{R}_\theta \mathbf{R}_\alpha \mathbf{F}_{ei} \mathbf{R}_\alpha^{-1} \mathbf{R}_\theta^{-1} \mathbf{R}_\varphi^{-1} \ . \tag{2}$$

A positive angle $\alpha$ is assigned to the lower arm of the dimer and a negative angle $\alpha$ to the upper arm. Together with positive $\theta$ and $\varphi$, this convention describes a dimer whose apex is oriented perpendicular to the plane spanned by $\hat{z}$ and $\vec{n}$ and points in the direction of helix rotation when $\varepsilon$ and $\gamma$ are positive [S2].

The scattering amplitude tensor of a four-layer unit is obtained by summing the tensor form factors of the individual layers, accounting for both the dimer orientation in each layer and the phase shifts arising from scattering from layers separated by a distance $d_0$. The scattering amplitude tensor ($\mathbf{A}$) is obtained by adding form factors of all four-layer units. The key difference from the model considered previously is that the four-layer unit is no longer identical throughout the structure but varies along the z axis.

The variation is periodic: in successive four-layer units, the angles $\varphi_1$ and $\varphi_3$ increase by $\gamma$, whereas $\varphi_2$ and $\varphi_4$ decrease by the same amount. As a result, summation over a very large (effectively infinite) number of layers yields delta-function contributions, and the tensor elements are nonzero only for specific values of the scattering vector, whose magnitude ($q$) is given by

$$q = \frac{2\pi}{d_0} h \quad , \tag{3}$$

where $h$ is the Miller index. If $\varepsilon = \gamma = 0$, the basic periodicity is $4d_0$, thus one expects interference peaks at (00h) of $h$ equal to intiger multiples of $1/4$. The peak corresponding to $h = 1/4$ is denoted $q_4$. For $\varepsilon \neq 0$ and/or $\gamma \neq 0$, the $q_4$ peak splits. Tensor elements are different from zero only for $\pm 8\varepsilon \pm 2\gamma + 8\pi h = 2\pi i$ and $\pm 4\varepsilon \pm \gamma + 8\pi h = 2\pi i$, where $i$ is an integer, which means:

$$h = \frac{i}{4} \pm \frac{\varepsilon}{\pi} \pm \frac{\gamma}{4\pi} \tag{4}$$

and

$$h = \frac{i}{4} \pm \frac{\varepsilon}{2\pi} \pm \frac{\gamma}{8\pi} \quad . \tag{5}$$

In the present work, we focus on the case $i = 1$, corresponding to the splitting of the $q_4$ peak. The magnitudes of the scattering vectors related to the allowed peaks are denoted by $q_4 \pm q_\varepsilon \pm q_\gamma$, where $q_4 = \pi/(2d_0)$, $q_\varepsilon = \varepsilon/d_0$ and $q_\gamma = \gamma/(4d_0)$.

The scattering amplitude tensor $\underline{A}$ can be expressed in a general form as:

$$\underline{A} = \begin{pmatrix} f_{11} & f_{12} & f_{13} \\ f_{21} & f_{22} & f_{23} \\ f_{31} & f_{32} & f_{33} \end{pmatrix} \quad . \tag{6}$$

The tensor elements $f_{11}$, $f_{22}$, $f_{12}$ and $f_{21}$ are nonzero at $h = \frac{i}{4} \pm \frac{\varepsilon}{\pi} \pm \frac{\gamma}{4\pi}$, $f_{11}$ and $f_{22}$ also at $h = \frac{i}{4}$, $f_{33}$ only at $h = \frac{i}{4}$ , while $f_{13}, f_{31}, f_{23}$ and $f_{32}$ are different from zero only at $h = \frac{i}{4} \pm \frac{\varepsilon}{2\pi} \pm \frac{\gamma}{8\pi}$.

Once the scattering amplitude tensor has been determined, the peak intensity at a given value of $q$ is calculated using the procedure introduced by Dmitrienko [S3] and described in detail in the SI of [S1]. The incident light is assumed to be unpolarized. The scattered intensity is therefore calculated for both $\sigma$- and $\pi$-polarized incident light. The scattered light may be polarized either within the scattering plane ($\pi$-polarization) or perpendicular to it ($\sigma$-polarization), and both contributions are summed. We find that the scattered intensity vanishes for $h = \frac{1}{4}$ and $h = \frac{1}{4} \pm \frac{\varepsilon}{\pi} \pm \frac{\gamma}{4\pi}$. In contrast, the peak at $h = \frac{1}{4} \pm \frac{\varepsilon}{2\pi} \pm \frac{\gamma}{8\pi}$ is observed only in the crossed-polarization channels ($\sigma\pi$ and $\pi\sigma$).

We searched for a set of parameter values $C_2, C_3$ and $C_4$ for which the calculated ratios of the resonant peak intensities qualitatively reproduce the experimental observations, namely that the peak with the lowest $q$, i.e. $q_{mm} = q_4 - q_\varepsilon - q_\gamma$, has the highest intensity $I_{mm}$. The ratio $I_{mm}/I_{pp}$, where $I_{pp}$ is the intensity of the peak with the highest value of $q$, $q_{pp} = q_4 + q_\varepsilon + q_\gamma$, is approximately 4. The resonant peaks with $q$ between $q_{mm}$ and $q_{pp}$ are peaks $q_{mp} = q_4 - q_\varepsilon + q_\gamma$ and $q_{pm} = q_4 + q_\varepsilon - q_\gamma$ with intensities $I_{mp}$ and $I_{pm}$ respectively. These intensities are by a factor of 2 or more lower than the intensity $I_{pp}$.

The homo- and heterochirality of successive layers is expected to strongly affect the resonant peak intensities, and we therefore examined both cases. For heterochiral layers, where dimer tips point in opposite directions in neighboring layers, we could not find any set of parameters for which $I_{mm}/I_{pp} \approx 4$. The largest value of this ratio is approximately 1.6. In Fig. S7 we show some characteristic dependencies for the ratios $I_{mm}/I_{pp}$, $I_{pm}/I_{pp}$ and $I_{mp}/I_{pp}$ as a function of the tilt angle $\theta$ and parameter $C_3$. We observe that the ratio $I_{mm}I_{pp} \leq 1.6$. The situation is similar for variation of other parameters. Specifically, variations in $C_2$ and $C_4$ affect only the ratios $I_{mp}/I_{pp}$ and $I_{pm}/I_{pp}$, while $I_{mm}/I_{pp}$ remains unchanged.

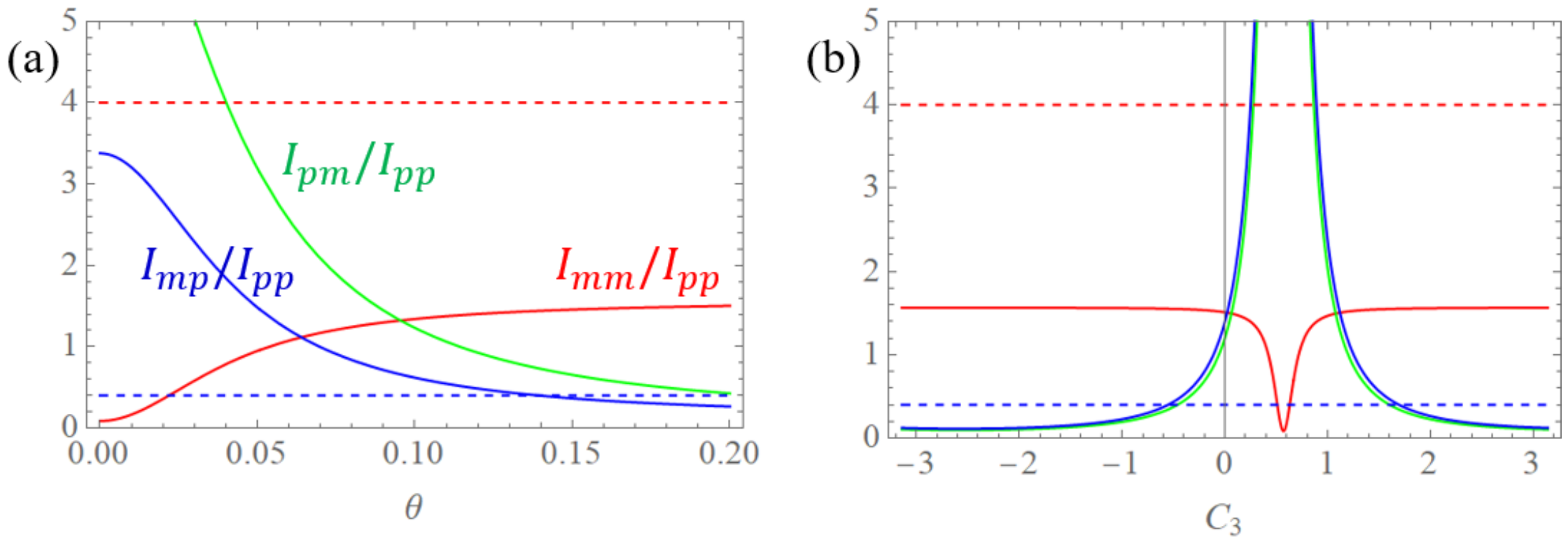

**Figure S7.** The ratios of the peak intensities as a function of (a) the tilt angle $\theta$ and (b) parameter $C_3$. (f) Unless shown as the variable on the horizontal axis, the parameter values are: $\theta = 10°$, $\alpha = 0.08$, $C_2 = 0$, $C_3 = \pi$, $C_4 = 0$, $\varepsilon = 0.62$, $\gamma = 1.14$. $I_{mm}$ is the resonant peak intensity at the scattering wavevector magnitude $q_{mm} = q_4 - q_\varepsilon - q_\gamma$, $I_{pp}$ the intensity at $q_{pp} = q_4 + q_\varepsilon + q_\gamma$, $I_{mp}$ at $q_{mp} = q_4 - q_\varepsilon + q_\gamma$ and $I_{pm}$ at $q_{pm} = q_4 + q_\varepsilon - q_\gamma$. Dashed red line: experimental value of $I_{mm}/I_{pp}$; dashed blue line: experimental value of $I_{pm}/I_{pp}$.

The homochiral case provides substantially better agreement with experiment. While several combinations of parameter values are possible, the simplest one that captures the major experimental features is the case with $C_2 = C_4 = 0$ and $C_3 = \pi$. In the homochiral case the ratio $I_{mm}/I_{pp}$ strongly depends on $\alpha$ and $\theta$ and is independent of the values of parameters $C_i$ $(i = 2,3,4)$. Intensities $I_{mm}$ and $I_{pp}$ depend on $C_3$, but they remain unchanged when $C_2$ and $C_4$ are varied. However, the ratio $I_{mm}/I_{pp}$ is independent of $C_3$ as well. Supplementary Fig. S8 indicates that $\alpha$ must be positive, meaning that the dimer tip points in the direction of the secondary helix (the helix due to $\varepsilon \neq 0$) and that the tilt angle is around 10°. We also observe that in the parameter regions where $I_{mm}/I_{pp} \approx 4$, $I_{mp}/I_{pp} < 1$ and $I_{pm}/I_{pp} < 1$, the blue curve ($I_{pm}/I_{pp}$) on graphs in Supplementary Figure 3 is always above the green curve ($I_{mp}/I_{pp}$), while experimentally we observe $I_{pm} > I_{mp}$. Thus, this experimental feature cannot be reproduced within the framework of the present simplified model.

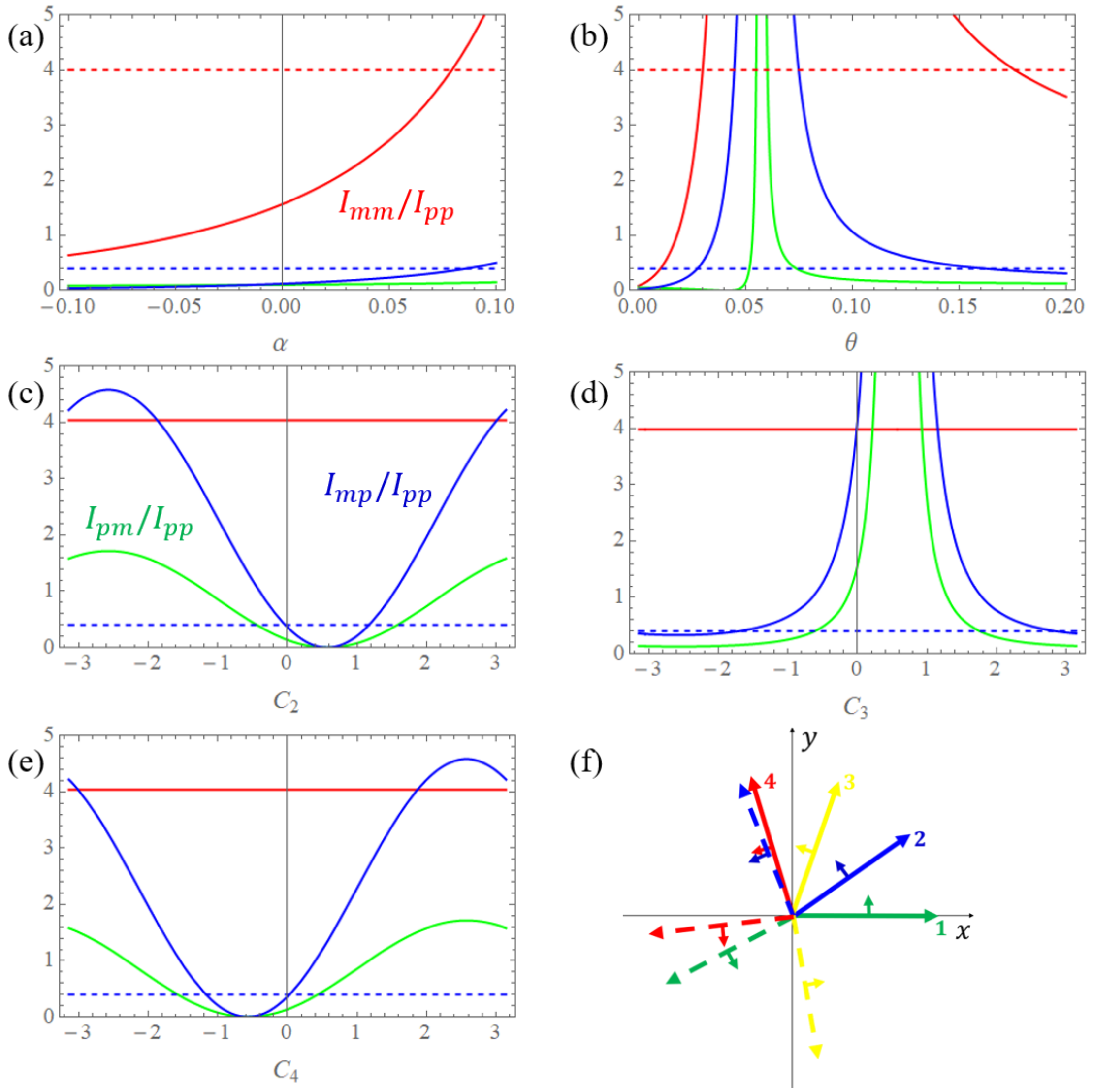

**Figure S8.** The ratios of the peak intensities as a function of (a) the angle $\alpha$, defining the apex angle of the dimer, (b) the tilt angle $\theta$, and parameters (c) $C_2$, (d) $C_3$ and (e) $C_4$. (f) Directions of the $\vec{c}$ vectors in the first two four-layer stacks ($j = 0$ and $j = 1$). Unless shown as the variable on the horizontal axis, the parameter values are: $\theta = 10°$, $\alpha = 0.08$, $C_2 = 0$, $C_3 = \pi$, $C_4 = 0$, $\varepsilon = 0.62$, $\gamma = 1.14$. $I_{mm}$ is the resonant peak intensity at the scattering wavevector magnitude $q_{mm} = q_4 - q_\varepsilon - q_\gamma$, $I_{pp}$ the intensity at $q_{pp} = q_4 + q_\varepsilon + q_\gamma$, $I_{mp}$ at $q_{mp} = q_4 - q_\varepsilon + q_\gamma$ and $I_{pm}$ at $q_{pm} = q_4 + q_\varepsilon - q_\gamma$. Dashed red line: experimental value of $I_{mm}/I_{pp}$; dashed blue line: experimental value of $I_{pm}/I_{pp}$.

## 3. Organic synthesis and analytical data

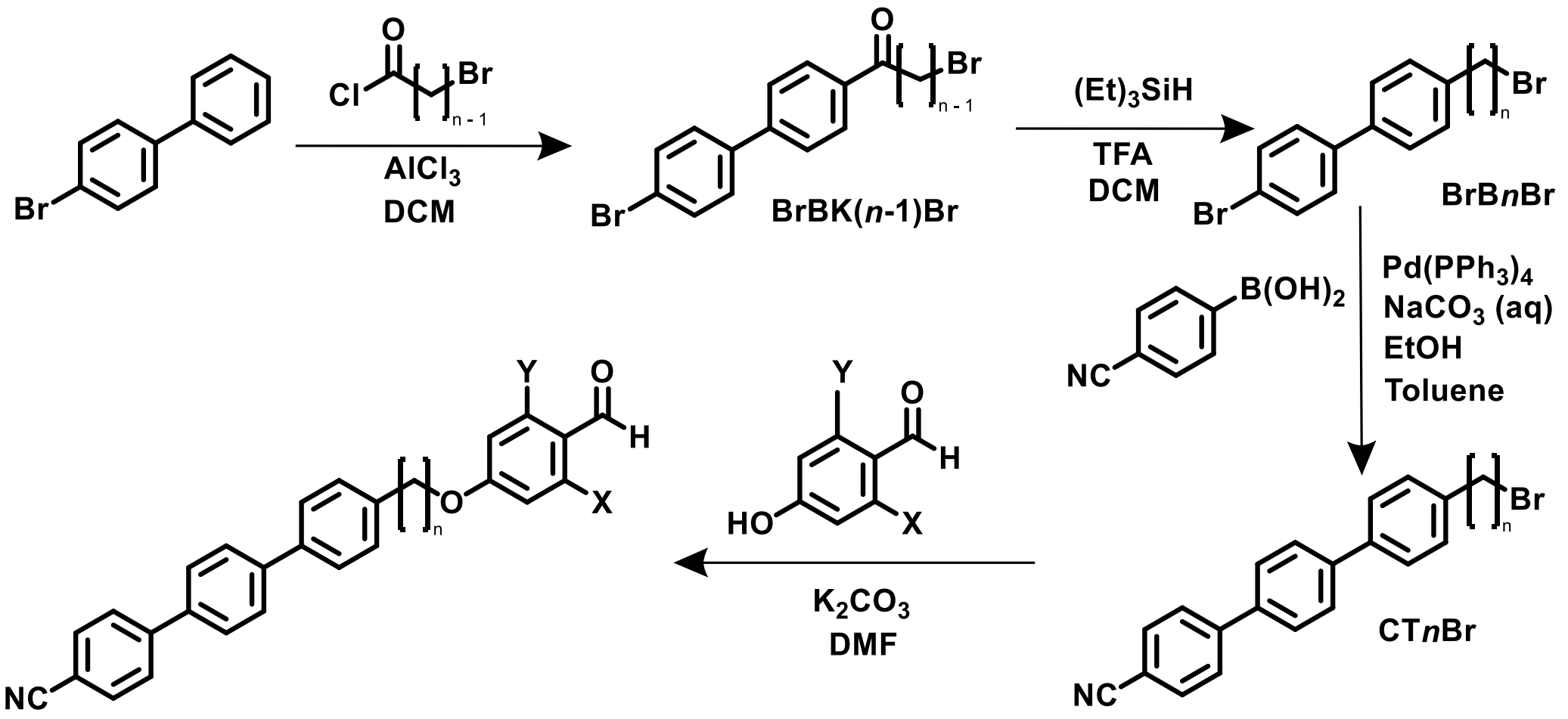

**Scheme S1**: Synthetic pathway used for the preparation of the 4-[4-[4-[6-(4-Formyl-X-phenoxy)alkyl]phenyl]phenyl]benzonitriles (CTnOXAH)

**ω-Bromo-1-(4'-bromo[1,1'-biphenyl]-4-yl)alkan-1-one** (BrBK(*n*-1)Br) (*n* = 4, 6)

To a reaction vessel sparged with argon, and wrapped in tinfoil, 4-bromobiphenyl (1 eq) in dry DCM (20 mL) was added in an ice bath. To this, $AlCl_3$ (1.1 eq) was added with stirring. The relevant bromoalkanyl chloride (1.1 eq) was added dropwise and the reaction mixture allowed to gradually come to room temperature gradually and proceed for 24 hr.

Once complete complete, the mixture was added slowly to water and extracted with DCM (3 x 100 mL). The organic layer was washed with sodium bicarbonate (1 x 100 mL) and brine (1 x 100 mL) and dried with $MgSO_4$. After filtration to remove the $MgSO_4$, the solvent was removed *in vacuo* and the crude product recrystallised from EtOH (400 mL) to give off-white crystals.

**ω-Bromo-1-(4'-bromo[1,1'-biphenyl]-4-yl)hexan-1-one** (**BrBK5Br)**

**Table S1**: Quantities of reagents used in the synthesis of the BrBK5Br intermediate

| | n / mol | m / g | V / mL |
|---|---|---|---|
| 4-bromobiphenyl | 0.0429 | 10.0 | - |
| $AlCl_3$ | 0.0472 | 6.29 | - |
| Bromohexanoyl chloride | 0.0472 | - | 7.22 |

Yield 13.3 g (75%). m.p. 85°C. $R_f$ (DCM) 0.74.

$^1$H NMR (400 MHz, $CDCl_3$) δ ppm: 8.03 (d, $J$ = 8.2 Hz, 2H), 7.64 (d, $J$ = 8.1 Hz, 2H), 7.59 (d, $J$ = 8.1 Hz, 2H), 7.49 (d, $J$ = 8.8 Hz, 2H), 3.44 (t, $J$ = 6.7 Hz, 2H), 3.02 (t, $J$ = 7.2 Hz, 2H), 1.93 (m, 2H), 1.80 (p, $J$ = 7.4 Hz, 2H), 1.56 (m, 2H). $^{13}$C NMR (101 MHz, $CDCl_3$) δ ppm: 199.59, 144.53, 138.92, 136.08, 132.25 (2C), 128.95 (2C), 128.87 (2C), 127.20 (2C), 122.79, 38.50, 33.77, 32.77, 28.02, 23.50. IR (νmax/$cm^{-1}$): 2938 ($sp^2$ hybridised C-H stretching), 2865 ($sp^2$ hybridised C-H stretching), 1909 (aromatic overtones), 1679 (ketone carbonyl stretching).

**ω-Bromo-1-(4'-bromo[1,1'-biphenyl]-4-yl)butan-1-one (BrBK3Br)**

**Table S2**: Quantities of reagents used in the synthesis of the BrBK3Br intermediate

| | n / mol | m / g | V / mL |
|---|---|---|---|
| 4-bromobiphenyl | 0.0297 | 6.91 | - |
| $AlCl_3$ | 0.0323 | 4.31 | - |
| Bromobutyryl chloride | 0.0267 | - | 3.10 |

Yield: 9.0 g, (88%). m.p. 113 °C. $R_f$ (DCM) 0.68.
$^1$H NMR (400 MHz, $CDCl_3$) δ ppm: 8.09 – 8.02 (m, 2H, Ar-H), 7.70 – 7.64 (m, 2H, Ar-H), 7.62 (m, 2H, Ar-H), 7.53 – 7.45 (m, 2H, Ar-H), 3.57 (t, $^3J$ = 6.3 Hz, 2H, ArC(=O)$CH_2CH_2\underline{CH_2}$Br), 3.22 (t, $^3J$ = 6.9 Hz, 2H, ArC(=O)$\underline{CH_2}CH_2CH_2$Br), 2.34 (p, $^3J$ = 6.6 Hz, 2H, ArC(=O)$CH_2\underline{CH_2}CH_2$Br). $^{13}$C NMR (101 MHz, $CDCl_3$) δ ppm: 198.31, 144.66, 138.74, 135.75, 132.14 (2C), 128.84 (2C), 128.75 (2C), 127.13 (2C), 122.73, 36.64, 33.64, 26.89.

**1-Bromo-6-(4'-bromobiphenyl-4-yl)alkane (BrB*n*Br) (*n* = 4, 6)**

In a flask flushed with argon, the relevant ω-Bromo-1-(4'-bromo[1,1'-biphenyl]-4-yl)alkan-1-one (BrBK5Br, BrBK3Br) (1 eq) and trifluoroacetic acid (TFA) (8 eq) were combined with the minimal amount of DCM while stirring at 0 °C. Dropwise, triethyl silane ($Et_3Si$) (4 eq) was added, and the reaction allowed to come gradually to room temperature. After 6 hr, the reaction was checked by TLC (DCM $R_f$ 0.88) and appeared to be complete.

The reaction mixture was then poured into a mixture of DCM (200 mL) and water (200 mL) and the organic layer extracted and washed with water (3 x 150 mL). The organic fraction was dried with $MgSO_4$ and after filtration to remove the $MgSO_4$, the solvent was removed *in vacuo*. The crude product was recrystallised from ethanol to give a white solid.

**1-Bromo-6-(4'-bromobiphenyl-4-yl)hexane (BrB6Br)**

**Table S3**: Quantities of reagents used in the synthesis of the BrB6Br intermediate

| | n / mol | m / g | V / mL |
|---|---|---|---|
| BrBK5Br | 0.0274 | 11.2 | - |
| Trifluoroacetic acid | 0.219 | - | 16.8 |
| Triethyl silane | 0.109 | - | 17.5 |

Yield 10.66 g (98%). m.p. 72 °C. $R_f$ (DCM) 0.88

$^1$H NMR (400 MHz, $CDCl_3$) δ ppm: 7.57 (d, *J* = 8.7 Hz, 2H), 7.48 (m, 4H), 7.27 (d, *J* = 8.3 Hz, 2H), 3.43 (t, *J* = 6.8 Hz, 2H), 2.68 (t, *J* = 7.6 Hz, 2H), 1.89 (p, *J* = 6.8 Hz, 2H), 1.69 (p, *J* = 7.6 Hz, 2H), 1.51 (m, 2H), 1.41 (m, 2H). $^{13}$C NMR (101 MHz, $CDCl_3$) δ ppm: 142.32, 140.17, 137.56, 131.94 (2C), 129.09 (2C), 128.70 (2C), 126.96 (2C), 121.34, 58.63, 35.56, 34.08, 32.86, 31.33, 28.53, 28.16, 18.59. IR (νmax/cm$^{-1}$): 2931 (sp$^2$ hybridised C-H stretching), 2855 (sp$^2$ hybridised C-H stretching), 1903 (aromatic overtones).

**1-Bromo-6-(4'-bromobiphenyl-4-yl)butane (BrB4Br)**

**Table S4**: Quantities of reagents used in the synthesis of the BrB4Br intermediate

| | n / mol | m / g | V / mL |
|---|---|---|---|
| BrBK3Br | 0.014 | 5.30 | - |
| Trifluoroacetic acid | 0.084 | - | 6.43 |
| Triethyl silane | 0.042 | - | 6.70 |

Yield 3.2 g (62%). m.p. 101 °C. $R_f$ (DCM) 0.82.

$^1$H NMR (400 MHz, $CDCl_3$) δ ppm: 7.57 – 7.52 (m, 2H, Ar-H), 7.48 (m, 4H, Ar-H), 7.24 (m, 2H, Ar-H), 3.44 (t, $^3J$ = 6.6 Hz, 2H, $ArCH_2CH_2CH_2\underline{CH_2}Br$), 2.69 (t, $^3J$ = 7.5 Hz, 2H, $Ar\underline{CH_2}CH_2CH_2CH_2Br$), 1.93 (m, 2H, $ArCH_2CH_2\underline{CH_2}CH_2Br$), 1.87 – 1.74 (m, 2H, $ArCH_2\underline{CH_2}CH_2CH_2Br$). $^{13}$C NMR (101 MHz, $CDCl_3$) δ ppm: 141.41, 139.96, 137.69, 131.84 (2C), 128.96 (2C), 128.58 (2C), 126.94 (2C), 121.29, 34.61, 33.61, 32.24, 29.80.

**1-Bromo-6-(4'-cyanoterphenyl-4-yl)alkane (CT*n*Br) (*n* = 4, 6)**

The relevant 1-Bromo-6-(4'-bromobiphenyl-4-yl)alkane (BrB6Br, BrB4Br) (1 eq), 4-cyanophenylboronic acid (1.1 eq) and 2 M aqueous $NaCO_3$ (2.5 eq) were combined in ethanol (30 mL) and toluene (20 mL) with stirring and the reaction vessel sparged with argon. Once sparged, the reaction vessel was kept under argon and $Pd(PPh_3)_4$ (0.015 eq) was added and the reaction heated to 90 °C. The reaction was left to proceed for 50 hours.

Once completed, the reaction mixture was allowed to gradually come to room temperature. To the cooled reaction mixture, 2 M HCl was added dropwise until gas stopped being produced, then solvent was removed *in vacuo* until only water remained. Using ethyl acetate (150 mL), the product was extracted and the resulting organic layer washed with water (3 x 150 mL) and brine (1 x 100 mL). The organic fraction was dried with $MgSO_4$, after filtration to remove the $MgSO_4$, the solvent was removed *in vacuo* to give an off-white solid. The crude product was purified by column chromatography and the solvent removed *in vacuo*. The product was then recrystalised from ethanol (200 mL) and ethyl acetate (20 mL) to give a white solid.

**1-Bromo-6-(4'-cyanoterphenyl-4-yl)hexane (CT6Br)**

**Table S5**: Quantities of reagents used in the synthesis of the CT6Br intermediate

| | n / mol | m / g | V / mL |
|---|---|---|---|
| BrB6Br | 0.013 | 5.00 | - |
| 4-cyanophenylboronic acid | 0.014 | 2.06 | - |
| $NaCO_3$ | - | - | 15 |
| $Pd(PPh_3)_4$ | 0.195 x$10^{-3}$ | 0.23 | - |

Yield 3.09 g (57%). m.p. 102 °C, $T_{NI}$ = 145 °C. $R_f$ (8;2, DCM:Petroleum ether (40-60) 0.73.

$^1$H NMR (400 MHz, $CDCl_3$) δ ppm: 7.74 (m, 4H, appt s), 7.71 (d, *J* = 8.6 Hz, 2H, appt dt), 7.66 (d, *J* = 8.6 Hz, 2H, appt dt), 7.56 (d, *J* = 8.2 Hz, 2H, appt dt), 7.28 (d, *J* = 8.3 Hz, 2H), 3.42 (t, *J* = 6.8 Hz, 2H), 2.68 (t, *J* = 7.5 Hz, 2H), 1.88 (dt, *J* = 7.6, 6.8 Hz, 2H), 1.69 (p, *J* = 7.5 Hz, 2H), 1.49 (m, 2H), 1.40 (m, 2H). $^{13}$C NMR (101 MHz, $CDCl_3$) δ ppm: 145.36, 142.46, 141.63, 137.81, 137.75, 132.79 (2C), 129.13 (2C), 127.76 (2C), 127.71 (2C), 127.68 (2C), 127.10 (2C), 119.12, 111.00, 35.60, 34.09, 32.87, 31.36, 28.55, 28.17. IR (νmax/$cm^{-1}$): 2928 ($sp^2$ hybridised C-H stretching), 2855 ($sp^2$ hybridised C-H stretching), 2225 (nitrile stretching), 1916 (aromatic overtones).

**1-Bromo-6-(4'-cyanoterphenyl-4-yl)butane (CT4Br)**

**Table S6**: Quantities of reagents used in the synthesis of the CT4Br intermediate.

| | n / mol | m / g | V / mL |
|---|---|---|---|
| BrB4Br | 0.009 | 3.10 | - |
| 4-cyanophenylboronic acid | 0.010 | 1.41 | - |
| $NaCO_3$ | - | - | 7 |
| $Pd(PPh_3)_4$ | $0.087 \times 10^{-3}$ | 0.10 | - |

Yield: 0.88 g (26%). m.p. 106 °C, $T_{N\text{-}I}$ 239 °C $R_f$ (DCM) 0.78.

$^1$H NMR (400 MHz, $CDCl_3$) δ ppm: 7.78 – 7.63 (m, 8H, Ar-H), 7.62 – 7.54 (m, 2H, Ar-H), 7.33 – 7.26 (m, 2H, Ar-H), 3.45 (t, $^3J$ = 6.6 Hz, 2H, $ArCH_2CH_2CH_2\underline{CH_2}Br$), 2.71 (t, $^3J$ = 7.5 Hz, 2H, $Ar\underline{CH_2}CH_2CH_2CH_2Br$), 2.00 – 1.77 (m, 4H, $ArCH_2\underline{CH_2}\underline{CH_2}CH_2Br$). $^{13}$C NMR (101 MHz, $CDCl_3$) δ ppm: 145.20, 141.54, 141.40, 137.86, 137.74, 132.67 (2C), 129.00 (2C), 127.64 (2C), 127.61 (2C), 127.56 (2C), 127.07 (2C), 126.79, 118.99, 110.89, 34.65, 33.64, 32.26, 29.82.

**4-[4-[4-[6-(4-Formyl-X-phenoxy)alkyl]phenyl]phenyl]benzonitriles (CT*n*OXAH)**

A mixture of the relevant 1-Bromo-6-(4’-cyanoterphenyl-4-yl)alkane (CT4Br, CT6Br) (1 eq), the relevant substituted 4-hydroxybenzaldehyde (1.1 eq) with potassium carbonate (2 eq) were combined in DMF (quantities detailed in Table S7). The reaction mixture was heated to 90 °C with stirring and allowed to proceed for 18 hr.

Once TLC indicated that the reaction was complete, it was cooled to room temperature and added to water (10x the volume of DMF used) resulting in the formation of a white precipitate. The solid was collected by filtration to give an off white solid. The crude product was purified by column chromatography and recrystallised to give the target compound.

**Table S7**: Quantities of reagents used in the synthesis of the CTnOXAH intermediates

| | Target Compound | | | Quantities | | | | | | |
|---|---|---|---|---|---|---|---|---|---|---|
| | | | | CT*n*Br | | Substituted 4-hydroxy benzaldehyde | | $K_2CO_3$ | | DMF |
| | X | Y | *n* | m / g | n / mmol | m / g | n / mmol | m / g | n / mmol | V / mL |
| CT4O2MeAH | Me | H | 4 | 0.88 | 2.08 | 0.28 | 2.06 | 0.58 | 4.20 | 10 |
| CT6O2MeAH | Me | H | 6 | 1.89 | 4.53 | 0.68 | 4.98 | 1.25 | 9.06 | 20 |
| CT6O2FAH | F | H | 6 | 2.00 | 4.78 | 0.74 | 5.26 | 1.32 | 9.56 | 25 |
| CT6O2,6dFAH | F | F | 6 | 2.00 | 4.78 | 0.83 | 5.26 | 1.32 | 9.56 | 25 |

**4-{[6-(4'-Cyanoterphenyl-4''-yl)butyl]oxy}-2-methylbenzaldehyde (CT4O2MeAH)**

X = Me, Y = H. $R_f$ (6:4 DCM:Petroleum Ether) 0.26. Recrystallised from ethanol. Yield: 0.66 g (73 %).

$^1$H NMR (400 MHz, $CDCl_3$) δ ppm: 10.11 (s, 1H, C(=O)$\underline{H}$), 7.79 – 7.66 (m, 8H, Ar-H), 7.61 – 7.54 (m, 2H, Ar-H), 7.35 – 7.27 (m, 2H, Ar-H), 6.83 (dd, $^3J$ = 8.5, 2.5 Hz, 1H, Ar-H), 6.73 (d, $^3J$ = 2.4 Hz, 1H, Ar-H), 4.10 – 4.02 (m, 2H, $ArCH_2CH_2CH_2\underline{CH_2}O$-Ar), 2.80 – 2.72 (m, 2H, $Ar\underline{CH_2}CH_2CH_2CH_2O$-Ar), 2.64 (s, 3H, $Ar\underline{CH_3}$), 1.88 (m, $^3J$ = 3.9 Hz, 4H, $ArCH_2\underline{CH_2CH_2}CH_2O$).

**4-{[6-(4'-Cyanoterphenyl-4''-yl)hexyl]oxy}-2-methylbenzaldehyde (CT6O2MeAH)**

X = Me, Y = H. $R_f$ (8:2 DCM:Petroleum Ether) 0.33. Recrystallised from ethanol. Yield 1.86 g (86 %). m.p. 116 °C, $T_{N-I}$ 178 °C.

$^1$H NMR (400 MHz, $CDCl_3$) δ ppm: 10.11 (s, 1H, Ar-$\underline{C(=O)H}$), 7.74 (m, 5H, Ar-H), 7.70 (d, $^3J$ = 8.6 Hz, 2H, Ar-H), 7.66 (d, $^3J$ = 8.6 Hz, 2H, Ar-H), 7.56 (d, $^3J$ = 8.2 Hz, 2H, Ar-H), 7.29 (d, $^3J$ = 8.2 Hz, 2H, Ar-H), 6.82 (dd, $^3J$ = 8.6 Hz, $^4J$ = 2.5 Hz, 1H, Ar-H), 6.73 (d, $^4J$ = 2.5 Hz, 1H, Ar-H), 4.02 (t, $^3J$ = 6.4 Hz, 2H, Ar-$O\underline{CH_2}CH_2CH_2CH_2CH_2CH_2$-Ar), 2.69 (t, $^3J$ = 7.7 Hz, 2H, Ar-$OCH_2CH_2CH_2CH_2CH_2\underline{CH_2}$-Ar), 2.64 (s, 3H, Ar-$\underline{CH_3}$), 1.82 (m, 2H, Ar-$OCH_2\underline{CH_2}CH_2CH_2CH_2CH_2$-Ar), 1.71 (p, $^3J$ = 7.7 Hz, 2H, Ar-$OCH_2CH_2CH_2CH_2\underline{CH_2}CH_2$-Ar), 1.49 (m, 4H, Ar-$OCH_2CH_2\underline{CH_2CH_2}CH_2CH_2$-Ar). $^{13}$C NMR (101 MHz, $CDCl_3$) δ ppm: 191.29, 163.37, 145.33, 143.40, 142.49, 141.58, 137.81, 137.72, 134.89, 132.78 (2C), 129.13 (2C), 127.92, 127.73 (2C), 127.71 (2C), 127.67 (2C), 127.07 (2C), 119.11, 117.60, 112.03, 111.00, 68.23, 35.62, 31.43, 29.14, 29.05, 25.98, 20.03. IR (νmax/$cm^{-1}$): 2938 ($sp^2$ hybridised C-H stretching), 2857 ($sp^2$ hybridised C-H stretching), 2225 (nitrile stretching), 1915 (aromatic overtones), 1672 (aldehyde C=O stretching), 1598 (Aryl C=C stretching).

**4-{[6-(4'-Cyanoterphenyl-4''-yl)hexyl]oxy}-2-fluorobenzaldehyde (CT6O2FAH)**

X = F, Y = H. TLC (DCM $R_f$ 0.51), Recrystallised from ethanol and toluene (1:1). Yield 1.78 g (78 %). m.p. 148 °C, $T_{NI}$ 193 °C.

$^1$H NMR (400 MHz, $CDCl_3$) δ ppm: 10.20 (s, 1H, Ar-$\underline{C(=O)H}$), 7.81 (dd, appt t, $^3J_{H-H}$ = 8.5 Hz, $^4J_{H-F}$ = 8.5 Hz, 1H), 7.74 (m, appt s, 4H, Ar-H), 7.70 (d, $^3J_{H-H}$ = 8.4 Hz, 2H, Ar-H), 7.66 (d, $^3J_{H-H}$ = 8.4 Hz, 2H, Ar-H), 7.56 (d, $^3J_{H-H}$ = 8.1 Hz, 2H, Ar-H), 7.28 (d, $^3J_{H-H}$ = 8.1 Hz, 2H, Ar-H), 6.76 (dd, $^3J_{H-H}$ = 8.5 Hz, $^4J_{H-H}$ = 2.3 Hz, 1H), 6.62 (dd, $^3J_{H-F}$ = 12.5 Hz, $^4J_{H-H}$ = 2.3 Hz,

1H), 4.02 (t, $^3J_{H-H}$ = 6.4 Hz, 2H, Ar-O$CH_2CH_2CH_2(CH_2)_3$-Ar), 2.69 (t, $^3J_{H-H}$ = 7.6 Hz, 2H, Ar-$CH_2CH_2CH_2(CH_2)_3$O-Ar), 1.83 (m, 2H, Ar-O$CH_2CH_2CH_2(CH_2)_3$-Ar), 1.71 (p, $^3J_{H-H}$ = 7.6 Hz, 2H, Ar-$CH_2CH_2CH_2(CH_2)_3$O-Ar), 1.47 (m, 4H, Ar-O$CH_2CH_2CH_2CH_2CH_2CH_2$-Ar). $^{13}$C NMR (101 MHz, $CDCl_3$) δ ppm: 186.07 (d, $^3J_{C-F}$ = 6.0 Hz), 166.44 (d, $^1J_{C-F}$ = 258.3 Hz), 165.90 (d, $^3J_{C-F}$ = 11.9 Hz), 145.34, 142.43, 141.57, 137.84, 137.77, 132.79 (2C), 130.23 (d, $^3J_{C-F}$ = 3.9 Hz), 129.13 (2C), 127.74 (2C), 127.72 (2C), 127.68 (2C), 127.10 (2C), 119.11, 117.84 (d, $^2J_{C-F}$ = 8.4 Hz), 111.78 (d, $^4J_{C-F}$ = 2.8 Hz), 111.02, 101.90 (d, $^2J_{C-F}$ = 24.0 Hz), 68.96, 35.62, 31.41, 29.02, 28.96, 25.94. $^{19}$F NMR (376 MHz, $CDCl_3$) δ ppm: -119.23. IR (νmax/cm$^{-1}$): 2928 (sp$^2$ hybridised C-H stretching), 2854 (sp$^2$ hybridised C-H stretching), 2225 (nitrile stretching), 1916 (aromatic overtones), 1677 (aldehyde C=O stretching), 1606 (Aryl C=C stretching).

**4-{[6-(4’-Cyanoterphenyl-4’’-yl)hexyl]oxy}-2,6-difluorobenzaldehyde (CT6O2,6dFAH)**

X = F, Y = F. TLC (DCM $R_f$ 0.52), Recrystallised from ethanol and toluene (1:9). Yield 1.11 g (47 %). m.p. 157 °C, $T_{NI}$ 188 °C.

$^1$H NMR (400 MHz, $CDCl_3$) δ ppm: 10.18 (s, 1H, Ar-C(=O)H), 7.74 (m, appt s, 4H, Ar-H), 7.70 (d, $^3J_{H-H}$ = 8.4 Hz, 2H, Ar-H), 7.66 (d, $^3J_{H-H}$ = 8.4 Hz, 2H, Ar-H), 7.56 (d, $^3J_{H-H}$ = 8.1 Hz, 2H, Ar-H), 7.28 (d, $^3J_{H-H}$ = 8.1 Hz, 2H, Ar-H), 6.47 (d, $^3J_{H-F}$ = 10.9 Hz, 2H, Ar-H), 4.00 (t, $^3J_{H-H}$ = 6.4 Hz, 2H, Ar-O$CH_2CH_2CH_2(CH_2)_3$-Ar), 2.69 (t, $^3J_{H-H}$ = 7.7 Hz, 2H, Ar-$CH_2CH_2CH_2(CH_2)_3$O-Ar), 1.82 (m, 2H, Ar-O$CH_2CH_2CH_2(CH_2)_3$-Ar), 1.71 (p, $^3J_{H-H}$ = 7.6 Hz, 2H, Ar-$CH_2CH_2CH_2(CH_2)_3$O-Ar), 1.48 (m, 4H, Ar-O$CH_2CH_2CH_2CH_2CH_2CH_2$-Ar). $^{13}$C NMR (101 MHz, $CDCl_3$) δ ppm: 183.61 (t, $^3J_{C-F}$ = 4.5 Hz), 165.57 (t, $^3J_{C-F}$ = 15.1 Hz), 164.82 (dd, $^1J_{C-F}$ = 261.1 Hz, $^3J_{C-F}$ = 9.0 Hz, 2C), 145.32, 142.37, 141.54, 137.83, 137.77, 132.78 (2C), 129.11 (2C), 127.72 (4C), 127.67 (2C), 127.09 (2C), 119.11, 110.99, 108.17 (t, $^2J_{C-F}$ = 12.1 Hz), 99.16 (dd, $^2J_{C-F}$ = 25.8 Hz, $^4J_{C-F}$ = 1.9 Hz, 2C), 69.40, 35.59, 31.37, 28.97, 28.79, 25.87. $^{19}$F NMR (376 MHz, $CDCl_3$) δ ppm: -113.11 (2F). IR (νmax/cm$^{-1}$): 2946 (sp$^2$ hybridised C-H stretching), 2854 (sp$^2$ hybridised C-H stretching), 2226 (nitrile stretching), 1912 (aromatic overtones), 1697 (aldehyde C=O stretching), 1627 (Aryl C=C stretching).

**$3^4$-{6-(4-(((4’-Alkylphenyl)imino)methyl)-3-methylphenoxy)butyl}[$1^1$,$2^1$:$2^4$,$3^1$-terphenyl]-$1^4$-carbonitrile (CT4O2Me.*m*)**

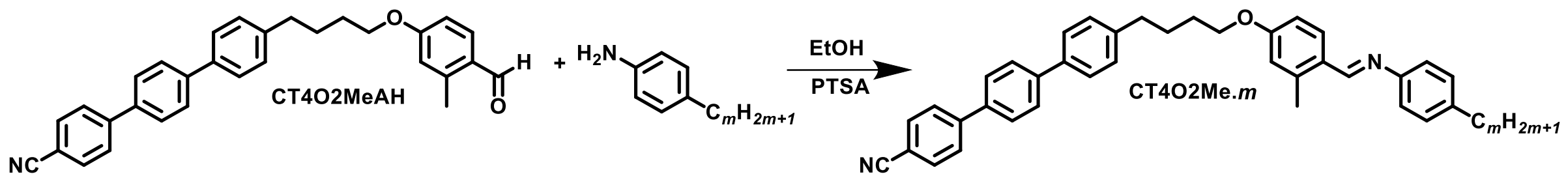

Scheme S2: Synthetic pathway used in the preparation of the CT4O2Me.*m* series

CT4O2MeAH (1 eq, 0.10 g, 0.224 mmol) and the relevant 4-alkylaniline (1.4 eq, 0.314 mmol, detailed in Table S8), were combined in ethanol (20 mL) and heated to reflux. Once at temperature, a catalytic amount of PTSA was added, and the reaction was allowed to proceed for 3 hrs. Once complete, the reaction was allowed to cool to room temperature resulting in the formation of a precipitate which was collected by filtration and recrystallised from ethanol (10 mL) to yield the target compound as a white solid.

**Table S8**: Quantities of 4-alkylaniline used in the synthesis of the CT4O2Me.*m* series.

| *m* | m / g | n / mmol |
|---|---|---|
| 1 | 0.034 | 0.314 |
| 2 | 0.038 | 0.314 |
| 3 | 0.043 | 0.314 |
| 4 | 0.048 | 0.314 |
| 5 | 0.052 | 0.314 |
| 6 | 0.057 | 0.314 |
| 7 | 0.062 | 0.314 |
| 8 | 0.067 | 0.314 |
| 9 | 0.071 | 0.314 |
| 10 | 0.076 | 0.314 |
| 12 | 0.085 | 0.314 |
| 14 | 0.095 | 0.314 |
| 16 | 0.104 | 0.314 |
| 18 | 0.114 | 0.314 |

**CT4O2Me.1**

White solid. Yield: 36 %.

$^{1}$H NMR (400 MHz, $CDCl_3$) δ ppm: 8.66 (s, 1H, CH=N), 8.04 (d, *J* = 8.7 Hz, 1H, Ar-H), 7.79 – 7.63 (m, 8H, Ar-H), 7.61 – 7.55 (m, 2H, Ar-H), 7.32 (d, *J* = 7.8 Hz, 2H, Ar-H), 7.19 (d, *J* = 7.9 Hz, 2H, Ar-H), 7.11 (d, *J* = 7.9 Hz, 2H, Ar-H), 6.82 (dd, *J* = 8.7, 2.6 Hz, 1H, Ar-H), 6.73

(d, *J* = 2.6 Hz, 1H, Ar-H), 4.04 (t, *J* = 5.6 Hz, 2H, Ar-$CH_2(CH_2)_2\underline{CH_2}$-O), 2.78 (t, *J* = 6.8 Hz, 2H, Ar-$\underline{CH_2}(CH_2)_2CH_2$-O), 2.55 (s, 3H, Ar$\underline{CH_3}$), 2.38 (s, 3H, Ar$\underline{CH_3}$), 1.89 (p, *J* = 3.5 Hz, 4H, Ar-$CH_2\underline{(CH_2)_2}CH_2$-O). $^{13}$C NMR (101 MHz, $CDCl_3$) δ ppm: 161.25, 157.83, 150.56, 145.30, 141.99, 141.55, 140.65, 137.87, 137.79, 135.31, 132.76 (2C), 129.83 (2C), 129.72, 129.15 (2C), 127.74 (2C), 127.70 (2C), 127.65 (2C), 127.41, 127.13 (2C), 120.93, 119.10, 116.64, 112.58, 110.97, 67.84, 35.35, 28.93, 27.87, 21.10, 19.72. Elemental Analysis for $C_{38}H_{34}N_2O$ Calculated: C = 85.36 % H = 6.41 % N = 5.24 %; Found C = 84.86 % H = 6.65 % N = 4.98 %.

**CT4O2Me.2**

Off-white solid. Yield: 32 %.

$^1$H NMR (400 MHz, $CDCl_3$) δ ppm: 8.67 (s, 1H, $\underline{CH}$=N), 8.03 (d, *J* = 8.7 Hz, 1H, Ar-H), 7.78 – 7.63 (m, 8H, Ar-H), 7.62 – 7.54 (m, 2H, Ar-H), 7.36 – 7.28 (m, 2H, Ar-H), 7.26 – 7.18 (m, 2H, Ar-H), 7.17 – 7.09 (m, 2H, Ar-H), 6.82 (dd, *J* = 8.7, 2.6 Hz, 1H, Ar-H), 6.73 (d, *J* = 2.5 Hz, 1H, Ar-H), 4.08 – 4.01 (m, 2H, Ar-$CH_2(CH_2)_2\underline{CH_2}$-O), 2.77 (m, *J* = 5.2 Hz, 2H, Ar-$\underline{CH_2}(CH_2)_2CH_2$-O), 2.68 (m, *J* = 7.6 Hz, 2H, Ar-$\underline{CH_2}CH_3$, 2.55 (s, 3H, Ar-$\underline{CH_3}$), 1.88 (p, *J* = 3.2 Hz, 4H, Ar-$CH_2\underline{(CH_2)_2}CH_2$-O), 1.26 (t, *J* = 7.6 Hz, 3H, Ar-$CH_2\underline{CH_3}$). $^{13}$C NMR (101 MHz, $CDCl_3$) δ ppm: 161.15, 157.77, 150.67, 145.22, 141.90, 141.66, 141.46, 140.55, 137.78, 137.70, 132.66 (2C), 129.64, 129.05 (2C), 128.54 (2C), 127.64 (2C), 127.60 (2C), 127.55 (2C), 127.34, 127.04 (2C), 120.89, 119.00, 116.54, 112.48, 110.87, 67.74, 35.25, 28.83, 28.43, 27.77, 19.62, 15.74. HRMS: (TOF ESI+) (m/z): $[M+H]^+$ for $C_{39}H_{37}N_2O$ Calculated: 549.2906; Found: 549.2888 (3.3 ppm difference).

**CT4O2Me.3**

Off-white solid. Yield: 45 %.

$^1$H NMR (400 MHz, $CDCl_3$) δ ppm: 8.67 (s, 1H, $\underline{CH}$=N), 8.04 (d, *J* = 8.6 Hz, 1H, Ar-H), 7.76 – 7.63 (m, 8H, Ar-H), 7.58 (d, *J* = 7.9 Hz, 2H, Ar-H), 7.32 (d, *J* = 7.9 Hz, 2H, Ar-H), 7.20 (d, *J* = 8.1 Hz, 2H, Ar-H), 7.13 (d, *J* = 8.3 Hz, 2H, Ar-H), 6.82 (dd, *J* = 8.7, 2.5 Hz, 1H, Ar-H), 6.74 (d, *J* = 2.5 Hz, 1H, Ar-H), 4.08 – 4.01 (m, 2H, Ar-$CH_2(CH_2)_2\underline{CH_2}$-O), 2.81 – 2.73 (m, 2H, Ar-$\underline{CH_2}(CH_2)_2CH_2$-O), 2.61 (t, *J* = 7.6 Hz, 2H, Ar-$\underline{CH_2}CH_2CH_3$), 2.55 (s, 3H, Ar-$\underline{CH_3}$), 1.89 (q, *J* = 3.1 Hz, 4H, Ar-$CH_2\underline{(CH_2)_2}CH_2$-O), 1.67 (h, *J* = 7.4 Hz, 2H, Ar-$CH_2\underline{CH_2}CH_3$), 0.97 (t, *J* = 7.3 Hz, 3H, Ar-$CH_2CH_2\underline{CH_3}$). $^{13}$C NMR (101 MHz, $CDCl_3$) δ ppm: 161.15, 157.73, 150.65, 145.21, 141.90, 141.45, 140.56, 140.11, 137.77, 137.70, 132.66 (2C), 129.67, 129.15 (2C), 129.06 (2C), 127.64 (2C), 127.60 (2C), 127.55 (2C), 127.36, 127.04

(2C), 120.81, 119.01, 116.55, 112.48, 110.87, 67.74, 37.60, 35.25, 28.83, 27.78, 24.69, 19.64, 13.87. Elemental Analysis for $C_{40}H_{38}N_2O$ Calculated: C = 85.37 % H = 6.81 % N = 4.98 %; Found C = 85.50 % H = 6.42 % N = 4.89 %.

**CT4O2Me.4**

Off-white solid. Yield: 30 %.

$^{1}$H NMR (400 MHz, $CDCl_3$) δ ppm: 8.67 (s, 1H, CH=N), 8.03 (d, *J* = 8.7 Hz, 1H, Ar-H), 7.74 (s, 4H, Ar-H), 7.78 – 7.63 (m, 4H, Ar-H), 7.62 – 7.54 (m, 2H, Ar-H), 7.35 – 7.29 (m, 2H, Ar-H), 7.23 – 7.16 (m, 2H, Ar-H), 7.15 – 7.08 (m, 2H, Ar-H), 6.82 (dd, *J* = 8.7, 2.6 Hz, 1H, Ar-H), 6.73 (d, *J* = 2.6 Hz, 1H, Ar-H), 4.08 – 4.01 (m, 2H, Ar-$CH_2(CH_2)_2\underline{CH_2}$-O), 2.77 (m, *J* = 5.3 Hz, 2H, Ar-$\underline{CH_2}(CH_2)_2CH_2$-O), 2.67 – 2.59 (m, 2H, Ar-$\underline{CH_2}CH_2CH_2CH_3$), 2.55 (s, 3H, Ar-$\underline{CH_3}$), 1.88 (m, *J* = 4.4 Hz, 4H, Ar-$CH_2\underline{(CH_2)_2}CH_2$-O), 1.68 – 1.56 (m, 2H, Ar-$CH_2\underline{CH_2}CH_2CH_3$), 1.38 (h, *J* = 7.4 Hz, 2H, Ar-$CH_2CH_2\underline{CH_2}CH_3$), 0.95 (t, *J* = 7.3 Hz, 3H, Ar-$CH_2CH_2CH_2\underline{CH_3}$). $^{13}$C NMR (101 MHz, $CDCl_3$) δ ppm: 161.14, 157.72, 150.60, 145.22, 141.90, 141.46, 140.55, 140.33, 137.77, 137.70, 132.66 (2C), 129.65, 129.09 (2C), 129.05 (2C), 127.64 (2C), 127.60 (2C), 127.55 (2C), 127.36, 127.04 (2C), 120.81, 119.00, 116.54, 112.47, 110.87, 67.74, 35.25, 35.19, 33.78, 28.83, 27.77, 22.38, 19.63, 14.00. Elemental Analysis for $C_{41}H_{40}N_2O$ Calculated: C = 85.38 % H = 6.99 % N = 4.86 %; Found C = 85.72 % H = 6.65 % N = 4.76 %.

**CT4O2Me.5**

Off-white solid. Yield: 31 %.

$^{1}$H NMR (400 MHz, $CDCl_3$) δ ppm: 8.67 (s, 1H, CH=N), 8.04 (d, *J* = 8.7 Hz, 1H, Ar-H), 7.72 (d, *J* = 14.2 Hz, 6H, Ar-H), 7.67 (d, *J* = 8.4 Hz, 2H, Ar-H), 7.61 – 7.55 (m, 2H, Ar-H), 7.35 – 7.29 (m, 2H, Ar-H), 7.23 – 7.17 (m, 2H, Ar-H), 7.16 – 7.09 (m, 2H, Ar-H), 6.82 (dd, *J* = 8.7, 2.6 Hz, 1H, Ar-H), 6.74 (d, *J* = 2.5 Hz, 1H, Ar-H), 4.05 (m, *J* = 4.1 Hz, 2H, Ar-$CH_2(CH_2)_2\underline{CH_2}$-O), 2.81 – 2.73 (m, 2H, Ar-$\underline{CH_2}(CH_2)_2CH_2$-O), 2.67 – 2.59 (m, 2H, Ar-$\underline{CH_2}CH_2CH_2CH_2CH_3$), 2.55 (s, 3H, Ar-$\underline{CH_3}$), 1.89 (m, *J* = 3.9 Hz, 4H, Ar-$CH_2\underline{(CH_2)_2}CH_2$-O), 1.70 – 1.58 (m, 2H, Ar-$CH_2\underline{CH_2}CH_2CH_2CH_3$), 1.43 – 1.28 (m, 4H, Ar-$CH_2CH_2\underline{CH_2CH_2}CH_3$), 0.95 – 0.85 (m, 3H, Ar-$CH_2CH_2CH_2CH_2\underline{CH_3}$). $^{13}$C NMR (101 MHz, $CDCl_3$) δ 161.14, 157.72, 150.60, 145.21, 141.90, 141.45, 140.55, 140.38, 137.77, 137.70, 132.66 (2C), 129.66, 129.09 (2C), 129.06 (2C), 127.64 (2C), 127.60 (2C), 127.55 (2C), 127.36, 127.04 (2C), 120.83, 119.01, 116.55, 112.48, 110.87, 35.48, 35.25, 31.54, 31.32, 28.83, 27.78, 22.60,

19.64, 14.09. Elemental Analysis for $C_{42}H_{42}N_2O$ Calculated: C = 85.38 % H = 7.17 % N = 4.74 %; Found C = 85.22 % H = 7.37 % N = 4.39 %.

**CT4O2Me.6**

Off-white solid. Yield: 47 %.

$^1$H NMR (400 MHz, $CDCl_3$) δ ppm: 8.67 (s, 1H, CH=N), 8.03 (d, *J* = 8.7 Hz, 1H, Ar-H), 7.78 – 7.63 (m, 8H, Ar-H), 7.61 – 7.54 (m, 2H, Ar-H), 7.35 – 7.29 (m, 2H, Ar-H), 7.23 – 7.16 (m, 2H, Ar-H), 7.16 – 7.08 (m, 2H, Ar-H), 6.82 (dd, *J* = 8.7, 2.6 Hz, 1H, Ar-H), 6.73 (d, *J* = 2.6 Hz, 1H, Ar-H), 4.05 (q, *J* = 3.4 Hz, 2H, Ar-$CH_2(CH_2)_2\underline{CH_2}$-O), 2.77 (m, *J* = 5.4 Hz, 2H, Ar-$\underline{CH_2}(CH_2)_2CH_2$-O), 2.67 – 2.58 (m, 2H, Ar-$\underline{CH_2}CH_2(CH_2)_3CH_3$), 2.55 (s, 3H, Ar-$\underline{CH_3}$), 1.88 (m, *J* = 3.2 Hz, 4H, Ar-$CH_2\underline{(CH_2)_2}CH_2$-O), 1.69 – 1.57 (m, 2H, Ar-$CH_2\underline{CH_2}(CH_2)_3CH_3$), 1.42 – 1.23 (m, 6H, Ar-$CH_2CH_2\underline{(CH_2)_3}CH_3$), 0.94 – 0.86 (m, 3H, Ar-$CH_2CH_2(CH_2)_3\underline{CH_3}$). $^{13}$C NMR (101 MHz, $CDCl_3$) δ 161.14, 157.71, 150.59, 145.21, 141.90, 141.46, 140.54, 140.38, 137.77, 137.70, 132.66 (2C), 129.65, 129.08 (2C), 129.05 (2C), 127.64 (2C), 127.60 (2C), 127.55 (2C), 127.36, 127.04 (2C), 120.82, 119.00, 116.54, 112.47, 110.87, 67.74, 35.52, 35.25, 31.77, 31.60, 29.01, 28.83, 27.78, 22.65, 19.64, 14.14. Elemental Analysis for $C_{43}H_{44}N_2O$ Calculated: C = 85.39 % H = 7.33 % N = 4.63 %; Found C = 85.32 % H = 6.97 % N = 4.51 %.

**CT4O2Me.7**

Off-white solid. Yield: 30 %.

$^1$H NMR (400 MHz, $CDCl_3$) δ ppm: 8.67 (s, 1H, CH=N), 8.03 (d, *J* = 8.7 Hz, 1H, Ar-H), 7.78 – 7.63 (m, 8H, Ar-H), 7.62 – 7.54 (m, 2H, Ar-H), 7.36 – 7.28 (m, 2H, Ar-H), 7.23 – 7.16 (m, 2H, Ar-H), 7.14 – 7.06 (m, 2H, Ar-H), 6.82 (dd, *J* = 8.7, 2.6 Hz, 1H, Ar-H), 6.73 (d, *J* = 2.5 Hz, 1H, Ar-H), 4.08 – 4.01 (m, 2H, Ar-$CH_2(CH_2)_2\underline{CH_2}$-O), 2.81 – 2.73 (m, 2H, Ar-$\underline{CH_2}(CH_2)_2CH_2$-O), 2.67 – 2.58 (m, 2H, Ar-$\underline{CH_2}CH_2(CH_2)_4CH_3$), 2.55 (s, 3H, Ar-$\underline{CH_3}$), 1.88 (p, *J* = 3.3 Hz, 4H, Ar-$CH_2\underline{(CH_2)_2}CH_2$-O), 1.62 (m, *J* = 7.0 Hz, 2H, Ar-$CH_2\underline{CH_2}(CH_2)_4CH_3$), 1.39 – 1.23 (m, 8H, Ar-$CH_2CH_2\underline{(CH_2)_4}CH_3$), 0.93 – 0.85 (m, 3H, Ar-$CH_2CH_2(CH_2)_4\underline{CH_3}$). $^{13}$C NMR (101 MHz, $CDCl_3$) δ ppm: 161.14, 157.71, 150.59, 145.21, 141.90, 141.46, 140.55, 140.38, 137.77, 137.70, 132.66, 129.65, 129.08 (2C), 129.05 (2C), 127.64 (2C), 127.60 (2C), 127.55 (2C), 127.36, 127.04 (2C), 120.82, 119.00, 116.55, 112.47, 110.87, 67.74, 35.52, 35.25, 31.86, 31.64, 29.31, 29.23, 28.83, 27.78, 22.70, 19.64, 14.14. HRMS: (TOF ESI+) (m/z): $[M+H]^+$ for $C_{44}H_{47}N_2O$ Calculated: 619.3688; Found: 619.3658 (4.8 ppm difference).

**CT4O2Me.8**

Off-white solid. Yield: 28 %.

$^{1}$H NMR (400 MHz, $CDCl_3$) δ ppm: 8.66 (s, 1H, CH=N), 8.03 (d, *J* = 8.7 Hz, 1H, Ar-H), 7.78 – 7.63 (m, 8H, Ar-H), 7.61 – 7.54 (m, 2H, Ar-H), 7.36 – 7.28 (m, 2H, Ar-H), 7.19 (d, *J* = 8.3 Hz, 2H, Ar-H), 7.15 – 7.07 (m, 2H, Ar-H), 6.82 (dd, *J* = 8.7, 2.6 Hz, 1H, Ar-H), 6.73 (d, *J* = 2.5 Hz, 1H, Ar-H), 4.05 (m, *J* = 3.3 Hz, 2H, Ar-$CH_2(CH_2)_2\underline{CH_2}$-O), 2.76 (m, *J* = 5.2 Hz, 2H, Ar-$\underline{CH_2}(CH_2)_2CH_2$-O), 2.67 – 2.58 (m, 2H, Ar-$\underline{CH_2}CH_2(CH_2)_5CH_3$) 2.55 (s, 3H, Ar-$\underline{CH_3}$), 1.89 (m, *J* = 4.0 Hz, 4H, Ar-$CH_2\underline{(CH_2)_2}CH_2$-O), 1.69 – 1.56 (m, 2H, Ar-$CH_2\underline{CH_2}(CH_2)_5CH_3$), 1.44 – 1.15 (m, 10H, Ar-$CH_2CH_2\underline{(CH_2)_5}CH_3$), 0.93 – 0.85 (m, 3H, Ar-$CH_2CH_2(CH_2)_5\underline{CH_3}$). $^{13}$C NMR (101 MHz, $CDCl_3$) δ ppm: 161.13, 157.71, 150.59, 145.22, 141.89, 141.46, 140.54, 140.38, 137.78, 137.71, 132.66, 129.64, 129.07 (2C), 129.05 (2C), 127.64 (2C), 127.60 (2C), 127.55 (2C), 127.36, 127.04 (2C), 120.81, 119.00, 116.54, 112.46, 110.88, 67.74, 35.52, 35.24, 31.92, 31.64, 29.52, 29.35, 29.30, 28.83, 27.78, 22.70, 19.63, 14.14. Elemental Analysis for $C_{45}H_{48}N_2O$ Calculated: C = 85.40 % H = 7.64 % N = 4.43 %; Found C = 85.49 % H = 7.24 % N = 4.27 %.

**CT4O2Me.9**

Off-white solid. Yield: 31 %.

$^{1}$H NMR (400 MHz, $CDCl_3$) δ ppm: 8.67 (d, *J* = 1.9 Hz, 1H, CH=N), 8.03 (dd, *J* = 8.7, 2.0 Hz, 1H, Ar-H), 7.76 – 7.66 (m, 6H, Ar-H), 7.70 – 7.63 (m, 2H, Ar-H), 7.58 (dd, *J* = 8.1, 2.0 Hz, 2H, Ar-H), 7.32 (dd, *J* = 8.3, 2.0 Hz, 2H, Ar-H), 7.23 – 7.16 (m, 2H, Ar-H), 7.16 – 7.08 (m, 2H, Ar-H), 6.82 (dd, *J* = 8.7, 2.5 Hz, 1H, Ar-H), 6.74 (d, *J* = 2.4 Hz, 1H, Ar-H), 4.04 (d, *J* = 5.7 Hz, 2H, Ar-$CH_2(CH_2)_2\underline{CH_2}$-O), 2.78 (d, *J* = 6.9 Hz, 2H, Ar-$\underline{CH_2}(CH_2)_2CH_2$-O), 2.63 (t, *J* = 7.7 Hz, 2H, Ar-$\underline{CH_2}CH_2(CH_2)_6CH_3$), 2.55 (d, *J* = 1.9 Hz, 3H, Ar-$\underline{CH_3}$), 1.89 (m, *J* = 3.5 Hz, 4H, Ar-$CH_2\underline{(CH_2)_2}CH_2$-O), 1.62 (m, *J* = 7.6 Hz, 3H, Ar-$CH_2\underline{CH_2}(CH_2)_6CH_3$), 1.31 (m, 12H, Ar-$CH_2CH_2\underline{(CH_2)_6}CH_3$), 0.90 (t, *J* = 6.6 Hz, 3H, Ar-$CH_2CH_2(CH_2)_6\underline{CH_3}$). $^{13}$C NMR (101 MHz, $CDCl_3$) δ ppm: 161.15, 157.71, 150.60, 145.21, 141.90, 141.46, 140.55, 140.39, 137.77, 137.70, 132.66, 129.67, 129.09 (2C), 129.06 (2C), 127.64 (2C), 127.60 (2C), 127.55 (2C), 127.37, 127.04 (2C), 120.82, 119.00, 116.56, 112.48, 110.88, 67.75, 35.52, 35.25, 31.94, 31.64, 29.61, 29.57, 29.37, 28.84, 27.78, 22.72, 19.64, 14.16. HRMS: (TOF ESI+) (m/z): $[M+H]^+$ for $C_{46}H_{51}N_2O$ Calculated: 647.4001; Found: 647.3978 (3.6 ppm difference).

**CT4O2Me.10**

Off-white solid. Yield: 25 %.

[1]H NMR (400 MHz, $CDCl_3$) δ ppm: 8.67 (s, 1H, $\underline{CH}$=N), 8.03 (d, *J* = 8.7 Hz, 1H, Ar-H), 7.74 (s, 4H, Ar-H), 7.72 – 7.65 (m, 4H, Ar-H), 7.61 – 7.54 (m, 2H, Ar-H), 7.32 (d, *J* = 8.1 Hz, 2H, Ar-H), 7.23 – 7.16 (m, 2H, Ar-H), 7.16 – 7.08 (m, 2H, Ar-H), 6.82 (dd, *J* = 8.7, 2.6 Hz, 1H, Ar-H), 6.73 (d, *J* = 2.5 Hz, 1H, Ar-H), 4.10 – 4.01 (m, 2H, Ar-$CH_2(CH_2)_2\underline{CH_2}$-O), 2.81 – 2.73 (m, 2H, Ar-$\underline{CH_2}(CH_2)_2CH_2$-O), 2.67 – 2.58 (m, 2H, Ar-$\underline{CH_2}CH_2(CH_2)_7CH_3$), 2.55 (s, 3H, Ar-H), 1.89 (m, *J* = 3.1 Hz, 4H, Ar-$CH_2\underline{(CH_2)_2}CH_2$-O), 1.69 – 1.57 (m, 2H, Ar-$CH_2\underline{CH_2}(CH_2)_7CH_3$), 1.45 – 1.10 (m, 14H, Ar-$CH_2CH_2\underline{(CH_2)_7}CH_3$), 0.89 (t, *J* = 6.7 Hz, 3H, Ar-$CH_2CH_2(CH_2)_7\underline{CH_3}$). [13]C NMR (101 MHz, $CDCl_3$) δ ppm: 161.14, 157.71, 150.59, 145.21, 141.90, 141.46, 140.54, 140.39, 137.77, 137.70, 132.66, 129.65, 129.08 (2C), 129.06 (2C), 127.64 (2C), 127.60 (2C), 127.55 (2C), 127.36, 127.04 (2C), 120.82, 119.00, 116.55, 112.47, 110.88, 67.74, 35.52, 35.25, 31.94, 31.64, 29.66, 29.65, 29.57, 29.37, 29.35, 28.83, 27.78, 22.72, 19.64, 14.16. HRMS: (TOF ESI+) (m/z): $[M+H]^+$ for $C_{45}H_{54}N_2O$ Calculated: 661.4158; Found: 661.4147 (1.7 ppm difference).

**CT4O2Me.12**

Off-white solid. Yield: 26 %.

[1]H NMR (400 MHz, $CDCl_3$) δ ppm: 8.67 (s, 1H, $\underline{CH}$=N), 8.03 (d, *J* = 8.7 Hz, 1H, Ar-H), 7.76 – 7.64 (m, 8H, Ar-H), 7.62 – 7.54 (m, 2H, Ar-H), 7.35 – 7.26 (m, 2H, Ar-H), 7.19 (d, *J* = 8.4 Hz, 2H, Ar-H), 7.15 – 7.05 (m, 2H, Ar-H), 6.82 (dd, *J* = 8.7, 2.6 Hz, 1H, Ar-H), 6.73 (d, *J* = 2.5 Hz, 1H, Ar-H), 4.08 – 4.01 (m, 2H, Ar-$CH_2(CH_2)_2\underline{CH_2}$-O), 2.77 (m, *J* = 5.2 Hz, 2H, Ar-$\underline{CH_2}(CH_2)_2CH_2$-O), 2.67 – 2.58 (m, 2H, Ar-$\underline{CH_2}CH_2(CH_2)_9CH_3$), 2.55 (s, 3H, Ar-$\underline{CH_3}$), 1.88 (m, *J* = 3.4 Hz, 4H, Ar-$CH_2\underline{(CH_2)_2}CH_2$-O), 1.64 (q, *J* = 7.3 Hz, 2H, Ar-$CH_2\underline{CH_2}(CH_2)_9CH_3$), 1.27 (m, 18H, Ar-$CH_2CH_2\underline{(CH_2)_9}CH_3$), 0.89 (t, *J* = 6.8 Hz, 3H, Ar-$CH_2CH_2(CH_2)_9\underline{CH_3}$). [13]C NMR (101 MHz, $CDCl_3$) δ ppm: 161.14, 157.71, 150.59, 145.22, 141.89, 141.46, 140.54, 140.39, 137.78, 137.70, 132.66, 129.65, 129.08 (2C), 129.05 (2C), 127.64 (2C), 127.60 (2C), 127.55 (2C), 127.36, 127.04 (2C), 120.81, 119.00, 116.54, 112.47, 110.87, 67.74, 35.52, 35.25, 31.95, 31.64, 29.71, 29.67, 29.64, 29.56, 29.39, 29.35, 28.83, 27.78, 22.72, 19.63, 14.16. Elemental Analysis for $C_{49}H_{56}N_2O$ Calculated: C = 85.42 % H = 8.19 % N = 4.07 %; Found C = 85.31 % H = 7.98 % N = 4.02 %.

**CT4O2Me.14**

Off-white solid. Yield: 27 %.

$^1H$ NMR (400 MHz, $CDCl_3$) δ ppm: 8.66 (s, 1H, CH=N), 8.03 (d, *J* = 8.7 Hz, 1H, Ar-H), 7.77 – 7.65 (m, 8H, Ar-H), 7.61 – 7.52 (m, 2H, Ar-H), 7.36 – 7.28 (m, 2H, Ar-H), 7.23 – 7.15 (m, 2H, Ar-H), 7.15 – 7.07 (m, 2H, Ar-H), 6.82 (dd, *J* = 8.7, 2.6 Hz, 1H, Ar-H), 6.73 (d, *J* = 2.6 Hz, 1H, Ar-H), 4.08 – 4.01 (m, 2H, Ar-$CH_2(CH_2)_2\underline{CH_2}$-O), 2.76 (m, *J* = 5.6 Hz, 2H, Ar-$\underline{CH_2}(CH_2)_2CH_2$-O), 2.67 – 2.58 (m, 2H, Ar-$\underline{CH_2}CH_2(CH_2)_{11}CH_3$), 2.55 (s, 3H, Ar-$\underline{CH_3}$), 1.88 (m, *J* = 2.7 Hz, 4H, Ar-$CH_2\underline{(CH_2)_2}CH_2$-O), 1.70 – 1.59 (m, 2H, Ar-$CH_2\underline{CH_2}(CH_2)_{11}CH_3$), 1.43 – 1.17 (m, 22H, Ar-$CH_2CH_2\underline{(CH_2)_{11}}CH_3$), 0.93 – 0.84 (m, 3H, Ar-$\underline{CH_3}$). $^{13}C$ NMR (101 MHz, $CDCl_3$) δ ppm: 161.13, 157.70, 150.59, 145.22, 141.89, 141.46, 140.54, 140.39, 137.78, 137.71, 132.66, 129.65, 129.07 (2C), 129.05 (2C), 127.64 (2C), 127.60 (2C), 127.55 (2C), 127.36, 127.04 (2C), 120.81, 118.99, 116.54, 112.46, 110.88, 67.74, 35.52, 35.24, 31.95, 31.64, 29.73, 29.71, 29.68, 29.64, 29.56, 29.39, 29.35, 28.83, 27.78, 22.72, 19.63, 14.15. Elemental Analysis for $C_{51}H_{60}N_2O$ Calculated: C = 85.43 % H = 8.43 % N = 3.91 %; Found C = 85.55 % H = 8.25 % N = 3.84 %.

**CT4O2Me.16**

Off-white solid. Yield: 25 %.

$^1H$ NMR (400 MHz, $CDCl_3$) δ ppm: 8.66 (s, 1H, CH=N), 8.03 (d, *J* = 8.7 Hz, 1H, Ar-H), 7.80 – 7.61 (m, 8H, Ar-H), 7.62 – 7.52 (m, 2H, Ar-H), 7.36 – 7.28 (m, 2H, Ar-H), 7.23 – 7.15 (m, 2H, Ar-H), 7.15 – 7.07 (m, 2H, Ar-H), 6.82 (dd, *J* = 8.7, 2.6 Hz, 1H, Ar-H), 6.73 (d, *J* = 2.5 Hz, 1H, Ar-H), 4.08 – 4.01 (m, 2H, Ar-$CH_2(CH_2)_2\underline{CH_2}$-O), 2.76 (m, *J* = 5.4 Hz, 2H, Ar-$\underline{CH_2}(CH_2)_2CH_2$-O), 2.67 – 2.58 (m, 2H, Ar-$\underline{CH_2}CH_2(CH_2)_{13}CH_3$), 2.55 (s, 3H, Ar-$\underline{CH_3}$), 1.88 (m, 4H, Ar-$CH_2\underline{(CH_2)_2}CH_2$-O), 1.70 – 1.58 (m, 2H, Ar-$CH_2\underline{CH_2}(CH_2)_{13}CH_3$), 1.26 (s, 26H, Ar-$CH_2CH_2\underline{(CH_2)_{13}}CH_3$), 0.92 – 0.84 (m, 3H, Ar-$CH_2CH_2(CH_2)_{13}\underline{CH_3}$). $^{13}C$ NMR (101 MHz, $CDCl_3$) δ ppm: 161.13, 157.70, 150.58, 145.22, 141.89, 141.46, 140.54, 140.39, 137.78, 137.71, 132.66, 129.64, 129.07 (2C), 129.05 (2C), 127.64 (2C), 127.60 (2C), 127.55 (2C), 127.36, 127.04 (2C), 120.81, 118.99, 116.54, 112.46, 110.88, 67.74, 35.52, 35.24, 31.95, 31.64, 29.73, 29.71, 29.69, 29.64, 29.56, 29.39, 29.35, 28.83, 27.78, 22.72, 19.63, 14.15. Elemental Analysis for $C_{53}H_{64}N_2O$ Calculated: C = 85.43 % H = 8.66 % N = 3.76 %; Found C = 85.10 % H = 9.05 % N = 3.63 %.

**CT4O2Me.18**

Off-white solid. Yield: 25 %.

$^1$H NMR (400 MHz, $CDCl_3$) δ ppm: 8.67 (s, 1H, $\underline{CH}$=N), 8.03 (d, *J* = 8.7 Hz, 1H, Ar-H), 7.77 – 7.61 (m, 8H, Ar-H), 7.64 – 7.54 (m, 2H, Ar-H), 7.35 – 7.28 (m, 2H, Ar-H), 7.19 (d, *J* = 8.4 Hz, 2H, Ar-H), 7.16 – 7.09 (m, 2H, Ar-H), 6.82 (m, 1H, Ar-H), 6.73 (s, 1H, Ar-H), 4.04 (m, *J* = 4.2 Hz, 2H, Ar-$CH_2(CH_2)_2\underline{CH_2}$-O), 2.77 (q, *J* = 5.4 Hz, 2H), 2.67 – 2.58 (m, 2H, Ar-$\underline{CH_2}CH_2(CH_2)_{15}CH_3$), 2.55 (s, 3H, Ar-$\underline{CH_3}$), 1.88 (m, *J* = 3.4 Hz, 4H, Ar-$CH_2\underline{(CH_2)_2}CH_2$-O), 1.69 – 1.57 (m, 2H, Ar-$CH_2\underline{CH_2}(CH_2)_{15}CH_3$), 1.27 (s, 30H, Ar-$CH_2CH_2\underline{(CH_2)_{15}}CH_3$), 0.89 (t, *J* = 6.8 Hz, 3H, Ar-$CH_2CH_2(CH_2)_{15}\underline{CH_3}$). $^{13}$C NMR (101 MHz, $CDCl_3$) δ ppm: 161.14, 157.70, 150.59, 145.22, 141.89, 141.46, 140.54, 140.39, 137.77, 137.70, 132.66, 129.65, 129.08 (2C), 129.05 (2C), 127.64 (2C), 127.60 (2C), 127.55 (2C), 127.36, 127.04 (2C), 120.81, 118.99, 116.54, 112.46, 110.88, 67.74, 35.52, 35.25, 31.96, 31.64, 29.73, 29.72, 29.69, 29.65, 29.57, 29.39, 29.36, 28.83, 27.78, 22.72, 19.63, 14.16. Elemental Analysis for $C_{55}H_{68}N_2O$ Calculated: C = 85.44 % H = 8.87 % N = 3.62 %; Found C = 85.65 % H = 8.58 % N = 3.57 %.

### 3$^4$-{6-(4-(((4'-Alkylthiophenyl)imino)methyl)-3-methylphenoxy)hexyl}[1$^1$,2$^1$:2$^4$,3$^1$-terphenyl]-1$^4$-carbonitrile (CT6O2Me.S*m*)

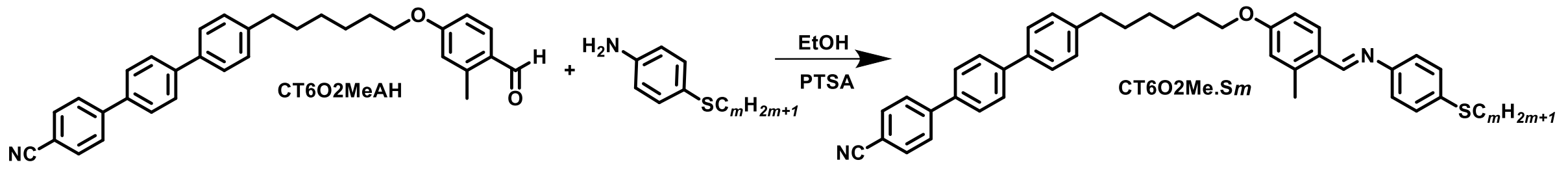

**Scheme S3**: Synthetic pathway for the synthesis of the CT6O2Me.S*m* series

CT6O2MeAH (1 eq, 0.15 g, 0.317 mmol) and the relevant 4-alkylthioaniline (1.4 eq, 0.449 mmol, detailed in Table S9), were combined in ethanol (20 mL) and heated to reflux. Once at temperature, a catalytic amount of PTSA was added, and the reaction was allowed to proceed for 3 hrs. Once complete, the reaction was allowed to cool to room temperature resulting in the formation of a precipitate which was collected by filtration and recrystallised from acetonitrile (10 mL) to yield the target compound as a pale yellow solid.

**Table S9**: Quantities of 4-alkylthioaniline used in the synthesis of the CT6O2Me.Sm series

| *m* | m / g | n / mmol |
|---|---|---|
| 1 | 0.063 | 0.449 |
| 2 | 0.069 | 0.449 |
| 3 | 0.075 | 0.449 |
| 4 | 0.081 | 0.449 |
| 5 | 0.088 | 0.449 |
| 6 | 0.094 | 0.449 |
| 7 | 0.100 | 0.449 |
| 8 | 0.107 | 0.449 |
| 9 | 0.113 | 0.449 |
| 10 | 0.119 | 0.449 |
| 11 | 0.125 | 0.449 |
| 12 | 0.132 | 0.449 |
| 13 | 0.138 | 0.449 |
| 15 | 0.151 | 0.449 |
| 16 | 0.157 | 0.449 |
| 17 | 0.163 | 0.449 |
| 18 | 0.169 | 0.449 |

**CT6O2Me.S1**

Pale yellow solid. Yield: 75 %.

$^{1}$H NMR (400 MHz, $CDCl_3$) δ ppm: 8.65 (s, 1H, $\underline{CH}$=N), 8.02 (d, $^{3}J$ = 8.7 Hz, 1H, Ar-H), 7.73 (m, appt d, 4H, Ar-H), 7.69 (d, $^{3}J$ = 8.6 Hz, 2H, Ar-H), 7.65 (d, $^{3}J$ = 8.6 Hz, 2H, Ar-H), 7.55 (d, $^{3}J$ = 8.2 Hz, 2H, Ar-H), 7.29 (m, 4H, Ar-H), 7.14 (d, $^{3}J$ = 8.7 Hz, 2H, Ar-H), 6.81 (dd, $^{3}J$ = 8.7 Hz, $^{4}J$ = 2.6 Hz, 1H, Ar-H), 6.73 (d, $^{4}J$ = 2.6 Hz, 1H, Ar-H), 4.01 (t, $^{3}J$ = 6.5 Hz, 2H, $Ar(CH_2)_5\underline{CH_2}O$), 2.69 (t, $^{3}J$ = 7.6 Hz, 2H, Ar-$\underline{CH_2}(CH_2)_5O$), 2.55 (s, 3H, Ar-$\underline{CH_3}$), 2.51 (s, 3H, Ar-S$\underline{CH_3}$), 1.82 (, 2H, $Ar(CH_2)_4\underline{CH_2}CH_2O$), 1.71 (p, $^{3}J$ = 7.6 Hz, 2H, Ar-$CH_2\underline{CH_2}(CH_2)_4O$), 1.53 (m, 2H, Ar-$CH_2CH_2CH_2\underline{CH_2}CH_2CH_2O$), 1.44 (m, 2H, Ar-$CH_2CH_2\underline{CH_2}CH_2CH_2CH_2O$). $^{13}$C NMR (101 MHz, $CDCl_3$) δ ppm: 161.49, 158.04, 150.55, 145.34, 142.54, 141.62, 140.80, 137.77, 137.70, 135.26, 132.78 (2C), 129.88, 129.15 (2C), 128.05 (2C), 127.75 (2C), 127.70 (2C), 127.66 (2C), 127.21, 127.07 (2C), 121.71 (2C), 119.12, 116.72, 112.65, 110.98, 68.02, 35.60, 31.43, 29.23, 29.03, 25.99, 19.77, 16.73. IR ($\nu$max/cm$^{-1}$): 2917 (sp$^{2}$ hybridised C-H stretching), 2850 (sp$^{2}$ hybridised C-H stretching), 2226 (nitrile stretching), 1918 (aromatic overtones), 1595 (imine stretching), 1576 (Aryl C=C stretching). HRMS: (TOF ESI$^{+}$)(m/z): $[M+H]^{+}$ Calculated for $C_{40}H_{39}N_2OS$ = 595.2783; Found 595.2802 (+3.2 ppm).

**CT6O2Me.S2**

Pale yellow solid. Yield: 82 %.

$^{1}$H NMR (400 MHz, $CDCl_3$) δ ppm: 8.65 (s, 1H, $\underline{CH}$=N), 8.02 (d, $^{3}J$ = 8.7 Hz, 1H, Ar-H), 7.73 (m, appt d 4H, Ar-H), 7.70 (d, $^{3}J$ = 8.6 Hz, 2H, Ar-H), 7.65 (d, $^{3}J$ = 8.6 Hz, 2H, Ar-H), 7.55 (d, $^{3}J$ = 8.2 Hz, 2H, Ar-H), 7.36 (d, $^{3}J$ = 8.5 Hz, 2H, Ar-H), 7.29 (d, $^{3}J$ = 8.2 Hz, 2H, Ar-H), 7.12 (d, $^{3}J$ = 8.5 Hz, 2H, Ar-H), 6.82 (dd, $^{3}J$ = 8.7 Hz, $^{4}J$ = 2.6 Hz, 1H, Ar-H), 6.73 (d, $^{4}J$ = 2.6 Hz, 1H, Ar-H), 4.01 (t, $^{3}J$ = 6.5 Hz, 2H, $Ar(CH_2)_5\underline{CH_2}O$), 2.95 (q, $^{3}J$ = 7.3 Hz, 2H, Ar-$S\underline{CH_2}CH_3$), 2.69 (t, $^{3}J$ = 7.6 Hz, 2H, Ar-$\underline{CH_2}(CH_2)_5O$), 2.55 (s, 3H, Ar-$\underline{CH_3}$), 1.81 (m, 2H, $Ar(CH_2)_4\underline{CH_2}CH_2O$), 1.71 (p, $^{3}J$ = 7.6 Hz, 2H, Ar-$CH_2\underline{CH_2}(CH_2)_4O$), 1.54 (m, 2H, Ar-$CH_2CH_2CH_2\underline{CH_2}CH_2CH_2O$), 1.46 (m, 2H, Ar-$CH_2CH_2\underline{CH_2}CH_2CH_2CH_2O$),1.32 (t, $^{3}J$ = 7.3 Hz, 3H, Ar-$SCH_2\underline{CH_3}$). $^{13}$C NMR (101 MHz, $CDCl_3$) δ ppm: 161.51, 158.24, 151.26, 145.33, 142.53, 141.61, 140.85, 137.77, 137.69, 133.22, 132.78 (2C), 130.73 (2C), 129.94, 129.14 (2C), 127.74 (2C), 127.70 (2C), 127.66 (2C), 127.18, 127.07 (2C), 121.66 (2C), 119.12, 116.73, 112.64, 110.98, 68.02, 35.60, 31.43, 29.23, 29.04, 28.58, 25.99, 19.78, 14.62. IR (νmax/$cm^{-1}$): 2921 ($sp^2$ hybridised C-H stretching), 2852 ($sp^2$ hybridised C-H stretching), 2221 (nitrile stretching), 1912 (aromatic overtones), 1596 (imine stretching), 1577 (aryl C=C stretching). HRMS: (TOF $ESI^+$)(m/z): $[M+H]^+$ Calculated for $C_{41}H_{41}N_2OS$ = 609.2940; Found: 609.2943 (+0.5 ppm).

**CT6O2Me.S3**

Pale yellow solid. Yield: 22 %.

$^{1}$H NMR (400 MHz, $CDCl_3$) δ ppm: 8.65 (s, 1H, $\underline{CH}$=N), 8.02 (d, $^{3}J$ = 8.7 Hz, 1H, Ar-H), 7.73 (m, appt d, 4H, Ar-H), 7.70 (d, $^{3}J$ = 8.6 Hz, 2H, Ar-H), 7.66 (d, $^{3}J$ = 8.6 Hz, 2H, Ar-H), 7.55 (d, $^{3}J$ = 8.2 Hz, 2H, Ar-H), 7.36 (d, $^{3}J$ = 8.5 Hz, 2H, Ar-H), 7.29 (d, $^{3}J$ = 8.2 Hz, 2H, Ar-H), 7.12 (d, $^{3}J$ = 8.5 Hz, 2H, Ar-H), 6.82 (dd, $^{3}J$ = 8.7 Hz, $^{4}J$ = 2.6 Hz, 1H, Ar-H), 6.73 (d, $^{4}J$ = 2.6 Hz, 1H, Ar-H), 4.01 (t, $^{3}J$ = 6.5 Hz, 2H, $Ar(CH_2)_5\underline{CH_2}O$), 2.90 (t, $^{3}J$ = 7.3 Hz, 2H, Ar-$SCH_2\underline{CH_2}CH_3$), 2.69 (t, $^{3}J$ = 7.7 Hz, 2H, Ar-$\underline{CH_2}(CH_2)_5O$), 2.55 (s, 3H, Ar-$\underline{CH_3}$), 1.81 (m, 2H, $Ar(CH_2)_4\underline{CH_2}CH_2O$), 1.69 (m, 4H, Ar-$CH_2\underline{CH_2}(CH_2)_4O$, Ar-$SCH_2\underline{CH_2}CH_3$),1.54 (m, 2H, Ar-$CH_2CH_2CH_2\underline{CH_2}CH_2CH_2O$), 1.44 (m, 2H, Ar-$CH_2CH_2\underline{CH_2}CH_2CH_2CH_2O$),1.03 (t, $^{3}J$ = 7.3 Hz, 3H, Ar-$SCH_2CH_2\underline{CH_3}$). $^{13}$C NMR (101 MHz, $CDCl_3$) δ ppm: 161.51, 158.20, 151.16, 145.33, 142.54, 141.62, 140.84, 137.77, 137.70, 133.58, 132.78 (2C), 130.64 (2C), 129.93, 129.15 (2C), 127.74 (2C), 127.70 (2C), 127.66 (2C), 127.19, 127.07 (2C), 121.65, 119.12, 116.73, 112.64, 110.98, 68.02, 36.56, 35.60, 31.44, 29.23, 29.04, 25.99, 22.73, 19.78, 13.57.

IR (νmax/$cm^{-1}$): 2921 ($sp^2$ hybridised C-H stretching), 2851 ($sp^2$ hybridised C-H stretching), 2223 (nitrile stretching), 1912 (aromatic overtones), 1597 (imine stretching), 1578 (aryl C=C stretching). HRMS: (TOF $ESI^+$)(m/z): $[M+H]^+$ Calculated for $C_{42}H_{43}N_2OS$ = 623.3096; Found 623.3097 (+0.2 ppm).

**CT6O2Me.S4**

Pale yellow solid. Yield: 50 %.

$^1$H NMR (400 MHz, $CDCl_3$) δ ppm: 8.65 (s, 1H, $\underline{CH}$=N), 8.02 (d, $^3J$ = 8.7 Hz, 1H, Ar-H), 7.73 (m, appt s, 4H, Ar-H), 7.70 (d, $^3J$ = 8.4 Hz, 2H, Ar-H), 7.66 (d, $^3J$ = 8.4 Hz, 2H, Ar-H), 7.55 (d, $^3J$ = 7.9 Hz, 2H, Ar-H), 7.36 (d, $^3J$ = 8.4 Hz, 2H, Ar-H), 7.29 (d, $^3J$ = 7.9 Hz, 2H, Ar-H), 7.12 (d, $^3J$ = 8.4 Hz, 2H, Ar-H), 6.82 (dd, $^3J$ = 8.7 Hz, $^4J$ = 2.6 Hz, 1H, Ar-H), 6.73 (d, $^4J$ = 2.6 Hz, 1H, Ar-H), 4.01 (t, $^3J$ = 6.5 Hz, 2H, $Ar(CH_2)_5\underline{CH_2}O$), 2.92 (t, $^3J$ = 7.4 Hz, 2H, Ar-$S\underline{CH_2}CH_2CH_2CH_3$), 2.69 (t, $^3J$ = 7.6 Hz, 2H, Ar-$\underline{CH_2}(CH_2)_5O$), 2.55 (s, 3H, Ar-$\underline{CH_3}$), 1.82 (m, 2H, $Ar(CH_2)_4\underline{CH_2}CH_2O$), 1.71 (p, $^3J$ = 7.6 Hz, 2H, Ar-$CH_2\underline{CH_2}(CH_2)_4O$), 1.63 (p, $^3J$ = 7.4 Hz, 2H, Ar-$SCH_2\underline{CH_2}CH_2CH_3$), 1.55 (m, 2H, Ar-$CH_2CH_2CH_2\underline{CH_2}CH_2CH_2O$), 1.46 (m, 4H, Ar-$CH_2CH_2\underline{CH_2}CH_2CH_2CH_2O$, Ar-$SCH_2CH_2\underline{CH_2}CH_3$), 0.93 (t, $^3J$ = 7.3 Hz, 3H, Ar-$SCH_2CH_2CH_2\underline{CH_3}$). $^{13}$C NMR (101 MHz, $CDCl_3$) δ ppm: 161.50, 158.18, 151.11, 145.33, 142.54, 141.61, 140.84, 137.77, 137.69, 133.68, 132.78 (2C), 130.51 (2C), 129.93, 129.14 (2C), 127.74 (2C), 127.70 (2C), 127.66 (2C), 127.19, 127.07 (2C), 121.65 (2C), 119.12, 116.72, 112.64, 110.98, 68.02, 35.60, 34.21, 31.44 (2C), 29.23, 29.04, 25.99, 22.08, 19.78, 13.80. IR (νmax/$cm^{-1}$): 2920 ($sp^2$ hybridised C-H stretching), 2851 ($sp^2$ hybridised C-H stretching), 2222 (nitrile stretching), 1912 (aromatic overtones), 1597 (imine stretching), 1577 (aryl C=C stretching). HRMS: (TOF $ESI^+$)(m/z): $[M+H]^+$ Calculated for $C_{43}H_{45}N_2OS$ = 637.3241; Found 637.3248 (-0.8 ppm).

**CT6O2Me.S5**

Pale yellow solid. Yield: 42 %.

$^{1}$H NMR (400 MHz, $CDCl_3$) δ ppm: 8.65 (s, 1H, $\underline{CH}$=N), 8.02 (d, $^{3}J$ = 8.7 Hz, 1H, Ar-H), 7.74 (m, appt s 4H, Ar-H), 7.70 (d, $^{3}J$ = 8.6 Hz, 2H, Ar-H), 7.66 (d, $^{3}J$ = 8.6 Hz, 2H, Ar-H), 7.55 (d, $^{3}J$ = 8.2 Hz, 2H, Ar-H), 7.35 (d, $^{3}J$ = 8.5 Hz, 2H, Ar-H), 7.29 (d, $^{3}J$ = 8.2 Hz, 2H, Ar-H), 7.12 (d, $^{3}J$ = 8.5 Hz, 2H, Ar-H), 6.82 (dd, $^{3}J$ = 8.7 Hz, $^{4}J$ = 2.6 Hz, 1H, Ar-H), 6.73 (d, $^{4}J$ = 2.6 Hz, 1H, Ar-H), 4.01 (t, $^{3}J$ = 6.5 Hz, 2H, $Ar(CH_2)_5\underline{CH_2}O$), 2.92 (t, $^{3}J$ = 7.4 Hz, 2H, Ar-$S\underline{CH_2}CH_2(CH_2)_2CH_3$), 2.69 (t, $^{3}J$ = 7.7 Hz, 2H, Ar-$\underline{CH_2}(CH_2)_5O$), 2.55 (s, 3H, Ar-$\underline{CH_3}$), 1.82 (m, 2H, $Ar(CH_2)_4\underline{CH_2}CH_2O$), 1.67 (m, 4H, Ar-$SCH_2\underline{CH_2}(CH_2)_2CH_3$, Ar-$CH_2\underline{CH_2}(CH_2)_4O$), 1.54 (m, 2H, Ar-$CH_2CH_2CH_2\underline{CH_2}CH_2CH_2O$), 1.37 (m, 6H, Ar-$CH_2CH_2\underline{CH_2}CH_2CH_2CH_2O$, Ar-$SCH_2CH_2\underline{(CH_2)_2}CH_3$), 0.90 (t, $^{3}J$ = 7.2 Hz, 3H, Ar-$SCH_2CH_2(CH_2)_2\underline{CH_3}$). $^{13}$C NMR (101 MHz, $CDCl_3$) δ ppm: 161.50, 158.18, 151.11, 145.33, 142.54, 141.61, 140.83, 137.77, 137.69, 133.69, 132.78 (2C), 130.50 (2C), 129.93, 129.14 (2C), 127.74 (2C), 127.70 (2C), 127.66 (2C), 127.19, 127.07 (2C), 121.65 (2C), 119.11, 116.72, 112.64, 110.98, 68.02, 35.61, 34.50, 31.44, 31.13, 29.23, 29.05 (2C), 25.99, 22.41, 19.78, 14.12. IR (vmax/cm$^{-1}$): 2920 (sp$^{2}$ hybridised C-H stretching), 2856 (sp$^{2}$ hybridised C-H stretching), 2225 (nitrile stretching), 1916 (aromatic overtones), 1601 (imine stretching). HRMS: (TOF ESI$^{+}$)(m/z): $[M+H]^{+}$ Calculated for $C_{44}H_{47}N_2OS$ = 651.3409; Found: 651.3409 (+0.0 ppm).

**CT6O2Me.S6**

Pale yellow solid. Yield: 51 %.

$^{1}$H NMR (400 MHz, $CDCl_3$) δ ppm: 8.65 (s, 1H, $\underline{CH}$=N), 8.02 (d, $^{3}J$ = 8.7 Hz, 1H, Ar-H), 7.73 (m, appt s, 4H, Ar-H), 7.70 (d, $^{3}J$ = 8.6 Hz, 2H, Ar-H), 7.66 (d, $^{3}J$ = 8.6 Hz, 2H, Ar-H), 7.56 (d, $^{3}J$ = 8.2 Hz, 2H, Ar-H), 7.35 (d, $^{3}J$ = 8.5 Hz, 2H, Ar-H), 7.29 (d, $^{3}J$ = 8.2 Hz, 2H, Ar-H), 7.12 (d, $^{3}J$ = 8.5 Hz, 2H, Ar-H), 6.82 (dd, $^{3}J$ = 8.7 Hz, $^{4}J$ = 2.5 Hz, 1H, Ar-H), 6.73 (d, $^{4}J$ = 2.5 Hz, 1H, Ar-H), 4.01 (t, $^{3}J$ = 6.5 Hz, 2H, $Ar(CH_2)_5\underline{CH_2}O$), 2.92 (t, $^{3}J$ = 7.4 Hz, 2H, Ar-$S\underline{CH_2}CH_2(CH_2)_3CH_3$), 2.69 (t, $^{3}J$ = 7.6 Hz, 2H, Ar-$\underline{CH_2}(CH_2)_5O$), 2.55 (s, 3H, Ar-$\underline{CH_3}$), 1.82 (m, 2H, $Ar(CH_2)_4\underline{CH_2}CH_2O$), 1.72 (m, 2H, Ar-$CH_2\underline{CH_2}(CH_2)_4O$), 1.65 (m, 2H, Ar-$SCH_2\underline{CH_2}(CH_2)_3CH_3$), 1.55 (m, 2H, Ar-$CH_2CH_2CH_2\underline{CH_2}CH_2CH_2O$), 1.44 (m, 4H, Ar-$CH_2CH_2\underline{CH_2}CH_2CH_2CH_2O$, Ar-$SCH_2CH_2\underline{CH_2}(CH_2)_2CH_3$), 1.30 (m, 4H, Ar-$SCH_2CH_2CH_2\underline{(CH_2)_2}CH_3$), 0.89 (t, $^{3}J$ = 6.9 Hz, 3H, Ar-$SCH_2CH_2(CH_2)_3\underline{CH_3}$). $^{13}$C NMR (101 MHz, $CDCl_3$) δ ppm: 161.50, 158.18, 151.11, 145.33, 142.54, 141.61, 140.83, 137.77,

137.70, 133.70, 132.78 (2C), 130.50 (2C), 129.93, 129.14 (2C), 127.74 (2C), 127.70 (2C), 127.66 (2C), 127.20, 127.07 (2C), 121.65 (2C), 119.12, 116.73, 112.64, 110.98, 68.02, 35.61, 34.53, 31.52, 31.44, 29.33, 29.23, 29.05, 28.65, 26.00, 22.69, 19.78, 14.17. IR (vmax/$cm^{-1}$): 2920 ($sp^2$ hybridised C-H stretching), 2854 ($sp^2$ hybridised C-H stretching), 2227 (nitrile stretching), 1889 (aromatic overtones), 1601 (imine stretching). HRMS: (TOF $ESI^+$)(m/z): $[M+H]^+$ Calculated for $C_{45}H_{49}N_2OS$ = 665.3566; Found 665.3577 (+1.7 ppm).

**CT6O2Me.S7**

Pale yellow solid. Yield: 36 %.

$^1H$ NMR (400 MHz, $CDCl_3$) δ ppm: 8.65 (s, 1H, $\underline{CH}$=N), 8.02 (d, $^3J$ = 8.7 Hz, 1H, Ar-H), 7.73 (m, appt d, 4H, Ar-H), 7.70 (d, $^3J$ = 8.6 Hz, 2H, Ar-H), 7.66 (d, $^3J$ = 8.6 Hz, 2H, Ar-H), 7.55 (d, $^3J$ = 8.2 Hz, 2H, Ar-H), 7.35 (d, $^3J$ = 8.5 Hz, 2H, Ar-H), 7.29 (d, $^3J$ = 8.2 Hz, 2H, Ar-H), 7.12 (d, $^3J$ = 8.5 Hz, 2H, Ar-H), 6.82 (dd, $^3J$ = 8.7 Hz, $^4J$ = 2.5 Hz, 1H, Ar-H), 6.73 (d, $^4J$ = 2.5 Hz, 1H, Ar-H), 4.01 (t, $^3J$ = 6.5 Hz, 2H, $Ar(CH_2)_5\underline{CH_2}O$), 2.92 (t, $^3J$ = 7.4 Hz, 2H, Ar-$S\underline{CH_2}CH_2(CH_2)_4CH_3$), 2.69 (t, $^3J$ = 7.6 Hz, 2H, Ar-$\underline{CH_2}(CH_2)_5O$), 2.55 (s, 3H, Ar-$\underline{CH_3}$), 1.82 (m, 2H, $Ar(CH_2)_4\underline{CH_2}CH_2O$), 1.72 (m, 2H, Ar-$CH_2\underline{CH_2}(CH_2)_4O$), 1.65 (m, 2H, Ar-$SCH_2\underline{CH_2}(CH_2)_4CH_3$), 1.54 (m, 2H, Ar-$CH_2CH_2CH_2\underline{CH_2}CH_2CH_2O$), 1.43 (m, 4H, Ar-$CH_2CH_2\underline{CH_2}CH_2CH_2CH_2O$, Ar-$SCH_2CH_2\underline{CH_2}(CH_2)_3CH_3$), 1.30 (m, 6H, Ar-$SCH_2CH_2CH_2\underline{(CH_2)_3}CH_3$), 0.89 (t, $^3J$ = 6.9 Hz, 3H, Ar-$SCH_2CH_2(CH_2)_4\underline{CH_3}$). $^{13}C$ NMR (101 MHz, $CDCl_3$) δ 161.50, 158.18, 151.11, 145.33, 142.54, 141.62, 140.84, 137.77, 137.70, 133.70, 132.78 (2C), 130.50 (2C), 129.93, 129.15 (2C), 127.74 (2C), 127.70 (2C), 127.66 (2C), 127.20, 127.07 (2C), 121.65 (2C), 119.12, 116.73, 112.64, 110.98, 68.02, 35.61, 34.53, 31.86, 31.44, 29.37, 29.24, 29.05, 29.00, 28.93, 26.00, 22.74, 19.78, 14.22. IR (vmax/$cm^{-1}$): 2920 ($sp^2$ hybridised C-H stretching), 2854 ($sp^2$ hybridised C-H stretching), 2227 (nitrile stretching), 1910 (aromatic overtones), 1601 (imine stretching), 1577 (aryl C=C stretching). HRMS: (TOF $ESI^+$)(m/z): $[M+H]^+$ Calculated for $C_{46}H_{51}N_2OS$ = 679.3722; Found 679.3700 (-3.2ppm).

**CT6O2Me.S8**

Pale yellow solid. Yield: 54 %.

$^1H$ NMR (400 MHz, $CDCl_3$) δ ppm: 8.65 (s, 1H, $\underline{CH}$=N), 8.02 (d, $^3J$ = 8.7 Hz, 1H, Ar-H), 7.73 (m, appt d, 4H, Ar-H), 7.70 (d, $^3J$ = 8.7 Hz, 2H, Ar-H), 7.66 (d, $^3J$ = 8.7 Hz, 2H, Ar-H), 7.55 (d, $^3J$ = 8.2 Hz, 2H, Ar-H), 7.35 (d, $^3J$ = 8.5 Hz, 2H, Ar-H), 7.29 (d, $^3J$ = 8.2 Hz, 2H, Ar-

H), 7.12 (d, $^3J$ = 8.5 Hz, 2H, Ar-H), 6.82 (dd, $^3J$ = 8.7 Hz, $^4J$ = 2.6 Hz, 1H, Ar-H), 6.73 (d, $^4J$ = 2.6 Hz, 1H, Ar-H), 4.01 (t, $^3J$ = 6.5 Hz, 2H, $Ar(CH_2)_5\underline{CH_2}O$), 2.91 (t, $^3J$ = 7.4 Hz, 2H, Ar-$S\underline{CH_2}CH_2(CH_2)_5CH_3$), 2.69 (t, $^3J$ = 7.6 Hz, 2H, Ar-$\underline{CH_2}(CH_2)_5O$), 2.55 (s, 3H, Ar-$\underline{CH_3}$), 1.82 (m, 2H, $Ar(CH_2)_4\underline{CH_2}CH_2O$), 1.72 (m, 2H, Ar-$CH_2\underline{CH_2}(CH_2)_4O$), 1.65 (m, 2H, Ar-$SCH_2\underline{CH_2}(CH_2)_5CH_3$), 1.54 (m, 2H, Ar-$CH_2CH_2CH_2\underline{CH_2}CH_2CH_2O$), 1.44 (m, 4H, Ar-$CH_2CH_2\underline{CH_2}CH_2CH_2CH_2O$, Ar-$SCH_2CH_2\underline{CH_2}(CH_2)_4CH_3$), 1.30 (m, 8H, Ar-$SCH_2CH_2CH_2\underline{(CH_2)_4}CH_3$), 0.88 (t, $^3J$ = 6.8 Hz, 3H, Ar-$SCH_2CH_2(CH_2)_5\underline{CH_3}$). $^{13}$C NMR (101 MHz, $CDCl_3$) δ 161.51, 158.18, 151.12, 145.35, 142.55, 141.63, 140.84, 137.78, 137.71, 133.71, 132.79 (2C), 130.51 (2C), 129.93, 129.15 (2C), 127.75 (2C), 127.71 (2C), 127.67 (2C), 127.20, 127.08 (2C), 121.65 (2C), 119.12, 116.73, 112.65, 111.00, 68.03, 35.62, 34.54, 31.95, 31.45, 29.37, 29.32, 29.29, 29.24, 29.06, 28.97, 26.00, 22.79, 19.78, 14.25. IR (νmax/$cm^{-1}$): 2920 ($sp^2$ hybridised C-H stretching), 2851 ($sp^2$ hybridised C-H stretching), 2230 (nitrile stretching), 1909 (aromatic overtones), 1602 (imine stretching). HRMS: (TOF $ESI^+$)(m/z): $[M+H]^+$ Calculated for $C_{47}H_{53}N_2OS$ = 693.3879; Found 6933.3892 (+1.9 ppm).

**CT6O2Me.S9**

Pale yellow solid. Yield: 52 %.

$^1$H NMR (400 MHz, $CDCl_3$) δ ppm: 8.65 (s, 1H, $\underline{CH}$=N), 8.02 (d, $^3J$ = 8.7 Hz, 1H, Ar-H), 7.73 (m, appt d, 4H, Ar-H), 7.70 (d, $^3J$ = 8.6 Hz, 2H, Ar-H), 7.65 (d, $^3J$ = 8.6 Hz, 2H, Ar-H), 7.55 (d, $^3J$ = 8.2 Hz, 2H, Ar-H), 7.35 (d, $^3J$ = 8.5 Hz, 2H, Ar-H), 7.29 (d, $^3J$ = 8.2 Hz, 2H, Ar-H), 7.12 (d, $^3J$ = 8.5 Hz, 2H, Ar-H), 6.82 (dd, $^3J$ = 8.7 Hz, $^4J$ = 2.6 Hz, 1H, Ar-H), 6.73 (d, $^4J$ = 2.6 Hz, 1H, Ar-H), 4.01 (t, $^3J$ = 6.5 Hz, 2H, $Ar(CH_2)_5\underline{CH_2}O$), 2.91 (t, $^3J$ = 7.4 Hz, 2H, Ar-$S\underline{CH_2}CH_2(CH_2)_6CH_3$), 2.69 (t, $^3J$ = 7.6 Hz, 2H, Ar-$\underline{CH_2}(CH_2)_5O$), 2.55 (s, 3H, Ar-$\underline{CH_3}$), 1.81 (m, 2H, $Ar(CH_2)_4\underline{CH_2}CH_2O$), 1.71 (m, 2H, Ar-$CH_2\underline{CH_2}(CH_2)_4O$), 1.65 (m, 2H, Ar-$SCH_2\underline{CH_2}(CH_2)_6CH_3$), 1.54 (m, 2H, Ar-$CH_2CH_2CH_2\underline{CH_2}CH_2CH_2O$), 1.43 (m, 4H, Ar-$CH_2CH_2\underline{CH_2}CH_2CH_2CH_2O$, Ar-$SCH_2CH_2\underline{CH_2}(CH_2)_5CH_3$), 1.28 (m, 10H, Ar-$SCH_2CH_2CH_2\underline{(CH_2)_5}CH_3$), 0.88 (t, $^3J$ = 6.9 Hz, 3H, Ar-$SCH_2CH_2(CH_2)_6\underline{CH_3}$). $^{13}$C NMR (101 MHz, $CDCl_3$) δ 161.50, 158.17, 151.10, 145.33, 142.54, 141.62, 140.83, 137.77, 137.70, 133.71, 132.78 (2C), 130.50 (2C), 129.93, 129.14 (2C), 127.74 (2C), 127.70 (2C), 127.66 (2C), 127.20, 127.07 (2C), 121.65 (2C), 119.12, 116.73, 112.64, 110.99, 68.02, 35.61, 34.53, 32.01, 31.44, 29.61, 29.39, 29.37, 29.33, 29.24, 29.05, 28.97, 26.00, 22.81, 19.78, 14.26. IR (νmax/$cm^{-1}$): 2920 ($sp^2$ hybridised C-H stretching), 2848 ($sp^2$ hybridised C-H stretching),

2232 (nitrile stretching), 1909 (aromatic overtones), 1601 (imine stretching). HRMS: (TOF $ESI^+$)(m/z): $[M+H]^+$ Calculated for $C_{48}H_{55}N_2OS$ = 707.4035; Found 707.4063 (+4.0 ppm).

**CT6O2Me.S10**

Pale yellow solid. Yield: 72 %.

$^1H$ NMR (400 MHz, $CDCl_3$) δ ppm: 8.65 (s, 1H, $\underline{CH}$=N), 8.02 (d, $^3J$ = 8.7 Hz, 1H, Ar-H), 7.73 (m, appt d, 4H, Ar-H), 7.70 (d, $^3J$ = 8.6 Hz, 2H, Ar-H), 7.66 (d, $^3J$ = 8.6 Hz, 2H, Ar-H), 7.55 (d, $^3J$ = 8.2 Hz, 2H, Ar-H), 7.35 (d, $^3J$ = 8.5 Hz, 2H, Ar-H), 7.29 (d, $^3J$ = 8.2 Hz, 2H, Ar-H), 7.12 (d, $^3J$ = 8.5 Hz, 2H, Ar-H), 6.82 (dd, $^3J$ = 8.7 Hz, $^4J$ = 2.6 Hz, 1H, Ar-H), 6.73 (d, $^4J$ = 2.6 Hz, 1H, Ar-H), 4.01 (t, $^3J$ = 6.5 Hz, 2H, $Ar(CH_2)_5\underline{CH_2}O$), 2.91 (t, $^3J$ = 7.4 Hz, 2H, $Ar\text{-}S\underline{CH_2}CH_2(CH_2)_7CH_3$), 2.69 (t, $^3J$ = 7.6 Hz, 2H, $Ar\text{-}\underline{CH_2}(CH_2)_5O$), 2.55 (s, 3H, $Ar\text{-}\underline{CH_3}$), 1.81 (m, 2H, $Ar(CH_2)_4\underline{CH_2}CH_2O$), 1.72 (m, 2H, $Ar\text{-}CH_2\underline{CH_2}(CH_2)_4O$), 1.65 (m, 2H, $Ar\text{-}SCH_2\underline{CH_2}(CH_2)_7CH_3$), 1.55 (m, 2H, $Ar\text{-}CH_2CH_2CH_2\underline{CH_2}CH_2CH_2O$), 1.44 (m, 4H, $Ar\text{-}CH_2CH_2\underline{CH_2}CH_2CH_2CH_2O$, $Ar\text{-}SCH_2CH_2\underline{CH_2}(CH_2)_6CH_3$), 1.27 (m, 12H, $Ar\text{-}SCH_2CH_2CH_2\underline{(CH_2)_6}CH_3$), 0.88 (t, $^3J$ = 6.7 Hz, 3H, $Ar\text{-}SCH_2CH_2(CH_2)_7\underline{CH_3}$). $^{13}C$ NMR (101 MHz, $CDCl_3$) δ ppm: 161.50, 158.17, 151.11, 145.34, 142.54, 141.62, 140.83, 137.78, 137.70, 133.71, 132.78 (2C), 130.50 (2C), 129.93, 129.15 (2C), 127.75 (2C), 127.70 (2C), 127.67 (2C), 127.20, 127.08 (2C), 121.65 (2C), 119.12, 116.73, 112.64, 110.99, 68.03, 35.61, 34.54, 32.04, 31.45, 29.69, 29.66, 29.45, 29.37, 29.33, 29.24, 29.06, 28.97, 26.00, 22.82, 19.78, 14.27. IR ($\nu max/cm^{-1}$): 2920 ($sp^2$ hybridised C-H stretching), 2850 ($sp^2$ hybridised C-H stretching), 2231 (nitrile stretching), 1909 (aromatic overtones), 1602 (imine stretching), 1577 (aryl C=C stretching). HRMS: (TOF $ESI^+$)(m/z): $[M+H]^+$ Calculated for $C_{49}H_{57}N_2OS$ = 721.4192; Found 721.4192 (+1.5 ppm).

**CT6O2Me.S11**

Pale yellow solid. Yield: 82 %.

$^1H$ NMR (400 MHz, $CDCl_3$) δ ppm: 8.65 (s, 1H, $\underline{CH}$=N), 8.02 (d, $^3J$ = 8.7 Hz, 1H, Ar-H), 7.73 (m, 4H, Ar-H), 7.70 (d, $^3J$ = 8.6 Hz, 2H, Ar-H), 7.66 (d, $^3J$ = 8.6 Hz, 2H, Ar-H), 7.56 (d, $^3J$ = 8.2 Hz, 2H, Ar-H), 7.35 (d, $^3J$ = 8.5 Hz, 2H, Ar-H), 7.29 (d, $^3J$ = 8.2 Hz, 2H, Ar-H), 7.12 (d, $^3J$ = 8.5 Hz, 2H, Ar-H), 6.82 (dd, $^3J$ = 8.7 Hz, $^4J$ = 2.6 Hz, 1H, Ar-H), 6.73 (d, $^3J$ = 2.6 Hz, 1H, Ar-H), 4.01 (t, $^3J$ = 6.5 Hz, 2H, $Ar(CH_2)_5\underline{CH_2}O$), 2.91 (t, $^3J$ = 7.4 Hz, 2H, $Ar\text{-}S\underline{CH_2}CH_2(CH_2)_8CH_3$), 2.69 (t, $^3J$ = 7.6 Hz, 2H, $Ar\text{-}\underline{CH_2}(CH_2)_5O$), 2.55 (s, 3H, $Ar\text{-}\underline{CH_3}$), 1.82

(m, 2H, $Ar(CH_2)_4\underline{CH_2}CH_2O$), 1.72 (m, 2H, $Ar\text{-}CH_2\underline{CH_2}(CH_2)_4O$), 1.65 (m, 2H, $Ar\text{-}SCH_2\underline{CH_2}(CH_2)_8CH_3$), 1.53 (m, 2H, $Ar\text{-}CH_2CH_2CH_2\underline{CH_2}CH_2CH_2O$), 1.44 (m, 4H, $Ar\text{-}CH_2CH_2\underline{CH_2}CH_2CH_2CH_2O$, $Ar\text{-}SCH_2CH_2\underline{CH_2}(CH_2)_7CH_3$), 1.27 (m, 14H, $Ar\text{-}SCH_2CH_2CH_2\underline{(CH_2)_7}CH_3$), 0.88 (t, $^3J$ = 6.7 Hz, 3H, $Ar\text{-}SCH_2CH_2(CH_2)_8\underline{CH_3}$). $^{13}C$ NMR (101 MHz, $CDCl_3$) δ ppm: 161.50, 158.17, 151.09, 145.33, 142.54, 141.61, 140.83, 137.77, 137.69, 133.71, 132.78 (2C), 130.49 (2C), 129.92, 129.14 (2C), 127.74 (2C), 127.70 (2C), 127.66 (2C), 127.19, 127.07 (2C), 121.65 (2C), 119.11, 116.72, 112.64, 110.98, 68.02, 35.61, 34.53, 32.05, 31.44, 29.75, 29.73, 29.66, 29.48, 29.36, 29.33, 29.24, 29.05, 28.97, 26.00, 22.83, 19.77, 14.26. IR (νmax/$cm^{-1}$): 2916 ($sp^2$ hybridised C-H stretching), 2849 ($sp^2$ hybridised C-H stretching), 2226 (nitrile stretching), 1922 (aromatic overtones), 1595 (imine stretching), 1577 (aryl C=C stretching). HRMS: (TOF $ESI^+$)(m/z): $[M+H]^+$ Calculated for $C_{50}H_{59}N_2OS$ = 735.4348; Found 735.4370 (+3.0 ppm).

**CT6O2Me.S12**

Pale yellow solid. Yield: 54 %.

$^1H$ NMR (400 MHz, $CDCl_3$) δ ppm: 8.65 (s, 1H, $\underline{CH}$=N), 8.02 (d, $^3J$ = 8.7 Hz, 1H, Ar-H), 7.73 (m, appt d, 4H, Ar-H), 7.70 (d, $^3J$ = 8.6 Hz, 2H, Ar-H), 7.65 (d, $^3J$ = 8.6 Hz, 2H, Ar-H), 7.56 (d, $^3J$ = 8.2 Hz, 2H, Ar-H), 7.35 (d, $^3J$ = 8.5 Hz, 2H, Ar-H), 7.29 (d, $^3J$ = 8.2 Hz, 2H, Ar-H), 7.12 (d, $^3J$ = 8.5 Hz, 2H, Ar-H), 6.82 (dd, $^3J$ = 8.7 Hz, $^4J$ = 2.6 Hz, 1H, Ar-H), 6.73 (d, $^3J$ = 2.6 Hz, 1H, Ar-H), 4.01 (t, $^3J$ = 6.4 Hz, 2H, $Ar(CH_2)_5\underline{CH_2}O$), 2.91 (t, $^3J$ = 7.3 Hz, 2H, $Ar\text{-}S\underline{CH_2}CH_2(CH_2)_9CH_3$), 2.69 (t, $^3J$ = 7.6 Hz, 2H, $Ar\text{-}\underline{CH_2}(CH_2)_5O$), 2.55 (s, 3H, $Ar\text{-}\underline{CH_3}$), 1.82 (m, 2H, $Ar(CH_2)_4\underline{CH_2}CH_2O$), 1.73 (m, 2H, $Ar\text{-}CH_2\underline{CH_2}(CH_2)_4O$), 1.66 (m, 2H, $Ar\text{-}SCH_2\underline{CH_2}(CH_2)_9CH_3$), 1.56 (m, 2H, $Ar\text{-}CH_2CH_2CH_2\underline{CH_2}CH_2CH_2O$), 1.45 (m, 4H, $Ar\text{-}CH_2CH_2\underline{CH_2}CH_2CH_2CH_2O$, $Ar\text{-}SCH_2CH_2\underline{CH_2}(CH_2)_8CH_3$), 1.27 (m, 16H, $Ar\text{-}SCH_2CH_2CH_2\underline{(CH_2)_8}CH_3$), 0.89 (t, $^3J$ = 6.7 Hz, 3H, $Ar\text{-}SCH_2CH_2(CH_2)_9\underline{CH_3}$). $^{13}C$ NMR (101 MHz, $CDCl_3$) δ ppm: 161.50, 158.17, 151.08, 145.32, 142.53, 141.60, 140.83, 137.76, 137.68, 133.71, 132.77 (2C), 130.48 (2C), 129.92, 129.14 (2C), 127.73 (2C), 127.69 (2C), 127.65 (2C), 127.19, 127.06 (2C), 121.64 (2C), 119.11, 116.72, 112.63, 110.97, 68.02, 35.60, 34.52, 32.05, 31.43, 29.79, 29.77, 29.73, 29.66, 29.48, 29.36, 29.33, 29.23, 29.05, 28.96, 25.99, 22.83, 19.77, 14.26. IR (νmax/$cm^{-1}$): 2918 ($sp^2$ hybridised C-H stretching), 2850 ($sp^2$ hybridised C-H stretching), 2229 (nitrile stretching), 1910 (aromatic overtones), 1601 (imine stretching), 1568 (aryl C=C stretching). HRMS: (TOF $ESI^+$)(m/z): $[M+H]^+$ Calculated for $C_{51}H_{61}N_2OS$ = 749.4505; Found 749.4528 (+3.1 ppm).

**CT6O2Me.S13**

Pale yellow solid. Yield: 27 %.

$^{1}$H NMR (400 MHz, $CDCl_3$) δ ppm: 8.65 (s, 1H,<u>CH</u>=N), 8.02 (d, $^{3}J$ = 8.7 Hz, 1H, Ar-H), 7.73 (m, appt d 4H, Ar-H), 7.70 (d, $^{3}J$ = 8.6 Hz, 2H, Ar-H), 7.66 (d, $^{3}J$ = 8.6 Hz, 2H, Ar-H), 7.55 (d, $^{3}J$ = 8.2 Hz, 2H, Ar-H), 7.35 (d, $^{3}J$ = 8.5 Hz, 2H, Ar-H), 7.29 (d, $^{3}J$ = 8.2 Hz, 2H, Ar-H), 7.12 (d, $^{3}J$ = 8.5 Hz, 2H, Ar-H), 6.82 (dd, $^{3}J$ = 8.7 Hz, $^{4}J$ = 2.6 Hz, 1H, Ar-H), 6.73 (d, $^{4}J$ = 2.6 Hz, 1H, Ar-H), 4.01 (t, $^{3}J$ = 6.5 Hz, 2H, Ar$(CH_2)_5\underline{CH_2}$O), 2.91 (t, $^{3}J$ = 7.4 Hz, 2H, Ar-S$\underline{CH_2}CH_2(CH_2)_{10}CH_3$), 2.69 (t, $^{3}J$ = 7.6 Hz, 2H, Ar-$\underline{CH_2}(CH_2)_5$O), 2.55 (s, 3H, Ar-$\underline{CH_3}$), 1.82 (m, 2H, Ar$(CH_2)_4\underline{CH_2}CH_2$O), 1.72 (p, $^{3}J$ = 7.6 Hz, 2H, Ar-$CH_2\underline{CH_2}(CH_2)_4$O), 1.65 (m, 2H, Ar-S$CH_2\underline{CH_2}(CH_2)_{10}CH_3$), 1.54 (m, 2H, Ar-$CH_2CH_2CH_2\underline{CH_2}CH_2CH_2$O), 1.44 (m, 4H, Ar-$CH_2CH_2\underline{CH_2}CH_2CH_2CH_2$O, Ar-S$CH_2CH_2\underline{CH_2}(CH_2)_9CH_3$), 1.26 (m, 18H, Ar-S$CH_2CH_2CH_2\underline{(CH_2)_9}CH_3$), 0.88 (t, $^{3}J$ = 6.7 Hz, 3H, Ar-S$CH_2CH_2(CH_2)_{10}\underline{CH_3}$). $^{13}$C NMR (101 MHz, $CDCl_3$) δ ppm: 161.51, 158.17, 151.11, 145.34, 142.55, 141.62, 140.83, 137.78, 137.71, 133.71, 132.79 (2C), 130.50 (2C), 129.93, 129.15 (2C), 127.75 (2C), 127.71 (2C), 127.67 (2C), 127.21, 127.08 (2C), 121.65 (2C), 119.12, 116.73, 112.64, 110.99, 68.03, 35.62, 34.54, 32.07, 31.45, 29.83, 29.79 (2C), 29.74, 29.67, 29.50, 29.37, 29.34, 29.24, 29.06, 28.98, 26.00, 22.84, 19.78, 14.27. IR (νmax/$cm^{-1}$): 2921 ($sp^2$ hybridised C-H stretching), 2849 ($sp^2$ hybridised C-H stretching), 2225 (nitrile stretching), 1908 (aromatic overtones), 1598 (imine stretching), 1579 (aryl C=C stretching). HRMS: (TOF $ESI^+$)(m/z): $[M+H]^+$ Calculated for $C_{52}H_{63}N_2OS$ = 763.4661; Found 763.4680 (+2.5 ppm).

**CT6O2Me.S15**

Pale yellow solid. Yield: 51 %.

$^{1}$H NMR (400 MHz, $CDCl_3$) δ ppm: 8.65 (s, 1H, <u>CH</u>=N), 8.02 (d, $^{3}J$ = 8.7 Hz, 1H, Ar-H), 7.74 (m, appt s 4H, Ar-H), 7.71 (d, $^{3}J$ = 8.6 Hz, 2H, Ar-H), 7.66 (d, $^{3}J$ = 8.6 Hz, 2H, Ar-H), 7.56 (d, $^{3}J$ = 8.2 Hz, 2H, Ar-H), 7.35 (d, $^{3}J$ = 8.5 Hz, 2H, Ar-H), 7.29 (d, $^{3}J$ = 8.2 Hz, 2H, Ar-H), 7.12 (d, $^{3}J$ = 8.5 Hz, 2H, Ar-H), 6.82 (dd, $^{3}J$ = 8.7 Hz, $^{4}J$ = 2.6 Hz, 1H, Ar-H), 6.73 (d, $^{4}J$ = 2.6 Hz, 1H, Ar-H), 4.01 (t, $^{3}J$ = 6.5 Hz, 2H, Ar$(CH_2)_5\underline{CH_2}$O), 2.91 (t, $^{3}J$ = 7.4 Hz, 2H, Ar-S$\underline{CH_2}CH_2(CH_2)_{12}CH_3$), 2.69 (t, $^{3}J$ = 7.7 Hz, 2H, Ar-$\underline{CH_2}(CH_2)_5$O), 2.55 (s, 3H, Ar-$\underline{CH_3}$), 1.82 (m, 2H, Ar$(CH_2)_4\underline{CH_2}CH_2$O), 1.72 (m, 2H, Ar-$CH_2\underline{CH_2}(CH_2)_4$O), 1.65 (m, 2H, Ar-S$CH_2\underline{CH_2}(CH_2)_{12}CH_3$), 1.55 (m, 2H, Ar-$CH_2CH_2CH_2\underline{CH_2}CH_2CH_2$O), 1.45 (m, 4H, Ar-$CH_2CH_2\underline{CH_2}CH_2CH_2CH_2$O, Ar-S$CH_2CH_2\underline{CH_2}(CH_2)_{11}CH_3$), 1.26 (s, 22H, Ar-

$SCH_2CH_2CH_2(\underline{CH_2})_{11}CH_3$), 0.88 (t, $^3J$ = 7.0 Hz, 3H, Ar-$SCH_2CH_2(CH_2)_{12}\underline{CH_3}$). $^{13}$C NMR (101 MHz, $CDCl_3$) δ ppm: 161.51, 158.17, 151.07, 145.33, 142.54, 141.61, 140.84, 137.76, 137.69, 133.73, 132.78 (2C), 130.48 (2C), 129.94, 129.14 (2C), 127.74 (2C), 127.69 (2C), 127.66 (2C), 127.18, 127.07 (2C), 121.65 (2C), 119.11, 116.73, 112.64, 110.98, 68.02, 35.61, 34.52, 32.07, 31.44, 29.84, 29.83, 29.82, 29.80 (2C), 29.74, 29.66, 29.50, 29.36, 29.33, 29.23, 29.05, 28.97, 25.99, 22.83, 19.77, 14.27. IR (νmax/$cm^{-1}$): 2919 ($sp^2$ hybridised C-H stretching), 2850 ($sp^2$ hybridised C-H stretching), 2224 (nitrile stretching), 1903 (aromatic overtones), 1598 (imine stretching), 1564 (aryl C=C stretching). HRMS: (TOF $ESI^+$)(m/z): $[M+H]^+$ Calculated for $C_{54}H_{67}N_2OS$ = 791.4974; Found 791.5004 (+3.8 ppm).

**CT6O2Me.S16**

Pale yellow solid. Yield: 16 %.

$^1$H NMR (400 MHz, $CDCl_3$) δ ppm: 8.65 (s, 1H, $\underline{CH}$=N), 8.02 (d, $^3J$ = 8.7 Hz, 1H, Ar-H), 7.73 (m, appt s 4H, Ar-H), 7.70 (d, $^3J$ = 8.6 Hz, 2H, Ar-H), 7.66 (d, $^3J$ = 8.6 Hz, 2H, Ar-H), 7.55 (d, $^3J$ = 8.2 Hz, 2H, Ar-H), 7.35 (d, $^3J$ = 8.5 Hz, 2H, Ar-H), 7.29 (d, $^3J$ = 8.2 Hz, 2H, Ar-H), 7.12 (d, $^3J$ = 8.5 Hz, 2H, Ar-H), 6.82 (dd, $^3J$ = 8.7 Hz, $^4J$ = 2.6 Hz, 1H, Ar-H), 6.73 (d, $^4J$ = 2.6 Hz, 1H, Ar-H), 4.01 (t, $^3J$ = 6.5 Hz, 2H, $Ar(CH_2)_5\underline{CH_2}O$), 2.91 (t, $^3J$ = 7.4 Hz, 2H, Ar-$S\underline{CH_2}CH_2(CH_2)_{13}CH_3$), 2.69 (t, $^3J$ = 7.6 Hz, 2H, Ar-$\underline{CH_2}(CH_2)_5O$), 2.55 (s, 3H, Ar-$\underline{CH_3}$), 1.83 (m, 2H, $Ar(CH_2)_4\underline{CH_2}CH_2O$), 1.72 (p, $^3J$ = 7.6 Hz, 2H, Ar-$CH_2\underline{CH_2}(CH_2)_4O$), 1.65 (m, 2H, Ar-$SCH_2\underline{CH_2}(CH_2)_{13}CH_3$), 1.54 (m, 2H, Ar-$CH_2CH_2CH_2\underline{CH_2}CH_2CH_2O$), 1.43 (m, 4H, Ar-$CH_2CH_2\underline{CH_2}CH_2CH_2CH_2O$, Ar-$SCH_2CH_2\underline{CH_2}(CH_2)_{12}CH_3$), 1.26 (m, 24H, Ar-$SCH_2CH_2CH_2(\underline{CH_2})_{12}CH_3$), 0.88 (t, $^3J$ = 6.9 Hz, 3H, Ar-$SCH_2CH_2(CH_2)_{13}\underline{CH_3}$). $^{13}$C NMR (101 MHz, $CDCl_3$) δ ppm: 161.50, 158.17, 151.11, 145.34, 142.54, 141.62, 140.83, 137.78, 137.70, 133.71, 132.78 (2C), 130.50 (2C), 129.93, 129.15 (2C), 127.75 (2C), 127.70 (2C), 127.67 (2C), 127.20, 127.08 (2C), 121.65 (2C), 119.12, 116.73, 112.64, 110.99, 68.03, 35.61, 34.54, 32.07, 31.45, 29.84 (3C), 29.82 , 29.80 (2C), 29.74, 29.67, 29.51, 29.37, 29.34, 29.24, 29.06, 28.98, 26.00, 22.84, 19.78, 14.27. IR (νmax/$cm^{-1}$): 2921 ($sp^2$ hybridised C-H stretching), 2850 ($sp^2$ hybridised C-H stretching), 2227 (nitrile stretching), 1918 (aromatic overtones), 1603 (imine stretching), 1557 (aryl C=C stretching). HRMS: (TOF $ESI^+$)(m/z): $[M+H]^+$ Calculated for $C_{55}H_{69}N_2OS$ = 805.5131; Found 805.5114 (-2.1 ppm).

**CT6O2Me.S17**

Pale yellow solid. Yield: 91%.

$^{1}$H NMR (400 MHz, $CDCl_3$) δ ppm: 8.65 (s, 1H, $\underline{CH}$=N), 8.02 (d, $^{3}J$ = 8.7 Hz, 1H, Ar-H), 7.73 (m, appt s, 4H, Ar-H), 7.71 (d, $^{3}J$ = 8.6 Hz, 2H, Ar-H), 7.66 (d, $^{3}J$ = 8.6 Hz, 2H, Ar-H), 7.55 (d, $^{3}J$ = 8.2 Hz, 2H, Ar-H), 7.35 (d, $^{3}J$ = 8.5 Hz, 2H, Ar-H), 7.29 (d, $^{3}J$ = 8.2 Hz, 2H, Ar-H), 7.12 (d, $^{3}J$ = 8.5 Hz, 2H, Ar-H), 6.82 (dd, $^{3}J$ = 8.7 Hz, $^{4}J$ = 2.6 Hz, 1H, Ar-H), 6.73 (d, $^{4}J$ = 2.6 Hz, 1H, Ar-H), 4.01 (t, $^{3}J$ = 6.5 Hz, 2H, $Ar(CH_2)_5\underline{CH_2}O$), 2.91 (t, $^{3}J$ = 7.4 Hz, 2H, Ar-$S\underline{CH_2}CH_2(CH_2)_{14}CH_3$), 2.69 (t, $^{3}J$ = 7.7 Hz, 2H, Ar-$\underline{CH_2}(CH_2)_5O$), 2.55 (s, 3H, Ar-$\underline{CH_3}$), 1.82 (m, 2H, $Ar(CH_2)_4\underline{CH_2}CH_2O$), 1.72 (m, 2H, Ar-$CH_2\underline{CH_2}(CH_2)_4O$), 1.64 (m, 2H, Ar-$SCH_2\underline{CH_2}(CH_2)_{14}CH_3$), 1.54 (m, 2H, Ar-$CH_2CH_2CH_2\underline{CH_2}CH_2CH_2O$), 1.44 (m, 4H, Ar-$CH_2CH_2\underline{CH_2}CH_2CH_2CH_2O$, Ar-$SCH_2CH_2\underline{CH_2}(CH_2)_{13}CH_3$), 1.26 (m, 26H, Ar-$SCH_2CH_2CH_2\underline{(CH_2)_{13}}CH_3$), 0.88 (t, $^{3}J$ = 6.8 Hz, 3H, Ar-$SCH_2CH_2(CH_2)_{14}\underline{CH_3}$). $^{13}$C NMR (101 MHz, $CDCl_3$) δ ppm: 161.50, 158.17, 151.11, 145.34, 142.54, 141.62, 140.83, 137.78, 137.70, 133.71, 132.78 (2C), 130.50 (2C), 129.94, 129.15 (2C), 127.75 (2C), 127.70 (2C), 127.66 (2C), 127.20, 127.08 (2C), 121.65 (2C), 119.12, 116.73, 112.64, 110.99, 68.03, 35.61, 34.54, 32.08, 31.45, 29.85 (4C), 29.83, 29.80 (2C), 29.74, 29.67, 29.51, 29.37, 29.34, 29.24, 29.06, 28.98, 26.00, 22.84, 19.78, 14.27. IR (νmax/$cm^{-1}$): 2917 ($sp^2$ hybridised C-H stretching), 2848 ($sp^2$ hybridised C-H stretching), 2224 (nitrile stretching), 1910 (aromatic overtones), 1599 (imine stretching), 1568 (aryl C=C stretching). HRMS: (TOF $ESI^+$)(m/z): $[M+H]^+$ Calculated for $C_{56}H_{71}N_2OS$ = 819.5287; Found 819.5280 (-0.9 ppm).

**CT6O2Me.S18**

Pale yellow solid. Yield: 9 %.

$^{1}$H NMR (400 MHz, $CDCl_3$) δ ppm: 8.64 (s, 1H, $\underline{CH}$=N), 8.02 (d, $^{3}J$ = 8.7 Hz, 1H, Ar-H), 7.73 (m, appt s, 4H, Ar-H), 7.70 (d, $^{3}J$ = 8.7 Hz, 2H, Ar-H), 7.65 (d, $^{3}J$ = 8.7 Hz, 2H, Ar-H), 7.55 (d, $^{3}J$ = 8.2 Hz, 2H, Ar-H), 7.35 (d, $^{3}J$ = 8.5 Hz, 2H, Ar-H), 7.28 (d, $^{3}J$ = 8.2 Hz, 2H, Ar-H), 7.12 (d, $^{3}J$ = 8.5 Hz, 2H, Ar-H), 6.81 (dd, $^{3}J$ = 8.7 Hz, $^{4}J$ = 2.6 Hz, 1H, Ar-H), 6.73 (d, $^{4}J$ = 2.6 Hz, 1H, Ar-H), 4.01 (t, $^{3}J$ = 6.5 Hz, 2H, $Ar(CH_2)_5\underline{CH_2}O$), 2.91 (t, $^{3}J$ = 7.4 Hz, 2H, Ar-$S\underline{CH_2}CH_2(CH_2)_{15}CH_3$), 2.69 (t, $^{3}J$ = 7.7 Hz, 2H, Ar-$\underline{CH_2}(CH_2)_5O$), 2.55 (s, 3H, Ar-$\underline{CH_3}$), 1.82 (m, 2H, $Ar(CH_2)_4\underline{CH_2}CH_2O$), 1.71 (m, 2H, Ar-$CH_2\underline{CH_2}(CH_2)_4O$), 1.64 (m, 2H, Ar-$SCH_2\underline{CH_2}(CH_2)_{15}CH_3$), 1.54 (m, 2H, Ar-$CH_2CH_2CH_2\underline{CH_2}CH_2CH_2O$), 1.44 (m, 4H, Ar-$CH_2CH_2\underline{CH_2}CH_2CH_2CH_2O$, Ar-$SCH_2CH_2\underline{CH_2}(CH_2)_{14}CH_3$), 1.25 (m, 28H, Ar-

$SCH_2CH_2CH_2(\underline{CH_2})_{14}CH_3$), 0.88 (t, $^3J$ = 6.7 Hz, 3H, Ar-$SCH_2CH_2(CH_2)_{15}\underline{CH_3}$). $^{13}$C NMR (101 MHz, $CDCl_3$) δ ppm: 161.51, 158.18, 151.11, 145.35, 142.55, 141.63, 140.84, 137.79, 137.71, 133.71, 132.79 (2C), 130.50 (2C), 129.93, 129.15 (2C), 127.76 (2C), 127.71 (2C), 127.67 (2C), 127.21, 127.08 (2C), 121.65 (2C), 119.12, 116.73, 112.65, 111.00, 68.03, 35.62, 34.54, 32.08, 31.45, 29.85 (4C), 29.84 (2C), 29.81 (2C), 29.75, 29.67, 29.51, 29.38, 29.34, 29.25, 29.06, 28.98, 26.01, 22.85, 19.79, 14.28. IR (νmax/cm$^{-1}$): 2918 (sp$^2$ hybridised C-H stretching), 2849 (sp$^2$ hybridised C-H stretching), 2223 (nitrile stretching), 1599 (imine stretching), 1568 (aryl C=C stretching). HRMS: (TOF ESI$^+$)(m/z): [M+H]$^+$ Calculated for $C_{57}H_{73}N_2OS$ = 833.5444; Found 833.5403 (-4.9 ppm).

**3$^4$-{6-(4-(((4'-Alkylphenyl)imino)methyl)-3-fluorophenoxy)hexyl}[1$^1$,2$^1$:2$^4$,3$^1$-terphenyl]-1$^4$-carbonitrile (CT6O2F.*m*)**

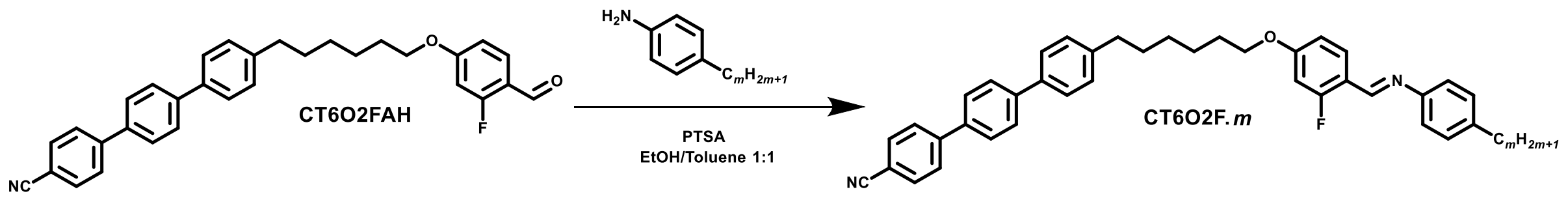

**Scheme S4**: Synthetic pathway used in the synthesis of the CT6O2F.*m* series

To a flask containing 4″-(6-(4-formyl-3-fluorophenoxy)hexyl)-[1,1′:4,1″-terphenyl]-4-carbonitrile (CT6O2FAH, 10.3.8) (1 eq, 0.10 g, 0.209 mmol) and 15 mL EtOH, 15 mL toluene, the appropriate 4-alkylaniline (2 eq, detailed in Table S10) was added and the reaction mixture heated to reflux. A catalytic amount of *p*-toluenesulfonic acid was added and the reaction allowed to proceed for 3 hr. The reaction mixture was cooled to room temperature, and off white precipitate formed which was collected by vacuum filtration. The crude product was recrystalised from 8 mL ethanol and 8 mL toluene to yield the target compound as a white solid.

**Table S10**: Quantities of 4-alkylanilines used in the synthesis of the CT6O2F.m series

| *m* | m / g | n / mmol |
|---|---|---|
| 1 | 0.045 | 0.419 |
| 2 | 0.051 | 0.419 |
| 3 | 0.057 | 0.419 |
| 4 | 0.060 | 0.419 |
| 5 | 0.068 | 0.419 |
| 6 | 0.074 | 0.419 |
| 7 | 0.080 | 0.419 |
| 8 | 0.086 | 0.419 |
| 9 | 0.092 | 0.419 |
| 10 | 0.098 | 0.419 |
| 12 | 0.110 | 0.419 |
| 14 | 0.121 | 0.419 |
| 16 | 0.133 | 0.419 |
| 18 | 0.145 | 0.419 |

**CT6O2F.1**

White solid. Yield: 79%.

$^{1}$H NMR (400 MHz, $CDCl_3$) δ ppm: 8 .68 (s, 1H, <u>CH</u>=N), 8.09 (dd, appt t, $^{3}J_{H-H}$ = 8.6 Hz, $^{4}J_{H-F}$ = 8.6 Hz, 1H, Ar-H), 7.73 (m, 4H, appt s, Ar-H), 7.70 (d, $^{3}J_{H-H}$ = 8.5 Hz, 2H, Ar-H), 7.66 (d, $^{3}J_{H-H}$ = 8.5 Hz, 2H, Ar-H), 7.56 (d, $^{3}J_{H-H}$ = 8.2 Hz, 2H, Ar-H), 7.29 (d, $^{3}J_{H-H}$ = 8.2 Hz, 2H, Ar-H), 7.19 (d, $^{3}J_{H-H}$ = 8.3 Hz, 2H, Ar-H), 7.13 (d, $^{3}J_{H-H}$ = 8.3 Hz, 2H, Ar-H), 6.77 (dd, $^{3}J_{H-H}$ = 8.8 Hz, $^{4}J_{H-H}$ = 2.4 Hz, 1H, Ar-H), 6.62 (dd, $^{3}J_{H-F}$ = 12.5 Hz, $^{4}J_{H-H}$ 2.4 Hz, 1H), 4.00 (t, $^{3}J_{H-H}$ = 6.4 Hz, 2H, Ar$(CH_2)_5\underline{CH_2}$O), 2.70 (t, $^{3}J_{H-H}$ = 7.6 Hz, 2H, Ar-$\underline{CH_2}(CH_2)_5$O), 2.37 (s, 3H, Ar-$\underline{CH_3}$), 1.82 (m, 2H, Ar$(CH_2)_4\underline{CH_2}CH_2$O), 1.72 (p, $^{3}J_{H-H}$ = 7.6 Hz, 2H, Ar-$CH_2\underline{CH_2}(CH_2)_4$O), 1.54 (m, 2H, Ar-$CH_2CH_2CH_2\underline{CH_2}CH_2CH_2$O), 1.46 (m, 2H, Ar-$CH_2CH_2\underline{CH_2}CH_2CH_2CH_2$O). $^{13}$C NMR (101 MHz, $CDCl_3$) δ ppm: 163.98 (d, $^{1}J_{C-F}$ = 253.1 Hz), 163.12 (d, $^{3}J_{C-F}$ = 11.5 Hz), 152.48 (d, $^{3}J_{C-F}$ = 4.2 Hz), 149.83, 145.33, 142.48, 141.59, 137.78, 137.72, 135.84, 132.77 (2C), 129.88 (2C), 129.14 (2C), 128.82 (d, $^{3}J_{C-F}$ = 4.4 Hz), 127.74 (2C), 127.70 (2C), 127.66 (2C), 127.08 (2C), 121.03 (2C), 119.12, 117.00 (d, $^{2}J_{C-F}$ = 9.5 Hz), 111.65 (d, $^{4}J_{C-F}$ = 2.6 Hz), 110.97, 101.69 (d, $^{2}J_{C-F}$ = 24.7 Hz), 68.60, 35.59, 31.39, 29.05, 28.98, 25.94, 21.14. $^{19}$F NMR (376 MHz, $CDCl_3$) δ ppm: -119.10 (s, 1F). IR

(νmax/cm$^{-1}$): 2926 (sp$^2$ hybridised C-H stretching), 2852 (sp$^2$ hybridised C-H stretching), 2226 (nitrile stretching), 1920 (aromatic overtones), 1612 (imine stretching), 1568 (aryl C=C stretching). HRMS: (TOF ESI$^+$)(m/z): [M+H]$^+$ Calculated for $C_{39}H_{36}N_2OF$ = 567.2812; Found 567.2880 (-2.1 ppm).

**CT6O2F.2**

White solid. Yield: 54%.

$^1$H NMR (400 MHz, $CDCl_3$) δ ppm: 8.69 (s, 1H, CH=N), 8.09 (dd, appt t, $^3J_{H\text{-}H}$ = 8.6 Hz, $^4J_{H\text{-}F}$ = 8.6 Hz, 1H, Ar-H), 7.73 (m, appt s, 4H, Ar-H), 7.70 (d, $^3J_{H\text{-}H}$ = 8.6 Hz, 2H, Ar-H), 7.65 (d, $^3J_{H\text{-}H}$ = 8.6 Hz, 2H, Ar-H), 7.56 (d, $^3J_{H\text{-}H}$ = 8.2 Hz, 2H, Ar-H), 7.29 (d, $^3J_{H\text{-}H}$ = 8.2 Hz, 2H, Ar-H), 7.22 (d, $^3J_{H\text{-}H}$ = 8.3 Hz, 2H, Ar-H), 7.15 (d, $^3J_{H\text{-}H}$ = 8.3 Hz, 2H, Ar-H), 6.77 (dd, $^3J_{H\text{-}H}$ = 8.6 Hz, $^4J_{H\text{-}H}$ = 2.4 Hz, 1H, Ar-H), 6.62 (dd, $^3J_{H\text{-}F}$ = 12.5 Hz, $^4J_{H\text{-}H}$ = 2.4 Hz, 1H, Ar-H), 4.00 (t, $^3J_{H\text{-}H}$ = 6.4 Hz, 2H, $Ar(CH_2)_5CH_2O$), 2.70 (t, $^3J_{H\text{-}H}$ = 7.7 Hz, 2H, $Ar\text{-}CH_2(CH_2)_5O$), 2.67 (q, $^3J_{H\text{-}H}$ = 7.6 Hz, 2H, $Ar\text{-}CH_2CH_3$), 1.82 (m, 2H, $Ar(CH_2)_4CH_2CH_2O$), 1.72 (p, $^3J_{H\text{-}H}$ = 7.7 Hz, 2H, $Ar\text{-}CH_2CH_2(CH_2)_4O$), 1.53 (m, 2H, $Ar\text{-}CH_2CH_2CH_2CH_2CH_2CH_2O$), 1.46 (m, 2H, $Ar\text{-}CH_2CH_2CH_2CH_2CH_2CH_2O$),1.26 (t, $^3J_{H\text{-}H}$ = 7.6 Hz, 3H, $Ar\text{-}CH_2CH_3$). $^{13}$C NMR (101 MHz, $CDCl_3$) δ ppm: 163.99 (d, $^1J_{C\text{-}F}$ = 253.5 Hz), 163.12 (d, $^3J_{C\text{-}F}$ = 11.6 Hz), 152.52 (d, $^3J_{C\text{-}F}$ = 4.2 Hz), 150.04, 145.34, 142.48, 142.26, 141.60, 137.78, 137.73, 132.77 (2C), 129.14 (2C), 128.83 (d, $^3J_{C\text{-}F}$ = 4.5 Hz), 128.68 (2C), 127.74 (2C), 127.71 (2C), 127.66 (2C), 127.08 (2C), 121.09 (2C), 119.12, 117.03 (d, $^2J_{C\text{-}F}$ = 9.5 Hz), 111.64 (d, $^4J_{C\text{-}F}$ = 2.8 Hz), 110.98, 101.69 (d, $^2J_{C\text{-}F}$ = 24.7 Hz), 68.60, 35.59, 31.40, 29.05, 28.99, 28.56, 25.95, 15.80. $^{19}$F NMR (376 MHz, $CDCl_3$) δ ppm: -119.11 (s, 1F). IR (νmax/cm$^{-1}$): 2930 (sp$^2$ hybridised C-H stretching), 2857 (sp$^2$ hybridised C-H stretching), 2222 (nitrile stretching), 1920 (aromatic overtones), 1621 (imine stretching), 1569 (aryl C=C stretching). HRMS: (TOF ESI$^+$)(m/z): [M+H]$^+$ Calculated for $C_{40}H_{38}N_2OF$ = 581.2968; Found 581.2944 (-4.1 ppm).

**CT6O2F.3**

White solid. Yield: 64%.

$^1$H NMR (400 MHz, $CDCl_3$) δ ppm: 8.69 (s, 1H, CH=N), 8.09 (dd, appt t, $^3J_{H\text{-}H}$ = 8.6 Hz, $^4J_{H\text{-}F}$ = 8.6 Hz, 1H, Ar-H), 7.73 (m, appt s, 4H, Ar-H), 7.70 (d, $^3J_{H\text{-}H}$ = 8.6 Hz, 2H, Ar-H), 7.66 (d, $^3J_{H\text{-}H}$ = 8.6 Hz, 2H, Ar-H), 7.56 (d, $^3J_{H\text{-}H}$ = 8.1 Hz, 2H, Ar-H), 7.29 (d, $^3J_{H\text{-}H}$ = 8.1 Hz, 2H, Ar-H), 7.19 (d, $^3J_{H\text{-}H}$ = 8.4 Hz, 2H, Ar-H), 7.15 (d, $^3J_{H\text{-}H}$ = 8.4 Hz, 2H, Ar-H), 6.77 (dd, $^3J_{H\text{-}H}$ = 8.6, 2.4 Hz, 1H, Ar-H), 6.62 (dd, $^3J_{H\text{-}F}$ = 12.5 Hz, $^4J_{H\text{-}H}$ = 2.4 Hz, 1H, Ar-H),4.00 (t, $^3J_{H\text{-}H}$ =

6.4 Hz, 2H, $Ar(CH_2)_5\underline{CH_2}O$), 2.69 (t, $^3J_{H\text{-}H}$ = 7.6 Hz, 2H, $Ar\text{-}\underline{CH_2}(CH_2)_5O$), 2.60 (t, $^3J_{H\text{-}H}$ = 7.6 Hz, 2H, $Ar\text{-}\underline{CH_2}CH_2CH_3$), 1.82 (m, 2H, $Ar(CH_2)_4\underline{CH_2}CH_2O$), 1.72 (p, $^3J_{H\text{-}H}$ = 7.6 Hz, 2H, $Ar\text{-}CH_2\underline{CH_2}(CH_2)_4O$), 1.66 (m, 2H, $Ar\text{-}CH_2\underline{CH_2}CH_3$), 1.55 (m, 2H, $Ar\text{-}CH_2CH_2CH_2\underline{CH_2}CH_2CH_2O$), 1.45 (m, 2H, $Ar\text{-}CH_2CH_2\underline{CH_2}CH_2CH_2CH_2O$), 0.96 (t, $^3J_{H\text{-}H}$ = 7.3 Hz, 3H, $Ar\text{-}CH_2CH_2\underline{CH_3}$). $^{13}$C NMR (101 MHz, $CDCl_3$) δ ppm: 163.86 (d, $^1J_{C\text{-}F}$ = 253.0 Hz), 162.99 (d, $^3J_{C\text{-}F}$ = 11.4 Hz), 152.36 (d, $^3J_{C\text{-}F}$ = 4.2 Hz), 149.91, 145.22, 142.36, 141.48, 140.59, 137.66, 137.61, 132.65 (2C), 129.17 (2C), 129.02 (2C), 128.71 (d, $^3J_{C\text{-}F}$ = 4.7 Hz), 127.62 (2C), 127.58 (2C), 127.54 (2C), 126.96 (2C), 120.88 (2C), 119.00, 116.91 (d, $^2J_{C\text{-}F}$ = 9.6 Hz), 111.52 (d, $^4J_{C\text{-}F}$ = 2.5 Hz), 110.86, 101.56 (d, $^2J_{C\text{-}F}$ = 24.7 Hz), 68.47, 37.60, 35.47, 31.27, 28.93, 28.87, 25.82, 24.63, 13.83. $^{19}$F NMR (376 MHz, $CDCl_3$) δ ppm: -119.12 (s, 1F). IR (νmax/cm$^{-1}$): 2927 (sp$^2$ hybridised C-H stretching), 2855 (sp$^2$ hybridised C-H stretching), 2224 (nitrile stretching), 1915 (aromatic overtones), 1621 (imine stretching), 1570 (aryl C=C stretching). HRMS: (TOF ESI$^+$)(m/z): [M+H]$^+$ Calculated for $C_{41}H_{40}N_2OF$ = 595.3125; Found 595.3096 (-4.9 ppm).

**CT6O2F.4**

White solid. Yield: 65%.

$^1$H NMR (400 MHz, $CDCl_3$) δ ppm: 8.68 (s, 1H, $\underline{CH}$=N), 8.09 (dd, appt t, $^3J_{H\text{-}H}$ = 8.6 Hz, $^4J_{H\text{-}F}$ 8.6 Hz, 1H, Ar-H), 7.73 (m, appt s, 4H, Ar-H), 7.70 (d, $^3J_{H\text{-}H}$ = 8.6 Hz, 2H, Ar-H), 7.66 (d, $^3J_{H\text{-}H}$ = 8.6 Hz, 2H, Ar-H), 7.56 (d, $^3J_{H\text{-}H}$ = 8.2 Hz, 2H, Ar-H), 7.29 (d, $^3J_{H\text{-}H}$ = 8.2 Hz, 2H, Ar-H), 7.19 (d, $^3J_{H\text{-}H}$ = 8.4 Hz, 2H, Ar-H), 7.14 (d, $^3J_{H\text{-}H}$ = 8.4 Hz, 2H, Ar-H), 6.77 (dd, $^3J_{H\text{-}H}$ = 8.6 Hz, $^4J_{H\text{-}H}$ = 2.3 Hz, 1H, Ar-H), 6.62 (dd, $^3J_{H\text{-}F}$ = 12.5 Hz, $^4J_{H\text{-}H}$ 2.3 Hz, 1H, Ar-H), 4.00 (t, $^3J_{H\text{-}H}$ = 6.4 Hz, 2H, $Ar(CH_2)_5\underline{CH_2}O$), 2.70 (t, $^3J_{H\text{-}H}$ = 7.6 Hz, 2H, $Ar\text{-}\underline{CH_2}(CH_2)_5O$), 2.63 (t, $^3J_{H\text{-}H}$ = 7.6 Hz, 2H, $Ar\text{-}\underline{CH_2}CH_2CH_2CH_3$), 1.83 (m, 2H, $Ar(CH_2)_4\underline{CH_2}CH_2O$), 1.72 (p, $^3J_{H\text{-}H}$ = 7.6 Hz, 2H, $Ar\text{-}CH_2\underline{CH_2}(CH_2)_4O$), 1.61 (m, 2H, $Ar\text{-}CH_2\underline{CH_2}CH_2CH_3$), 1.55 (m, 2H, $Ar\text{-}CH_2CH_2CH_2\underline{CH_2}CH_2CH_2O$), 1.46 (m, 2H, $Ar\text{-}CH_2CH_2\underline{CH_2}CH_2CH_2CH_2O$), 1.37 (h, $^3J_{H\text{-}H}$ = 7.3 Hz, 2H, $Ar\text{-}CH_2CH_2\underline{CH_2}CH_3$), 0.94 (t, $^3J_{H\text{-}H}$ = 7.3 Hz, 3H, $Ar\text{-}CH_2CH_2CH_2\underline{CH_3}$). $^{13}$C NMR (101 MHz, $CDCl_3$) δ ppm: 163.86 (d, $^1J_{C\text{-}F}$ = 253.2 Hz), 162.99 (d, $^3J_{C\text{-}F}$ = 11.2 Hz), 152.35 (d, $^3J_{C\text{-}F}$ = 4.1 Hz), 149.87, 145.22, 142.36, 141.48, 140.82, 137.66, 137.60, 132.65 (2C), 129.11 (2C), 129.02 (2C), 128.71 (d, $^3J_{C\text{-}F}$ = 4.6 Hz), 127.62 (2C), 127.58 (2C), 127.54 (2C), 126.96 (2C), 120.89 (2C), 119.00, 116.92 (d, $^2J_{C\text{-}F}$ = 9.6 Hz), 111.52 (d, $^4J_{C\text{-}F}$ = 2.7 Hz), 110.86, 101.56 (d, $^2J_{C\text{-}F}$ = 24.6 Hz), 68.48, 35.47, 35.21, 33.72, 31.28, 28.93, 28.87, 25.83, 22.37, 13.99. $^{19}$F NMR (376 MHz, $CDCl_3$) δ ppm: -119.12 (s, 1F). IR (νmax/cm$^{-1}$): 2926 (sp$^2$

hybridised C-H stretching), 2853 ($sp^2$ hybridised C-H stretching), 2223 (nitrile stretching), 1916 (aromatic overtones), 1620 (imine stretching), 1568 (aryl C=C stretching). HRMS: (TOF ESI$^+$)(m/z): $[M+H]^+$ Calculated for $C_{42}H_{42}N_2OF$ = 609.3281; Found 609.3251 (-4.9 ppm).

**CT6O2F.5**

White solid. Yield: 74%.

$^1$H NMR (400 MHz, $CDCl_3$) δ ppm: 8.69 (s, 1H, CH=N), 8.09 (dd, appt t, $^3J_{H-H}$ = 8.7 Hz, $^4J_{H-F}$ = 8.7 Hz, 1H, Ar-H), 7.73 (m, appt s, 4H, Ar-H), 7.70 (d, $^3J_{H-H}$ = 8.6 Hz, 2H, Ar-H), 7.66 (d, $^3J_{H-H}$ = 8.6 Hz, 2H, Ar-H), 7.56 (d, $^3J_{H-H}$ = 8.2 Hz, 2H, Ar-H), 7.29 (d, $^3J_{H-H}$ = 8.2 Hz, 2H, Ar-H), 7.20 (d, $^3J_{H-H}$ = 8.4 Hz, 2H, Ar-H), 7.15 (d, $^3J_{H-H}$ = 8.4 Hz, 2H, Ar-H), 6.77 (dd, $^3J_{H-H}$ = 8.7 Hz, $^4J_{H-H}$ = 2.4 Hz, 1H, Ar-H), 6.62 (dd, $^3J_{H-F}$ = 12.5 Hz, $^4J_{H-H}$ = 2.4 Hz, 1H, Ar-H), 4.00 (t, $^3J_{H-H}$ = 6.4 Hz, 2H, $Ar(CH_2)_5\underline{CH_2}O$), 2.70 (t, $^3J_{H-H}$ = 7.6 Hz, 2H, $Ar-\underline{CH_2}(CH_2)_5O$), 2.62 (t, $^3J_{H-H}$ = 7.6 Hz, 2H, $Ar-\underline{CH_2}CH_2(CH_2)_2CH_3$), 1.83 (m, 2H, $Ar(CH_2)_4\underline{CH_2}CH_2O$), 1.72 (p, $^3J_{H-H}$ = 7.6 Hz, 2H, $Ar-CH_2\underline{CH_2}(CH_2)_4O$), 1.63 (p, $^3J_{H-H}$ = 7.6 Hz, 2H, $Ar-CH_2\underline{CH_2}(CH_2)_2CH_3$), 1.54 (m, 2H, $Ar-CH_2CH_2CH_2\underline{CH_2}CH_2CH_2O$), 1.45 (m, 2H, $Ar-CH_2CH_2\underline{CH_2}CH_2CH_2CH_2O$), 1.34 (m, 4H, $Ar-CH_2CH_2\underline{(CH_2)_2}CH_3$), 0.94 (t, $^3J_{H-H}$ = 6.9 Hz, 3H, $Ar-CH_2CH_2(CH_2)_2\underline{CH_3}$). $^{13}$C NMR (101 MHz, $CDCl_3$) δ ppm: 163.98 (d, $^1J_{C-F}$ = 253.0 Hz), 163.11 (d, $^3J_{C-F}$ = 11.5 Hz), 152.47 (d, $^3J_{C-F}$ = 4.2 Hz), 149.99, 145.34, 142.48, 141.60, 140.98, 137.78, 137.73, 132.78 (2C), 129.23 (2C), 129.15 (2C), 128.83 (d, $^2J_{C-F}$ = 4.7 Hz), 127.74 (2C), 127.71 (2C), 127.67 (2C), 127.08 (2C), 121.02 (2C), 119.12, 117.04 (d, $^2J_{C-F}$ = 9.6 Hz), 111.64 (d, $^4J_{C-F}$ = 2.7 Hz), 110.98, 101.69 (d, $^2J_{C-F}$ = 24.7 Hz), 68.60, 35.61, 35.60, 31.64, 31.40, 31.38, 29.06, 29.00, 25.95, 22.70, 14.19. $^{19}$F NMR (376 MHz, $CDCl_3$) δ ppm: -119.11 (s, 1F). IR (νmax/$cm^{-1}$): 2929 ($sp^2$ hybridised C-H stretching), 2857 ($sp^2$ hybridised C-H stretching), 2223 (nitrile stretching), 1899 (aromatic overtones), 1620 (imine stretching), 1567 (aryl C=C stretching). HRMS: (TOF ESI$^+$)(m/z): $[M+H]^+$ Calculated for $C_{43}H_{44}N_2OF$ = 623.3438; Found 623.3414 (-3.9 ppm).

**CT6O2F.6**

White solid. Yield: 79%.

$^1$H NMR (400 MHz, $CDCl_3$) δ ppm: 8.69 (s, 1H, CH=N), 8.09 (dd, appt t, $^3J_{H-H}$ = 8.7 Hz, $^4J_{H-F}$ = 8.6 Hz, 1H, Ar-H), 7.73 (m, appt s, 4H, Ar-H), 7.70 (d, $^3J_{H-H}$ = 8.5 Hz, 2H, Ar-H), 7.66 (d, $^3J_{H-H}$ = 8.5 Hz, 2H, Ar-H), 7.56 (d, $^3J_{H-H}$ = 8.1 Hz, 2H, Ar-H), 7.29 (d, $^3J_{H-H}$ = 8.1 Hz, 2H,

Ar-H), 7.19 (d, $^3J_{H-H}$ = 8.4 Hz, 2H, Ar-H), 7.15 (d, $^3J_{H-H}$ = 8.4 Hz, 2H, Ar-H), 6.77 (dd, $^3J_{H-H}$ = 8.7 Hz, $^4J_{H-H}$ = 2.4 Hz, 1H, Ar-H), 6.62 (dd, $^3J_{H-F}$ = 12.5 Hz, $^4J_{H-H}$ = 2.4 Hz, 1H, Ar-H), 4.00 (t, $^3J_{H-H}$ = 6.4 Hz, 2H, $Ar(CH_2)_5\underline{CH_2}O$), 2.70 (t, $^3J_{H-H}$ = 7.6 Hz, 2H, $Ar\text{-}\underline{CH_2}(CH_2)_5O$), 2.62 (t, $^3J_{H-H}$ = 7.6 Hz, 2H, $Ar\text{-}\underline{CH_2}CH_2(CH_2)_3CH_3$), 1.83 (m, 2H, $Ar(CH_2)_4\underline{CH_2}CH_2O$), 1.72 (p, $^3J_{H-H}$ = 7.5 Hz, 2H, $Ar\text{-}CH_2\underline{CH_2}(CH_2)_4O$), 1.63 (m, 2H, $Ar\text{-}CH_2\underline{CH_2}(CH_2)_3CH_3$), 1.53 (m, 2H, $Ar\text{-}CH_2CH_2CH_2\underline{CH_2}CH_2CH_2O$), 1.46 (m, 2H, $Ar\text{-}CH_2CH_2\underline{CH_2}CH_2CH_2CH_2O$), 1.33 (m, 6H, $Ar\text{-}CH_2CH_2\underline{(CH_2)_3}CH_3$), 0.89 (t, $^3J_{H-H}$ = 6.6 Hz, 3H, $Ar\text{-}CH_2CH_2(CH_2)_3\underline{CH_3}$). $^{13}$C NMR (101 MHz, $CDCl_3$) δ ppm: 163.98 (d, $^1J_{C-F}$ = 253.1 Hz), 163.11 (d, $^3J_{C-F}$ = 11.5 Hz), 152.46 (d, $^3J_{C-F}$ = 4.2 Hz), 149.98, 145.33, 142.48, 141.60, 140.99, 137.78, 137.72, 132.77 (2C), 129.22 (2C), 129.14 (2C), 128.83 (d, $^3J_{C-F}$ = 4.6 Hz), 127.74 (2C), 127.70 (2C), 127.66 (2C), 127.08 (2C), 121.01 (2C), 119.12, 117.03 (d, $^2J_{C-F}$ = 9.5 Hz), 111.64 (d, $^4J_{C-F}$ = 2.7 Hz), 110.97, 101.68 (d, $^2J_{C-F}$ = 24.6 Hz), 68.60, 35.65, 35.59, 31.88, 31.66, 31.40, 29.12, 29.05, 29.00, 25.95, 22.76, 14.25. $^{19}$F NMR (376 MHz, $CDCl_3$) δ ppn: -119.10 (s, 1F). IR (νmax/$cm^{-1}$): 2918 ($sp^2$ hybridised C-H stretching), 2854 ($sp^2$ hybridised C-H stretching), 2227 (nitrile stretching), 1915 (aromatic overtones), 1622 (imine stretching), 1569 (aryl C=C stretching). HRMS: (TOF $ESI^+$)(m/z): $[M+H]^+$ Calculated for $C_{44}H_{46}N_2OF$ = 637.3594; Found 637.3575 (-3.0 ppm).

**CT6O2F.7**

White solid. Yield: 80%.

$^1$H NMR (400 MHz, $CDCl_3$) δ ppm: 8.69 (s, 1H, $\underline{CH}$=N), 8.09 (dd, appt t, $^3J_{H-H}$ = 8.7 Hz, $^4J_{H-F}$ = 8.5 Hz, 1H, Ar-H), 7.73 (m, appt s, 4H, Ar-H), 7.70 (d, $^3J_{H-H}$ = 8.4 Hz, 2H, Ar-H), 7.66 (d, $^3J_{H-H}$ = 8.4 Hz, 2H, Ar-H), 7.56 (d, $^3J_{H-H}$ = 8.1 Hz, 2H, Ar-H), 7.29 (d, $^3J_{H-H}$ = 8.1 Hz, 2H, Ar-H), 7.19 (d, $^3J_{H-H}$ = 8.4 Hz, 2H, Ar-H), 7.15 (d, $^3J_{H-H}$ = 8.4 Hz, 2H, Ar-H), 6.77 (dd, $^3J_{H-H}$ = 8.7 Hz, $^4J_{H-H}$ = 2.3 Hz, 1H, Ar-H), 6.62 (dd, $^3J_{H-F}$ = 12.5 Hz, $^4J_{H-H}$ = 2.3 Hz, 1H, Ar-H), 4.00 (t, $^3J_{H-H}$ = 6.4 Hz, 2H, $Ar(CH_2)_5\underline{CH_2}O$), 2.70 (t, $^3J_{H-H}$ = 7.7 Hz, 2H, $Ar\text{-}\underline{CH_2}(CH_2)_5O$), 2.62 (t, $^3J_{H-H}$ = 7.5 Hz, 2H, $Ar\text{-}\underline{CH_2}CH_2(CH_2)_4CH_3$), 1.83 (m, 2H, $Ar(CH_2)_4\underline{CH_2}CH_2O$), 1.72 (p, $^3J_{H-H}$ = 7.5 Hz, 2H, $Ar\text{-}CH_2\underline{CH_2}(CH_2)_4O$), 1.62 (m, 2H, $Ar\text{-}CH_2\underline{CH_2}(CH_2)_4CH_3$), 1.54 (m, 2H, $Ar\text{-}CH_2CH_2CH_2\underline{CH_2}CH_2CH_2O$), 1.47 (m, 2H, $Ar\text{-}CH_2CH_2\underline{CH_2}CH_2CH_2CH_2O$), 1.33 (m, 8H, $Ar\text{-}CH_2CH_2\underline{(CH_2)_4}CH_3$), 0.89 (t, $^3J_{H-H}$ = 6.7 Hz, 3H, $Ar\text{-}CH_2CH_2(CH_2)_4\underline{CH_3}$). $^{13}$C NMR (101 MHz, $CDCl_3$) δ ppm: 163.98 (d, $^1J_{C-F}$ = 253.2 Hz), 163.11 (d, $^3J_{C-F}$ = 11.5 Hz), 152.46 (d, $^3J_{C-F}$ = 4.1 Hz), 149.98, 145.34, 142.48, 141.60, 140.99, 137.78, 137.72, 132.77 (2C), 129.23 (2C), 129.14 (2C), 128.83 (d, $^3J_{C-F}$ = 4.6 Hz), 127.74 (2C), 127.70 (2C), 127.66 (2C),

127.08 (2C), 121.01 (2C), 119.12, 117.04 (d, $^{2}J_{C\text{-}F}$ = 9.6 Hz), 111.64 (d, $^{4}J_{C\text{-}F}$ = 2.7 Hz), 110.98, 101.68 (d, $^{2}J_{C\text{-}F}$ = 24.7 Hz), 68.60, 35.65, 35.60, 31.97, 31.70, 31.40, 29.41, 29.33, 29.06, 29.00, 25.95, 22.81, 14.25. $^{19}$F NMR (376 MHz, $CDCl_3$) δ ppm: -119.11 (s, 1F). IR (νmax/cm$^{-1}$): 2921 (sp$^{2}$ hybridised C-H stretching), 2847 (sp$^{2}$ hybridised C-H stretching), 2226 (nitrile stretching), 1900 (aromatic overtones), 1622 (imine stretching), 1570 (aryl C=C stretching). HRMS: (TOF ESI$^{+}$)(m/z): $[M+H]^{+}$ Calculated for $C_{45}H_{48}N_2OF$ = 651.3751; Found 651.3739 (-1.8 ppm).

**CT6O2F.8**

White solid. Yield: 72%.

$^{1}$H NMR (400 MHz, $CDCl_3$) δ 8.69 (s, 1H, $\underline{CH}$=N), 8.09 (dd, appt t, $^{3}J_{H\text{-}H}$ = 8.8 Hz, $^{4}J_{H\text{-}F}$ = 8.6 Hz, 1H, Ar-H), 7.73 (m, appt s, 4H, Ar-H), 7.70 (d, $^{3}J_{H\text{-}H}$ = 8.4 Hz, 2H, Ar-H), 7.66 (d, $^{3}J_{H\text{-}H}$ = 8.4 Hz, 2H, Ar-H), 7.56 (d, $^{3}J_{H\text{-}H}$ = 8.1 Hz, 2H, Ar-H), 7.29 (d, $^{3}J_{H\text{-}H}$ = 8.1 Hz, 2H, Ar-H), 7.19 (d, $^{3}J_{H\text{-}H}$ = 8.4 Hz, 2H, Ar-H), 7.15 (d, $^{3}J_{H\text{-}H}$ = 8.4 Hz, 2H, Ar-H), 6.77 (dd, $^{3}J_{H\text{-}H}$ = 8.8 Hz, $^{4}J_{H\text{-}H}$ = 2.4 Hz, 1H, Ar-H), 6.62 (dd, $^{3}J_{H\text{-}F}$ = 12.5 Hz, $^{4}J_{H\text{-}H}$ = 2.4 Hz, 1H, Ar-H), 4.00 (t, $^{3}J_{H\text{-}H}$ = 6.4 Hz, 2H, $Ar(CH_2)_5\underline{CH_2}O$), 2.70 (t, $^{3}J_{H\text{-}H}$ = 7.7 Hz, 2H, $Ar\text{-}\underline{CH_2}(CH_2)_5O$), 2.62 (t, $^{3}J_{H\text{-}H}$ = 7.5 Hz, 2H, $Ar\text{-}\underline{CH_2}CH_2(CH_2)_5CH_3$), 1.83 (m, 2H, $Ar(CH_2)_4\underline{CH_2}CH_2O$), 1.72 (p, $^{3}J_{H\text{-}H}$ = 7.6 Hz, 2H, $Ar\text{-}CH_2\underline{CH_2}(CH_2)_4O$), 1.62 (m, 2H, $Ar\text{-}CH_2\underline{CH_2}(CH_2)_5CH_3$), 1.54 (m, 2H, $Ar\text{-}CH_2CH_2CH_2\underline{CH_2}CH_2CH_2O$), 1.46 (m, 2H, $Ar\text{-}CH_2CH_2\underline{CH_2}CH_2CH_2CH_2O$), 1.31 (m, 10H, $Ar\text{-}CH_2CH_2\underline{(CH_2)_5}CH_3$), 0.89 (t, $^{3}J_{H\text{-}H}$ = 6.7 Hz, 3H, $Ar\text{-}CH_2CH_2(CH_2)_5\underline{CH_3}$). $^{13}$C NMR (101 MHz, $CDCl_3$) δ ppm: 163.98 (d, $^{1}J_{C\text{-}F}$ = 253.1 Hz), 163.11 (d, $^{3}J_{C\text{-}F}$ = 11.5 Hz), 152.46 (d, $^{3}J_{C\text{-}F}$ = 4.2 Hz), 149.97, 145.33, 142.48, 141.59, 140.99, 137.78, 137.72, 132.77 (2C), 129.23 (2C), 129.14 (2C), 128.83 (d, $^{3}J_{C\text{-}F}$ = 4.5 Hz), 127.74 (2C), 127.70 (2C), 127.66 (2C), 127.08 (2C), 121.01 (2C), 119.12, 117.04 (d, $^{2}J_{C\text{-}F}$ = 9.6 Hz), 111.64 (d, $^{4}J_{C\text{-}F}$ = 2.7 Hz), 110.97, 101.68 (d, $^{2}J_{C\text{-}F}$ = 24.7 Hz), 68.60, 35.65, 35.60, 32.03, 31.70, 31.40, 29.63, 29.46, 29.42, 29.06, 29.00, 25.95, 22.82, 14.26. $^{19}$F NMR (376 MHz, $CDCl_3$) δ ppm: -119.10 (s, 1F). IR (νmax/cm$^{-1}$): 2922 (sp$^{2}$ hybridised C-H stretching), 2845 (sp$^{2}$ hybridised C-H stretching), 2227 (nitrile stretching), 1908 (aromatic overtones), 1624 (imine stretching), 1570 (aryl C=C stretching). HRMS: (TOF ESI$^{+}$)(m/z): $[M+H]^{+}$ Calculated for $C_{46}H_{50}N_2OF$ = 665.3907; Found 665.3885 (-3.3 ppm).

**CT6O2F.9**

White solid. Yield: 63%.

$^{1}$H NMR (400 MHz, $CDCl_3$) δ ppm: 8.69 (s, 1H, $\underline{CH}$=N), 8.09 (dd, appt t, $^{3}J_{H\text{-}H}$ = 8.6 Hz, $^{4}J_{H\text{-}F}$ = 8.6 Hz, 1H, Ar-H), 7.73 (m, appt s, 4H, Ar-H), 7.70 (d, $^{3}J_{H\text{-}H}$ = 8.4 Hz, 2H, Ar-H), 7.66 (d, $^{3}J_{H\text{-}H}$ = 8.4 Hz, 2H, Ar-H), 7.56 (d, $^{3}J_{H\text{-}H}$ = 8.1 Hz, 2H, Ar-H), 7.29 (d, $^{3}J_{H\text{-}H}$ = 8.1 Hz, 2H, Ar-H), 7.19 (d, $^{3}J_{H\text{-}H}$ = 8.4 Hz, 2H, Ar-H), 7.15 (d, $^{3}J_{H\text{-}H}$ = 8.4 Hz, 2H, Ar-H), 6.77 (dd, $^{3}J_{H\text{-}H}$ = 8.6 Hz, $^{4}J_{H\text{-}H}$ = 2.4 Hz, 1H, Ar-H), 6.62 (dd, $^{3}J_{H\text{-}F}$ = 12.5 Hz, $^{4}J_{H\text{-}H}$ = 2.4 Hz, 1H, Ar-H), 4.00 (t, $^{3}J_{H\text{-}H}$ = 6.4 Hz, 2H, Ar$(CH_2)_5\underline{CH_2}$O), 2.70 (t, $^{3}J_{H\text{-}H}$ = 7.6 Hz, 2H, Ar-$\underline{CH_2}(CH_2)_5$O), 2.62 (t, $^{3}J_{H\text{-}H}$ = 7.8 Hz, 2H, Ar-$\underline{CH_2}CH_2(CH_2)_6CH_3$), 1.83 (m, 2H, Ar$(CH_2)_4\underline{CH_2}CH_2$O), 1.72 (p, $^{3}J_{H\text{-}H}$ = 7.6 Hz, 2H, Ar-$CH_2\underline{CH_2}(CH_2)_4$O), 1.62 (m, 2H, Ar-$CH_2\underline{CH_2}(CH_2)_6CH_3$), 1.55 (m, 2H, Ar-$CH_2CH_2CH_2\underline{CH_2}CH_2CH_2$O), 1.46 (m, 2H, Ar-$CH_2CH_2\underline{CH_2}CH_2CH_2CH_2$O), 1.30 (m, 12H, Ar-$CH_2CH_2\underline{(CH_2)_6}CH_3$), 0.89 (t, $^{3}J_{H\text{-}H}$ = 6.6 Hz, 3H, Ar-$CH_2CH_2(CH_2)_6\underline{CH_3}$). $^{13}$C NMR (101 MHz, $CDCl_3$) δ ppm: 163.98 (d, $^{1}J_{C\text{-}F}$ = 253.1 Hz), 163.11 (d, $^{3}J_{C\text{-}F}$ = 11.5 Hz), 152.46 (d, $^{3}J_{C\text{-}F}$ = 4.2 Hz), 149.98, 145.33, 142.48, 141.60, 140.99, 137.78, 137.72, 132.77 (2C), 129.23 (2C), 129.14 (2C), 128.83 (d, $^{3}J_{C\text{-}F}$ = 4.6 Hz), 127.74 (2C), 127.70 (2C), 127.66 (2C), 127.08 (2C), 121.01 (2C), 119.12, 117.04 (d, $^{2}J_{C\text{-}F}$ = 9.5 Hz), 111.64 (d, $^{4}J_{C\text{-}F}$ = 2.7 Hz), 110.98, 101.68 (d, $^{2}J_{C\text{-}F}$ = 24.7 Hz), 68.60, 35.65, 35.60, 32.04, 31.70, 31.40, 29.71, 29.67, 29.47, 29.45, 29.06, 29.00, 25.95, 22.82, 14.26. $^{19}$F NMR (376 MHz, $CDCl_3$) δ ppm: -119.11 (s, 1F). IR (νmax/cm$^{-1}$): 2921 (sp$^{2}$ hybridised C-H stretching), 2847 (sp$^{2}$ hybridised C-H stretching), 2231 (nitrile stretching), 1908 (aromatic overtones), 1626 (imine stretching), 1571 (aryl C=C stretching). HRMS: (TOF ESI$^{+}$)(m/z): $[M+H]^{+}$ Calculated for $C_{47}H_{52}N_2OF$ = 679.4064; Found 679.4033 (-4.6 ppm).

**CT6O2F.10**

White solid. Yield: 73%.

$^{1}$H NMR (400 MHz, $CDCl_3$) δ ppm: 8.69 (s, 1H, $\underline{CH}$=N), 8.09 (dd, $^{3}J_{H\text{-}H}$ = 8.6 Hz, $^{4}J_{H\text{-}F}$ = 8.6 Hz, 1H, Ar-H), 7.73 (m, appt s, 4H, Ar-H), 7.70 (d, $^{3}J_{H\text{-}H}$ = 8.3 Hz, 2H, Ar-H), 7.66 (d, $^{3}J_{H\text{-}H}$ = 8.3 Hz, 2H, Ar-H), 7.56 (d, $^{3}J_{H\text{-}H}$ = 8.1 Hz, 2H, Ar-H), 7.29 (d, $^{3}J_{H\text{-}H}$ = 8.1 Hz, 2H, Ar-H), 7.19 (d, $^{3}J_{H\text{-}H}$ = 8.4 Hz, 2H, Ar-H), 7.14 (d, $^{3}J_{H\text{-}H}$ = 8.4 Hz, 2H, Ar-H), 6.77 (dd, $^{3}J_{H\text{-}H}$ = 8.6 Hz, $^{4}J_{H\text{-}F}$ = 2.3 Hz, 1H, Ar-H), 6.62 (dd, $^{3}J_{H\text{-}F}$ = 12.4 Hz, $^{4}J_{H\text{-}H}$ = 2.3 Hz, 1H, Ar-H), 4.00 (t, $^{3}J_{H\text{-}H}$ = 6.4 Hz, 2H, Ar$(CH_2)_5\underline{CH_2}$O), 2.70 (t, $^{3}J_{H\text{-}H}$ = 7.7 Hz, 2H, Ar-$\underline{CH_2}(CH_2)_5$O), 2.62 (t, $^{3}J_{H\text{-}H}$ = 7.6 Hz, 2H, Ar-$\underline{CH_2}CH_2(CH_2)_7CH_3$), 1.83 (m, 2H, Ar$(CH_2)_4\underline{CH_2}CH_2$O), 1.72 (p, $^{3}J_{H\text{-}H}$ = 7.7 Hz, 2H, Ar-$CH_2\underline{CH_2}(CH_2)_4$O), 1.62 (p, $^{3}J_{H\text{-}H}$ = 7.6 Hz, 2H, Ar-$CH_2\underline{CH_2}(CH_2)_7CH_3$),

1.55 (m, 2H, Ar-$CH_2CH_2CH_2\underline{CH_2}CH_2CH_2O$), 1.47 (m, 2H, Ar-$CH_2CH_2\underline{CH_2}CH_2CH_2CH_2O$), 1.30 (m, 14H, Ar-$CH_2CH_2\underline{(CH_2)_7}CH_3$), 0.89 (t, $^3J_{H\text{-}H}$ = 6.8 Hz, 3H, Ar-$CH_2CH_2(CH_2)_7\underline{CH_3}$). $^{13}C$ NMR (101 MHz, $CDCl_3$) δ ppm: 163.98 (d, $^1J_{C\text{-}F}$ = 253.1 Hz), 163.11 (d, $^3J_{C\text{-}F}$ = 11.5 Hz), 152.46 (d, $^3J_{C\text{-}F}$ = 4.2 Hz), 149.98, 145.34, 142.49, 141.60, 141.00, 137.79, 137.73, 132.78 (2C), 129.23 (2C), 129.15 (2C), 128.83 (d, $^3J_{C\text{-}F}$ = 4.6 Hz), 127.74 (2C), 127.71 (2C), 127.67 (2C), 127.08 (2C), 121.01 (2C), 119.12, 117.04 (d, $^2J_{C\text{-}F}$ = 9.6 Hz), 111.64 (d, $^4J_{C\text{-}F}$ = 2.6 Hz), 110.98, 101.69 (d, $^2J_{C\text{-}F}$ = 24.6 Hz), 68.60, 35.65, 35.60, 32.05, 31.70, 31.41, 29.77, 29.76, 29.67, 29.49, 29.46, 29.06, 29.01, 25.96, 22.84, 14.27. $^{19}F$ NMR (376 MHz, $CDCl_3$) δ ppm: -119.12 (s, 1F). IR (vmax/$cm^{-1}$): 2919 ($sp^2$ hybridised C-H stretching), 2846 ($sp^2$ hybridised C-H stretching), 2234 (nitrile stretching), 1902 (aromatic overtones), 1625 (imine stretching), 1571 (aryl C=C stretching). HRMS: (TOF $ESI^+$)(m/z): $[M+H]^+$ Calculated for $C_{48}H_{54}N_2OF$ = 693.4220; Found 693.4196 (-3.5 ppm).

**CT6O2F.12**

White solid. Yield: 85%.

$^1H$ NMR (400 MHz, $CDCl_3$) δ ppm: 8.69 (s, 1H, $\underline{CH}$=N), 8.09 (dd, $^3J_{H\text{-}H}$ = 8.6 Hz, $^4J_{H\text{-}F}$ = 8.6 Hz, 1H, Ar-H), 7.73 (m, appt s, 4H, Ar-H), 7.70 (d, $^3J_{H\text{-}H}$ = 8.4 Hz, 2H, Ar-H), 7.66 (d, $^3J_{H\text{-}H}$ = 8.4 Hz, 2H, Ar-H), 7.56 (d, $^3J_{H\text{-}H}$ = 8.1 Hz, 2H, Ar-H), 7.29 (d, $^3J_{H\text{-}H}$ = 8.1 Hz, 2H, Ar-H), 7.19 (d, $^3J_{H\text{-}H}$ = 8.4 Hz, 2H, Ar-H), 7.14 (d, $^3J_{H\text{-}H}$ = 8.4 Hz, 2H, Ar-H), 6.77 (dd, $^3J_{H\text{-}H}$ = 8.6 Hz, $^4J_{H\text{-}H}$ = 2.4 Hz, 1H, Ar-H), 6.62 (dd, $^3J_{H\text{-}F}$ = 12.5 Hz, $^4J_{H\text{-}H}$ = 2.4 Hz, 1H, Ar-H), 4.00 (t, $^3J_{H\text{-}H}$ = 6.4 Hz, 2H, Ar$(CH_2)_5\underline{CH_2}O$), 2.70 (t, $^3J_{H\text{-}H}$ = 7.6 Hz, 2H, Ar-$\underline{CH_2}(CH_2)_5O$), 2.62 (t, $^3J_{H\text{-}H}$ = 7.6 Hz, 2H, Ar-$\underline{CH_2}CH_2(CH_2)_9CH_3$), 1.83 (m, 2H, Ar$(CH_2)_4\underline{CH_2}CH_2O$), 1.72 (p, $^3J_{H\text{-}H}$ = 7.6 Hz, 2H, Ar-$CH_2\underline{CH_2}(CH_2)_4O$), 1.62 (p, $^3J_{H\text{-}H}$ = 7.6 Hz, 2H, Ar-$CH_2\underline{CH_2}(CH_2)_9CH_3$), 1.53 (m, 2H, Ar-$CH_2CH_2CH_2\underline{CH_2}CH_2CH_2O$), 1.44 (m, 2H, Ar-$CH_2CH_2\underline{CH_2}CH_2CH_2CH_2O$), 1.27 (m, 18H, Ar-$CH_2CH_2\underline{(CH_2)_9}CH_3$), 0.89 (t, $^3J_{H\text{-}H}$ = 6.7 Hz, 3H, Ar-$CH_2CH_2(CH_2)_9\underline{CH_3}$). $^{13}C$ NMR (101 MHz, $CDCl_3$) δ ppm: 163.98 (d, $^1J_{C\text{-}F}$ = 253.0 Hz), 163.11 (d, $^3J_{C\text{-}F}$ = 11.3 Hz), 152.46 (d, $^3J_{C\text{-}F}$ = 4.3 Hz), 149.98, 145.34, 142.48, 141.60, 141.00, 137.78, 137.73, 132.77 (2C), 129.23 (2C), 129.14 (2C), 128.83 (d, $^3J_{C\text{-}F}$ = 4.6 Hz), 127.74 (2C), 127.70 (2C), 127.66 (2C), 127.08 (2C), 121.01 (2C), 119.12, 117.04 (d, $^2J_{C\text{-}F}$ = 9.5 Hz), 111.64 (d, $^4J_{C\text{-}F}$ = 2.7 Hz), 110.98, 101.68 (d, $^2J_{C\text{-}F}$ = 24.7 Hz), 68.60, 35.65, 35.60, 32.07, 31.70, 31.40, 29.83, 29.81, 29.79, 29.75, 29.67, 29.50, 29.46, 29.06, 29.01, 25.95, 22.84, 14.27. $^{19}F$ NMR (376 MHz, $CDCl_3$) δ ppm: -119.11 (s, 1F). IR (vmax/$cm^{-1}$): 2917 ($sp^2$ hybridised C-H stretching), 2847 ($sp^2$ hybridised C-H stretching), 2227 (nitrile stretching), 1908 (aromatic overtones),

1621 (imine stretching), 1569 (aryl C=C stretching). HRMS: (TOF ESI$^+$)(m/z): $[M+H]^+$ Calculated for $C_{50}H_{58}N_2OF$ = 721.4533; Found 721.4504 (-4.0 ppm).

**CT6O2F.14**

White solid. Yield: 42%.

$^1$H NMR (400 MHz, $CDCl_3$) δ ppm: 8.69 (s, 1H, $\underline{CH}$=N), 8.09 (dd, appt t, $^3J_{H-H}$ = 8.8 Hz, $^4J_{H-F}$ = 8.6 Hz, 1H, Ar-H), 7.73 (m, appt s, 4H, Ar-H), 7.70 (d, $^3J_{H-H}$ = 8.4 Hz, 2H, Ar-H), 7.66 (d, $^3J_{H-H}$ = 8.4 Hz, 2H, Ar-H), 7.56 (d, $^3J_{H-H}$ = 8.1 Hz, 2H, Ar-H), 7.29 (d, $^3J_{H-H}$ = 8.1 Hz, 2H, Ar-H), 7.19 (d, $^3J_{H-H}$ = 8.4 Hz, 2H, Ar-H), 7.15 (d, $^3J_{H-H}$ = 8.4 Hz, 2H, Ar-H), 6.77 (dd, $^3J_{H-H}$ = 8.8 Hz, $^4J_{H-H}$ = 2.4 Hz, 1H, Ar-H), 6.62 (dd, $^3J_{H-F}$ = 12.5 Hz, $^4J_{H-H}$ = 2.4 Hz, 1H, Ar-H), 4.00 (t, $^3J_{H-H}$ = 6.4 Hz, 2H, $Ar(CH_2)_5\underline{CH_2}O$), 2.70 (t, $^3J_{H-H}$ = 7.6 Hz, 2H, $Ar\text{-}\underline{CH_2}(CH_2)_5O$), 2.62 (t, $^3J_{H-H}$ = 7.7 Hz, 2H, $Ar\text{-}\underline{CH_2}CH_2(CH_2)_{11}CH_3$), 1.83 (m, 2H, $Ar(CH_2)_4\underline{CH_2}CH_2O$), 1.72 (p, $^3J_{H-H}$ = 7.6 Hz, 2H, $Ar\text{-}CH_2\underline{CH_2}(CH_2)_4O$), 1.61 (m, 2H, $Ar\text{-}CH_2\underline{CH_2}(CH_2)_{11}CH_3$), 1.54 (m, 2H, $Ar\text{-}CH_2CH_2CH_2\underline{CH_2}CH_2CH_2O$), 1.45 (m, 2H, $Ar\text{-}CH_2CH_2\underline{CH_2}CH_2CH_2CH_2O$), 1.26 (m, 22H, $Ar\text{-}CH_2CH_2\underline{(CH_2)_{11}}CH_3$), 0.89 (t, $^3J_{H-H}$ = 6.7 Hz, 3H, $Ar\text{-}CH_2CH_2(CH_2)_{11}\underline{CH_3}$). $^{13}$C NMR (101 MHz, $CDCl_3$) δ ppm: 163.98 (d, $^1J_{C-F}$ = 253.1 Hz), 163.11 (d, $^3J_{C-F}$ = 11.4 Hz), 152.45 (d, $^3J_{C-F}$ = 4.2 Hz), 149.98, 145.33, 142.48, 141.60, 141.00, 137.78, 137.72, 132.77 (2C), 129.23 (2C), 129.14 (2C), 128.83 (d, $^3J_{C-F}$ = 4.5 Hz), 127.74 (2C), 127.70 (2C), 127.66 (2C), 127.08 (2C), 121.01 (2C), 119.12, 117.04 (d, $^2J_{C-F}$ = 9.5 Hz), 111.64 (d, $^4J_{C-F}$ = 2.8 Hz), 110.98, 101.68 (d, $^2J_{C-F}$ = 24.7 Hz), 68.60, 35.65, 35.60, 32.07, 31.70, 31.40, 29.85, 29.83 (2C), 29.82, 29.81, 29.75, 29.67, 29.51, 29.46, 29.06, 29.01, 25.95, 22.84, 14.27. $^{19}$F NMR (376 MHz, $CDCl_3$) δ ppm: -119.11 (s, 1F). IR (νmax/$cm^{-1}$): 2919 ($sp^2$ hybridised C-H stretching), 2850 ($sp^2$ hybridised C-H stretching), 2228 (nitrile stretching), 1916 (aromatic overtones), 1624 (imine stretching), 1570 (aryl C=C stretching). HRMS: (TOF ESI$^+$)(m/z): $[M+H]^+$ Calculated for $C_{52}H_{62}N_2OF$ = 749.4846; Found 749.4835 (-1.5 ppm).

**CT6O2F.16**

White solid. Yield: 69%.

$^1$H NMR (400 MHz, $CDCl_3$) δ ppm: 8.69 (s, 1H, $\underline{CH}$=N), 8.09 (dd, appt t, $^3J_{H-H}$ = 8.8 Hz, $^4J_{H-F}$ 8.6 Hz, 1H, Ar-H), 7.73 (m, appt s, 4H, Ar-H), 7.70 (d, $^3J_{H-H}$ = 8.4 Hz, 2H, Ar-H), 7.66 (d, $^3J_{H-H}$ = 8.4 Hz, 2H, Ar-H), 7.56 (d, $^3J_{H-H}$ = 8.1 Hz, 2H, Ar-H), 7.29 (d, $^3J_{H-H}$ = 8.1 Hz, 2H, Ar-H), 7.19 (d, $^3J_{H-H}$ = 8.4 Hz, 2H, Ar-H), 7.14 (d, $^3J_{H-H}$ = 8.4 Hz, 2H, Ar-H), 6.77 (dd, $^3J_{H-H}$ = 8.8 Hz, $^4J_{H-H}$ = 2.4 Hz, 1H, Ar-H), 6.62 (dd, $^3J_{H-F}$ = 12.5 Hz, $^4J_{H-H}$ = 2.4 Hz, 1H, Ar-H),

4.00 (t, $^3J_{H\text{-}H}$ = 6.4 Hz, 2H, $Ar(CH_2)_5\underline{CH_2}O$), 2.70 (t, $^3J_{H\text{-}H}$ = 7.6 Hz, 2H, Ar-$\underline{CH_2}(CH_2)_5O$), 2.62 (t, $^3J_{H\text{-}H}$ = 7.9 Hz, 2H, Ar-$\underline{CH_2}CH_2(CH_2)_{13}CH_3$), 1.83 (m, 2H, $Ar(CH_2)_4\underline{CH_2}CH_2O$), 1.72 (p, $^3J_{H\text{-}H}$ = 7.6 Hz, 2H, Ar-$CH_2\underline{CH_2}(CH_2)_4O$), 1.62 (m, 2H, Ar-$CH_2\underline{CH_2}(CH_2)_{13}CH_3$), 1.56 (m, 2H, Ar-$CH_2CH_2CH_2\underline{CH_2}CH_2CH_2O$), 1.48 (m, 2H, Ar-$CH_2CH_2\underline{CH_2}CH_2CH_2CH_2O$), 1.26 (m, 26H, Ar-$CH_2CH_2\underline{(CH_2)_{13}}CH_3$), 0.88 (t, $^3J_{H\text{-}H}$ = 6.8 Hz, 3H, Ar-$CH_2CH_2(CH_2)_{13}\underline{CH_3}$). $^{13}C$ NMR (101 MHz, $CDCl_3$) δ ppm: 163.98 (d, $^1J_{C\text{-}F}$ = 253.1 Hz), 163.11 (d, $^3J_{C\text{-}F}$ = 11.5 Hz), 152.46 (d, $^3J_{C\text{-}F}$ = 4.1 Hz), 149.98, 145.34, 142.48, 141.60, 141.00, 137.78, 137.73, 132.77 (2C), 129.23 (2C), 129.14 (2C), 128.83 (d, $^3J_{C\text{-}F}$ = 4.7 Hz), 127.74 (2C), 127.71 (2C), 127.67 (2C), 127.08 (2C), 121.01 (2C), 119.12, 117.04 (d, $^2J_{C\text{-}F}$ = 9.5 Hz), 111.64 (d, $^4J_{C\text{-}F}$ = 2.7 Hz), 110.98, 101.69 (d, $^2J_{C\text{-}F}$ = 24.7 Hz), 68.60, 35.65, 35.60, 32.08, 31.71, 31.41, 29.85 (3C), 29.83 (2C), 29.81 (2C), 29.76, 29.67, 29.51, 29.46, 29.06, 29.01, 25.96, 22.84, 14.27. $^{19}F$ NMR (376 MHz, $CDCl_3$) δ ppm: -119.11 (s, 1F). IR (νmax/$cm^{-1}$): 2918 ($sp^2$ hybridised C-H stretching), 2850 ($sp^2$ hybridised C-H stretching), 2228 (nitrile stretching), 1916 (aromatic overtones), 1624 (imine stretching), 1570 (aryl C=C stretching). HRMS: (TOF $ESI^+$)(m/z): $[M+H]^+$ Calculated for $C_{54}H_{66}N_2OF$ = 777.5159; Found 777.5130 (-3.7 ppm).

**CT6O2F.18**

White solid. Yield: 70%.

$^1H$ NMR (400 MHz, $CDCl_3$) δ ppm: 8.69 (s, 1H, $\underline{CH}$=N), 8.09 (dd, appt t, $^3J_{H\text{-}H}$ = 8.6 Hz, $^4J_{H\text{-}F}$ 8.6 Hz, 1H, Ar-H), 7.73 (m, appt s, 4H, Ar-H), 7.70 (d, $^3J_{H\text{-}H}$ = 8.4 Hz, 2H, Ar-H), 7.66 (d, $^3J_{H\text{-}H}$ = 8.4 Hz, 2H, Ar-H), 7.56 (d, $^3J_{H\text{-}H}$ = 8.0 Hz, 2H, Ar-H), 7.29 (d, $^3J_{H\text{-}H}$ = 8.0 Hz, 2H, Ar-H), 7.19 (d, $^3J_{H\text{-}H}$ = 8.4 Hz, 2H, Ar-H), 7.14 (d, $^3J_{H\text{-}H}$ = 8.4 Hz, 2H, Ar-H), 6.77 (dd, $^3J_{H\text{-}H}$ = 8.6 Hz, $^4J_{H\text{-}H}$ = 2.4 Hz, 1H, Ar-H), 6.62 (dd, $^3J_{H\text{-}F}$ = 12.4 Hz, $^4J_{H\text{-}H}$ = 2.4 Hz, 1H, Ar-H), 4.00 (t, $^3J_{H\text{-}H}$ = 6.4 Hz, 2H, $Ar(CH_2)_5\underline{CH_2}O$), 2.70 (t, $^3J_{H\text{-}H}$ = 7.7 Hz, 2H, Ar-$\underline{CH_2}(CH_2)_5O$), 2.62 (t, $^3J_{H\text{-}H}$ = 7.6 Hz, 2H, Ar-$\underline{CH_2}CH_2(CH_2)_{15}CH_3$), 1.83 (m, 2H, $Ar(CH_2)_4\underline{CH_2}CH_2O$), 1.72 (p, $^3J_{H\text{-}H}$ = 7.7 Hz, 2H, Ar-$CH_2\underline{CH_2}(CH_2)_4O$), 1.62 (m, 2H, Ar-$CH_2\underline{CH_2}(CH_2)_{15}CH_3$), 1.54 (m, 2H, Ar-$CH_2CH_2CH_2\underline{CH_2}CH_2CH_2O$), 1.46 (m, 2H, Ar-$CH_2CH_2\underline{CH_2}CH_2CH_2CH_2O$), 1.26 (m, 30H, Ar-$CH_2CH_2\underline{(CH_2)_{15}}CH_3$), 0.88 (t, $^3J_{H\text{-}H}$ = 6.7 Hz, 3H, Ar-$CH_2CH_2(CH_2)_{15}\underline{CH_3}$). $^{13}C$ NMR (101 MHz, $CDCl_3$) δ ppm: 163.98 (d, $^1J_{C\text{-}F}$ = 253.1 Hz), 163.11 (d, $^3J_{C\text{-}F}$ = 11.4 Hz), 152.45 (d, $^3J_{C\text{-}F}$ = 4.2 Hz), 149.97, 145.34, 142.48, 141.60, 141.00, 137.78, 137.72, 132.77 (2C), 129.22 (2C), 129.14 (2C), 128.83 (d, $^3J_{C\text{-}F}$ = 4.6 Hz), 127.74 (2C), 127.70 (2C), 127.66 (2C), 127.08 (2C), 121.01 (2C), 119.11, 117.04 (d, $^2J_{C\text{-}F}$ = 9.5 Hz), 111.64 (d, $^4J_{C\text{-}F}$ = 2.7 Hz), 110.98, 101.68 (d, $^2J_{C\text{-}F}$ = 24.7 Hz), 68.60, 35.65, 35.60, 32.07, 31.70, 31.40, 29.85 (4C),

29.83 (2C), 29.82 (2C), 29.81, 29.76, 29.67, 29.51, 29.46, 29.06, 29.01, 25.95, 22.84, 14.27. $^{19}$F NMR (376 MHz, $CDCl_3$) δ ppm: -119.11 (s, 1F). IR (νmax/$cm^{-1}$): 2918 ($sp^2$ hybridised C-H stretching), 2850 ($sp^2$ hybridised C-H stretching), 2228 (nitrile stretching), 1916 (aromatic overtones), 1624 (imine stretching), 1571 (aryl C=C stretching). HRMS: (TOF $ESI^+$)(m/z): $[M+H]^+$ Calculated for $C_{56}H_{70}N_2OF$ = 805.5472; Found 805.5453 (-2.4 ppm).

**3$^4$-{6-(4-(((4'-Alkylphenyl)imino)methyl)-3,5-difluorophenoxy)hexyl}[1$^1$,2$^1$:2$^4$,3$^1$-terphenyl]-1$^4$-carbonitrile (CT6O2,6dF.*m*)**

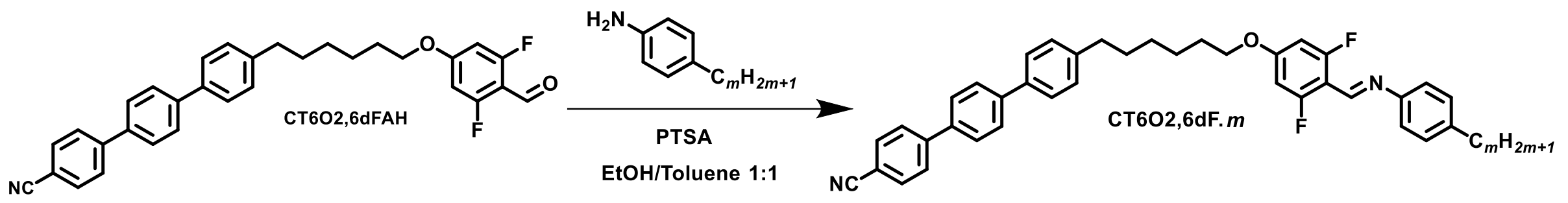

**Scheme S5**: Synthetic scheme for the CT6O2,6dF.*m* series.

To a flask containing 4″-(6-(4-formyl-3,5-difluorophenoxy)hexyl)-[1,1′:4,1″-terphenyl]-4-carbonitrile (CT6O2,6dFAH, 10.3.10) (1 eq, 0.08 g, 0.161 mmol) and 10 mL EtOH, 10 mL toluene, the appropriate 4-alkylaniline (1.4 eq, detailed in Table S11) was added and the reaction mixture heated to reflux. A catalytic amount of *p*-toluenesulfonic acid was added and the reaction allowed to proceed for 3 hr. The reaction mixture was cooled to room temperature, an off white precipitate formed which was collected by vacuum filtration. The crude product was recrystalised from 8 mL ethanol and 8 mL toluene to yield the target compound as a white solid.

**Table S11**: Quantities of 4-alkylanilines used in the synthesis of the CT6O2,6dF.m series

| *m* | m / g | n / mmol |
|---|---|---|
| 1 | 0.024 | 0.226 |
| 2 | 0.027 | 0.226 |
| 3 | 0.031 | 0.226 |
| 4 | 0.034 | 0.226 |
| 5 | 0.037 | 0.226 |
| 6 | 0.040 | 0.226 |
| 7 | 0.043 | 0.226 |
| 8 | 0.046 | 0.226 |
| 9 | 0.050 | 0.226 |
| 10 | 0.053 | 0.226 |
| 12 | 0.059 | 0.226 |
| 14 | 0.065 | 0.226 |
| 16 | 0.072 | 0.226 |
| 18 | 0.078 | 0.226 |

**CT6O2,6dF.1**

White solid. Yield: 47%.

$^1$H NMR (400 MHz, $CDCl_3$) δ ppm: 8.56 (s, 1H, $\underline{CH}$=N), 7.73 (m, appt s, 4H, Ar-H), 7.70 (d, $^3J_{H\text{-}H}$ = 8.3 Hz, 2H, Ar-H), 7.66 (d, $^3J_{H\text{-}H}$ = 8.3 Hz, 2H, Ar-H), 7.56 (d, $^3J_{H\text{-}H}$ = 8.0 Hz, 2H, Ar-H), 7.29 (d, $^3J_{H\text{-}H}$ = 8.0 Hz, 2H, Ar-H), 7.18 (d, $^3J_{H\text{-}H}$ = 8.2 Hz, 2H, Ar-H), 7.11 (d, $^3J_{H\text{-}H}$ = 8.2 Hz, 2H, Ar-H), 6.51 (d, $^3J_{H\text{-}F}$ = 10.7 Hz, 2H, Ar-H), 3.98 (t, $^3J_{H\text{-}H}$ = 6.4 Hz, 2H, Ar-O$\underline{CH_2}CH_2CH_2(CH_2)_3$-Ar), 2.69 (t, $^3J_{H\text{-}H}$ = 7.6 Hz, 2H, Ar-$\underline{CH_2}CH_2CH_2(CH_2)_3$O-Ar), 2.37 (s, 3H, Ar-$CH_3$), 1.82 (m, 2H, Ar-O$CH_2\underline{CH_2}CH_2(CH_2)_3$-Ar), 1.71 (p, $^3J_{H\text{-}H}$ = 7.6 Hz, 2H Ar-$CH_2\underline{CH_2}CH_2(CH_2)_3$O-Ar), 1.53 (m, 2H Ar-$CH_2CH_2\underline{CH_2}(CH_2)_3$O-Ar), 1.45 (m, 2H, Ar-O$CH_2CH_2\underline{CH_2}(CH_2)_3$-Ar). $^{13}$C NMR (101 MHz, $CDCl_3$) δ ppm: 163.18 (dd, $^1J_{C\text{-}F}$ = 256.3 Hz, $^3J_{C\text{-}F}$ = 9.6 Hz, 2C), 162.42 (t, $^3J_{C\text{-}F}$ = 14.7 Hz), 150.50 (m), 145.32, 142.42, 141.57, 137.78, 137.74, 136.03, 132.77 (2C), 129.84 C), 129.13 (2C), 127.72 (2C), 127.70 (2C), 127.66 (2C), 127.08 (2C), 120.90 (2C), 119.11, 115.43, 110.97, 107.15 (t, $^2J_{C\text{-}F}$ = 12.8 Hz), 99.09 (dd, $^2J_{C\text{-}F}$ = 26.8 Hz, $^4J_{C\text{-}F}$ = 2.0 Hz, 2C), 68.97, 35.58, 31.37, 28.96, 28.89, 25.89, 21.14. $^{19}$F NMR (376 MHz, $CDCl_3$) δ ppm: -111.03 (2F). IR (νmax/cm$^{-1}$): 2939 (sp$^2$ hybridised C-H stretching), 2219 (nitrile stretching), 1904 (aromatic overtones), 1621 (imine stretching).

HRMS: (TOF $ESI^+$)(m/z): $[M+H]^+$ Calculated for $C_{39}H_{35}N_2OF_2$ = 585.2717; Found 585.2692 (-4.3 ppm).

**CT6O2,6dF.2**

White solid. Yield: 53%.

$^1$H NMR (400 MHz, $CDCl_3$) δ ppm: 8.57 (s, 1H, $\underline{CH}$=N), 7.73 (m, appt s, 4H, Ar-H), 7.70 (d, $^3J_{H-H}$ = 8.4 Hz, 2H, Ar-H), 7.66 (d, $^3J_{H-H}$ = 8.4 Hz, 2H, Ar-H), 7.56 (d, $^3J_{H-H}$ = 8.1 Hz, 2H, Ar-H), 7.29 (d, $^3J_{H-H}$ = 8.1 Hz, 2H, Ar-H), 7.21 (d, $^3J_{H-H}$ = 8.3 Hz, 2H, Ar-H), 7.14 (d, $^3J_{H-H}$ = 8.3 Hz, 2H, Ar-H), 6.51 (d, $^3J_{H-F}$ = 10.7 Hz, 2H, Ar-H), 3.98 (t, $^3J_{H-H}$ = 6.4 Hz, 2H, Ar-$O\underline{CH_2}CH_2CH_2(CH_2)_3$-Ar), 2.68 (m, 4H, Ar-$\underline{CH_2}CH_2CH_2(CH_2)_3$O-Ar, Ar-$\underline{CH_2}CH_3$), 1.82 (m, 2H, Ar-$OCH_2\underline{CH_2}CH_2(CH_2)_3$-Ar), 1.71 (p, $^3J_{H-H}$ = 7.6 Hz, 2H, Ar-$OCH_2\underline{CH_2}CH_2(CH_2)_3$O-Ar), 1.49 (m, 4H, Ar-$OCH_2CH_2\underline{CH_2}\underline{CH_2}CH_2CH_2$-Ar), 1.25 (t, $^3J_{H-H}$ = 7.6 Hz, 3H, Ar-$CH_2\underline{CH_3}$). $^{13}$C NMR (101 MHz, $CDCl_3$) δ ppm: 163.19 (dd, $^1J_{C-F}$ = 256.3 Hz, $^3J_{C-F}$ = 9.7 Hz, 2C), 162.42 (t, $^3J_{C-F}$ = 14.7 Hz), 150.62 (m), 145.33, 142.46, 142.43, 141.58, 137.80, 137.75, 132.78 (2C), 129.13 (2C), 128.64 (2C), 127.73 (2C), 127.71 (2C), 127.67 (2C), 127.09 (2C), 120.97 (2C), 119.12, 115.41, 110.98, 107.19 (t, $^2J_{C-F}$ = 13.0 Hz), 99.10 (dd, $^2J$ = 26.5 Hz, $^4J_{C-F}$ = 2.1 Hz, 2C), 68.98, 35.59, 31.38, 28.97, 28.90, 28.56, 25.90, 15.79. $^{19}$F NMR (376 MHz, $CDCl_3$) δ ppm: -111.04 (2F). IR (νmax/$cm^{-1}$): 2923 ($sp^2$ hybridised C-H stretching), 2222 (nitrile stretching), 1915 (aromatic overtones), 1627 (imine stretching). HRMS: (TOF $ESI^+$)(m/z): $[M+H]^+$ Calculated for $C_{40}H_{37}N_2OF_2$ = 599.2874; Found 599.2852 (-3.7 ppm).

**CT6O2,6dF.3**

White solid. Yield: 43%.

$^1$H NMR (400 MHz, $CDCl_3$) δ ppm: 8.57 (s, 1H, $\underline{CH}$=N), 7.74 (m, appt s, 4H, Ar-H), 7.70 (d, $^3J_{H-H}$ = 8.4 Hz, 2H, Ar-H), 7.66 (d, $^3J_{H-H}$ = 8.4 Hz, 2H, Ar-H), 7.56 (d, $^3J_{H-H}$ = 8.1 Hz, 2H, Ar-H), 7.29 (d, $^3J_{H-H}$ = 8.1 Hz, 2H, Ar-H), 7.19 (d, $^3J_{H-H}$ = 8.3 Hz, 2H, Ar-H), 7.13 (d, $^3J_{H-H}$ = 8.3 Hz, 2H, Ar-H), 6.51 (d, $^3J_{H-F}$ = 10.6 Hz, 2H, Ar-H), 3.98 (t, $^3J_{H-H}$ = 6.4 Hz, 2H, Ar-$O\underline{CH_2}CH_2CH_2(CH_2)_3$-Ar), 2.69 (t, $^3J_{H-H}$ = 7.7 Hz, 2H, Ar-$\underline{CH_2}CH_2CH_2(CH_2)_3$O-Ar), 2.60 (t, $^3J_{H-H}$ = 7.4 Hz, 2H, Ar-$\underline{CH_2}CH_2CH_3$), 1.82 (m, 2H, Ar-$OCH_2\underline{CH_2}CH_2(CH_2)_3$-Ar), 1.72 (m, 2H, Ar-$CH_2\underline{CH_2}CH_2(CH_2)_3$O-Ar), 1.65 (m, 2H, Ar-$CH_2\underline{CH_2}CH_3$), 1.52 (m, 4H, Ar-$OCH_2CH_2\underline{CH_2}\underline{CH_2}CH_2CH_2$-Ar), 0.95 (t, $^3J_{H-H}$ = 7.3 Hz, 3H, Ar-$CH_2CH_2\underline{CH_3}$). $^{13}$C NMR (101 MHz, $CDCl_3$) δ ppm: 163.19 (dd, $^1J_{C-F}$ = 255.9 Hz, $^3J_{C-F}$ = 9.6 Hz, 2C), 162.41 (t, $^3J_{C-F}$ = 14.8 Hz), 150.59 (m), 145.34, 142.43, 141.59, 140.91, 137.80, 137.76, 132.78 (2C), 129.26

(2C), 129.14 (2C), 127.74 (2C), 127.71 (2C), 127.67 (2C), 127.09 (2C), 120.88 (2C), 119.12, 115.39, 110.98, 107.20 (t, $^{2}J_{C\text{-}F}$ = 13.1 Hz), 99.10 (dd, $^{2}J_{C\text{-}F}$ = 26.6 Hz, $^{4}J_{C\text{-}F}$ = 1.9 Hz, 2C), 68.98, 37.71, 35.59, 31.38, 28.97, 28.90, 25.90, 24.74, 13.94. $^{19}$F NMR (376 MHz, $CDCl_3$) δ ppm: -111.04 (2F). IR (νmax/cm$^{-1}$): 2926 ($sp^2$ hybridised C-H stretching), 2864 ($sp^2$ hybridised C-H stretching), 2221 (nitrile stretching), 1915 (aromatic overtones), 1627 (imine stretching). HRMS: (TOF ESI$^+$)(m/z): $[M+H]^+$ Calculated for $C_{41}H_{39}N_2OF_2$ = 613.3030; Found 613.3006 (-3.9 ppm).

**CT6O2,6dF.4**

White solid. Yield: 42%.

$^{1}$H NMR (400 MHz, $CDCl_3$) δ ppm: 8.57 (s, 1H, $\underline{CH}$=N), 7.73 (m, appt s, 4H, Ar-H), 7.70 (d, $^{3}J_{H\text{-}H}$ = 8.4 Hz, 2H, Ar-H), 7.66 (d, $^{3}J_{H\text{-}H}$ = 8.4 Hz, 2H, Ar-H), 7.56 (d, $^{3}J_{H\text{-}H}$ = 8.0 Hz, 2H, Ar-H), 7.29 (d, $^{3}J_{H\text{-}H}$ = 8.0 Hz, 2H, Ar-H), 7.19 (d, $^{3}J_{H\text{-}H}$ = 8.3 Hz, 2H, Ar-H), 7.13 (d, $^{3}J_{H\text{-}H}$ = 8.3 Hz, 2H, Ar-H), 6.51 (d, $^{3}J_{H\text{-}F}$ = 10.7 Hz, 2H, Ar-H), 3.98 (t, $^{3}J_{H\text{-}H}$ = 6.4 Hz, 2H, Ar-$O\underline{CH_2}CH_2CH_2(CH_2)_3$-Ar), 2.69 (t, $^{3}J_{H\text{-}H}$ = 7.7 Hz, 2H, Ar-$\underline{CH_2}CH_2CH_2(CH_2)_3O$-Ar), 2.62 (t, $^{3}J_{H\text{-}H}$ = 7.7 Hz, 2H, Ar-$\underline{CH_2}CH_2CH_2CH_3$), 1.82 (m, 2H, Ar-$OCH_2\underline{CH_2}CH_2(CH_2)_3$-Ar), 1.71 (p, $^{3}J_{H\text{-}H}$ = 7.7 Hz, 2H, Ar-$CH_2\underline{CH_2}CH_2(CH_2)_3O$-Ar), 1.61 (m, 2H, Ar-$CH_2\underline{CH_2}CH_2CH_3$), 1.49 (m, 4H, Ar-$OCH_2CH_2\underline{CH_2}\underline{CH_2}CH_2CH_2$-Ar), 1.36 (h, $^{3}J_{H\text{-}H}$ = 7.3 Hz, 2H, Ar-$CH_2CH_2\underline{CH_2}CH_3$), 0.94 (t, $^{3}J_{H\text{-}H}$ = 7.3 Hz, 3H, Ar-$CH_2CH_2CH_2\underline{CH_3}$). $^{13}$C NMR (101 MHz, $CDCl_3$) δ ppm: 163.19 (dd, $^{1}J_{C\text{-}F}$ = 256.2 Hz, $^{3}J_{C\text{-}F}$ = 9.7 Hz, 2C), 162.41 (t, $^{3}J_{C\text{-}F}$ = 14.4 Hz), 150.55 (m), 145.33, 142.43, 141.58, 141.13, 137.79, 137.75, 132.77 (2C), 129.19 (2C), 129.13 (2C), 127.73 (2C), 127.71 (2C), 127.67 (2C), 127.09 (2C), 120.89 (2C), 119.11, 115.40, 110.98, 107.20 (t, $^{2}J_{C\text{-}F}$ = 13.3 Hz), 99.09 (dd, $^{2}J_{C\text{-}F}$ = 26.8 Hz, $^{4}J_{C\text{-}F}$ = 1.8 Hz, 2C), 68.97, 35.59, 35.32, 33.82, 31.38, 28.97, 28.90, 25.90, 22.47, 14.10. $^{19}$F NMR (376 MHz, $CDCl_3$) δ ppm: -111.04 (2F). IR (νmax/cm$^{-1}$): 2925 ($sp^2$ hybridised C-H stretching), 2855 ($sp^2$ hybridised C-H stretching), 2219 (nitrile stretching), 1925 (aromatic overtones), 1621 (imine stretching). HRMS: (TOF ESI$^+$)(m/z): $[M+H]^+$ Calculated for $C_{42}H_{41}N_2OF_2$ = 627.3187; Found 627.3156 (-4.9 ppm).

**CT6O2,6dF.5**

White solid. Yield: 61%.

$^{1}$H NMR (400 MHz, $CDCl_3$) δ ppm: 8.57 (s, 1H, $\underline{CH}$=N), 7.73 (m, appt s, 4H, Ar-H), 7.70 (d, $^{3}J_{H\text{-}H}$ = 8.4 Hz, 2H, Ar-H), 7.66 (d, $^{3}J_{H\text{-}H}$ = 8.4 Hz, 2H, Ar-H), 7.56 (d, $^{3}J_{H\text{-}H}$ = 8.1 Hz, 2H, Ar-H), 7.29 (d, $^{3}J_{H\text{-}H}$ = 8.1 Hz, 2H, Ar-H), 7.19 (d, $^{3}J_{H\text{-}H}$ = 8.3 Hz, 2H, Ar-H), 7.13 (d, $^{3}J_{H\text{-}H}$ = 8.3 Hz, 2H, Ar-H), 6.51 (d, $^{3}J_{H\text{-}F}$ = 10.7 Hz, 2H, Ar-H), 3.98 (t, $^{3}J_{H\text{-}H}$ = 6.4 Hz, 2H, Ar-$O\underline{CH_2}CH_2CH_2(CH_2)_3$-Ar), 2.69 (t, $^{3}J_{H\text{-}H}$ = 7.7 Hz, 2H, Ar-$\underline{CH_2}CH_2CH_2(CH_2)_3O$-Ar), 2.62 (t, $^{3}J_{H\text{-}H}$ = 7.7 Hz, 2H, Ar-$\underline{CH_2}CH_2CH_2CH_2CH_3$), 1.82 (m, 2H, Ar-$OCH_2\underline{CH_2}CH_2(CH_2)_3$-Ar), 1.72 (p, $^{3}J_{H\text{-}H}$ = 7.7 Hz, 2H, Ar-$CH_2\underline{CH_2}CH_2(CH_2)_3O$-Ar), 1.61 (m, 2H, Ar-$CH_2\underline{CH_2}CH_2CH_2CH_3$), 1.47 (m, 4H, Ar-$OCH_2CH_2\underline{CH_2}\underline{CH_2}CH_2CH_2$-Ar), 1.33 (m, 4H, Ar-$CH_2CH_2\underline{CH_2}\underline{CH_2}CH_3$), 0.90 (t, $^{3}J_{H\text{-}H}$ = 6.8 Hz, 3H, Ar-$CH_2CH_2CH_2CH_2\underline{CH_3}$). $^{13}$C NMR (101 MHz, $CDCl_3$) δ ppm: 163.19 (dd, $^{1}J_{C\text{-}F}$ = 256.2 Hz, $^{3}J_{C\text{-}F}$ = 9.7 Hz, 2C), 162.40 (t, $^{3}J_{C\text{-}F}$ = 14.6 Hz), 150.55 (m), 145.33, 142.43, 141.58, 141.17, 137.79, 137.75, 132.77 (2C), 129.19 (2C), 129.13 (2C), 127.73 (2C), 127.71 (2C), 127.66 (2C), 127.09 (2C), 120.89 (2C), 119.11, 115.34, 110.98, 107.20 (t, $^{2}J_{C\text{-}F}$ = 13.0 Hz), 99.09 (dd, $^{2}J_{C\text{-}F}$ = 26.6 Hz, $^{4}J_{C\text{-}F}$ = 1.9 Hz, 2C), 68.97, 35.60 (2C), 31.62, 31.38, 31.35, 28.98, 28.90, 25.90, 22.69, 14.19. $^{19}$F NMR (376 MHz, $CDCl_3$) δ ppm: -111.04 (2F). IR (νmax/$cm^{-1}$): 2923 ($sp^2$ hybridised C-H stretching), 2855 ($sp^2$ hybridised C-H stretching), 2219 (nitrile stretching), 1919 (aromatic overtones), 1624 (imine stretching). HRMS: (TOF $ESI^+$)(m/z): $[M+H]^+$ Calculated for $C_{43}H_{43}N_2OF_2$ = 641.3343; Found 641.3314 (-4.5 ppm).

**CT6O2,6dF.6**

White solid. Yield: 42%.

$^{1}$H NMR (400 MHz, $CDCl_3$) δ ppm: 8.57 (s, 1H, $\underline{CH}$=N), 7.73 (m, appt s, 4H, Ar-H), 7.70 (d, $^{3}J_{H\text{-}H}$ = 8.6 Hz, 2H, Ar-H), 7.66 (d, $^{3}J_{H\text{-}H}$ = 8.6 Hz, 2H, Ar-H), 7.56 (d, $^{3}J_{H\text{-}H}$ = 8.2 Hz, 2H, Ar-H), 7.29 (d, $^{3}J_{H\text{-}H}$ = 8.2 Hz, 2H, Ar-H), 7.19 (d, $^{3}J_{H\text{-}H}$ = 8.4 Hz, 2H, Ar-H), 7.13 (d, $^{3}J_{H\text{-}H}$ = 8.4 Hz, 2H, Ar-H), 6.51 (d, $^{3}J_{H\text{-}F}$ = 10.6 Hz, 2H, Ar-H), 3.97 (t, $^{3}J_{H\text{-}H}$ = 6.4 Hz, 2H, Ar-$O\underline{CH_2}CH_2CH_2(CH_2)_3$-Ar), 2.69 (t, $^{3}J_{H\text{-}H}$ = 7.6 Hz, 2H, Ar-$\underline{CH_2}CH_2CH_2(CH_2)_3O$-Ar), 2.62 (t, $^{3}J_{H\text{-}H}$ = 7.8 Hz, 2H, Ar-$\underline{CH_2}CH_2(CH_2)_3CH_3$), 1.82 (m, 2H, Ar-$OCH_2\underline{CH_2}CH_2(CH_2)_3$-Ar), 1.71 (p, $^{3}J_{H\text{-}H}$ = 7.6 Hz, 2H, Ar-$CH_2\underline{CH_2}CH_2(CH_2)_3O$-Ar), 1.61 (m, 2H, Ar-$CH_2\underline{CH_2}(CH_2)_3CH_3$), 1.49 (m, 4H, Ar-$OCH_2CH_2\underline{CH_2}\underline{CH_2}CH_2CH_2$-Ar), 1.32 (m, 6H, Ar-$CH_2CH_2\underline{(CH_2)_3}CH_3$), 0.89 (t, $^{3}J_{H\text{-}H}$ = 6.7 Hz, 3H, Ar-$CH_2CH_2(CH_2)_3\underline{CH_3}$). $^{13}$C NMR (101 MHz, $CDCl_3$) δ ppm: 163.18 (dd, $^{1}J_{C\text{-}F}$ = 256.2 Hz, $^{3}J_{C\text{-}F}$ = 9.8 Hz, 2C), 162.41 (t, $^{3}J_{C\text{-}F}$ = 14.7 Hz), 150.54 (m), 145.33, 142.43, 141.57, 141.19, 137.78, 137.74, 132.77 (2C), 129.19 (2C), 129.13 (2C), 127.72 (2C),

127.70 (2C), 127.66 (2C), 127.08 (2C), 120.89 (2C), 119.11, 110.96, 107.18 (t, ${}^{2}J_{C\text{-}F}$ = 13.1 Hz), 99.09 (dd, ${}^{2}J_{C\text{-}F}$ = 26.4 Hz, ${}^{4}J_{C\text{-}F}$ = 1.8 Hz, 2C), 68.97, 35.63, 35.59, 31.86, 31.63, 31.38, 29.09, 28.97, 28.90, 25.89, 22.75, 14.23. ${}^{19}$F NMR (376 MHz, $CDCl_3$) δ ppm: -111.03 (2F). IR (νmax/cm$^{-1}$): 2924 (sp$^2$ hybridised C-H stretching), 2854 (sp$^2$ hybridised C-H stretching), 2219 (nitrile stretching), 1919 (aromatic overtones), 1625 (imine stretching).

HRMS: (TOF ESI$^+$)(m/z): [M+H]$^+$ Calculated for $C_{44}H_{45}N_2OF_2$ = 655.3500; Found 655.3472 (-4.3 ppm).

**CT6O2,6dF.7**

White solid. Yield: 60%.

${}^{1}$H NMR (400 MHz, $CDCl_3$) δ ppm: 8.57 (s, 1H, <u>CH</u>=N), 7.74 (m, appt s, 4H, Ar-H), 7.70 (d, ${}^{3}J_{H\text{-}H}$ = 8.4 Hz, 2H, Ar-H), 7.66 (d, ${}^{3}J_{H\text{-}H}$ = 8.4 Hz, 2H, Ar-H), 7.56 (d, ${}^{3}J_{H\text{-}H}$ = 8.0 Hz, 2H, Ar-H), 7.29 (d, ${}^{3}J_{H\text{-}H}$ = 8.0 Hz, 2H, Ar-H), 7.19 (d, ${}^{3}J_{H\text{-}H}$ = 8.3 Hz, 2H, Ar-H), 7.13 (d, ${}^{3}J_{H\text{-}H}$ = 8.3 Hz, 2H, Ar-H), 6.51 (d, ${}^{3}J_{H\text{-}F}$ = 10.8 Hz, 2H, Ar-H), 3.98 (t, ${}^{3}J_{H\text{-}H}$ = 6.4 Hz, 2H, Ar-O$\underline{CH_2}$$CH_2CH_2(CH_2)_3$-Ar), 2.69 (t, ${}^{3}J_{H\text{-}H}$ = 7.6 Hz, 2H, Ar-$\underline{CH_2}$$CH_2CH_2(CH_2)_3$O-Ar), 2.62 (t, ${}^{3}J_{H\text{-}H}$ = 7.6 Hz, 2H, Ar-$\underline{CH_2}$$CH_2(CH_2)_4CH_3$), 1.82 (m, 2H, Ar-O$CH_2$$\underline{CH_2}$$CH_2(CH_2)_3$-Ar), 1.71 (p, ${}^{3}J_{H\text{-}H}$ = 7.6 Hz, 2H, Ar-$CH_2$$\underline{CH_2}$$CH_2(CH_2)_3$O-Ar), 1.61 (m, 2H, Ar-$CH_2$$\underline{CH_2}$$(CH_2)_4CH_3$), 1.50 (m, 4H, Ar-O$CH_2CH_2$$\underline{CH_2CH_2}$$CH_2CH_2$-Ar), 1.30 (m, 8H, Ar-$CH_2CH_2$$\underline{(CH_2)_4}$$CH_3$), 0.88 (t, ${}^{3}J_{H\text{-}H}$ = 6.7 Hz, 3H, Ar-$CH_2CH_2(CH_2)_4$$\underline{CH_3}$). ${}^{13}$C NMR (101 MHz, $CDCl_3$) δ ppm: 163.19 (dd, ${}^{1}J_{C\text{-}F}$ = 256.2 Hz, ${}^{3}J_{C\text{-}F}$ = 9.8 Hz, 2C), 162.40 (t, ${}^{3}J_{C\text{-}F}$ = 14.5 Hz), 150.54 (m), 145.33, 142.43, 141.58, 141.18, 137.80, 137.75, 132.78 (2C), 129.19 (2C), 129.13 (2C), 127.73 (2C), 127.71 (2C), 127.67 (2C), 127.09 (2C), 120.89 (2C), 119.11, 110.98, 107.20 (t, ${}^{2}J_{C\text{-}F}$ = 13.1 Hz), 99.10 (dd, ${}^{2}J_{C\text{-}F}$ = 26.4 Hz, ${}^{4}J_{C\text{-}F}$ = 1.9 Hz), 68.98, 35.63, 35.59, 31.97, 31.68, 31.38, 29.39, 29.32, 28.98, 28.91, 25.90, 22.81, 14.25. ${}^{19}$F NMR (376 MHz, $CDCl_3$) δ ppm: -111.04 (2F). IR (νmax/cm$^{-1}$): 2925 (sp$^2$ hybridised C-H stretching), 2853 (sp$^2$ hybridised C-H stretching), 2220 (nitrile stretching), 1914 (aromatic overtones), 1624 (imine stretching). HRMS: (TOF ESI$^+$)(m/z): [M+H]$^+$ Calculated for $C_{45}H_{47}N_2OF_2$ = 669.3656; Found 669.3633 (-3.4 ppm).

**CT6O2,6dF.8**

White solid. Yield: 47%.

$^{1}$H NMR (400 MHz, $CDCl_3$) δ ppm: 8.57 (s, 1H, $\underline{CH}$=N), 7.73 (m, appt s, 4H, Ar-H), 7.70 (d, $^{3}J_{H-H}$ = 8.4 Hz, 2H, Ar-H), 7.66 (d, $^{3}J_{H-H}$ = 8.4 Hz, 2H, Ar-H), 7.56 (d, $^{3}J_{H-H}$ = 8.0 Hz, 2H, Ar-H), 7.29 (d, $^{3}J_{H-H}$ = 8.0 Hz, 2H, Ar-H), 7.19 (d, $^{3}J_{H-H}$ = 8.1 Hz, 2H, Ar-H), 7.13 (d, $^{3}J_{H-H}$ = 8.1 Hz, 2H, Ar-H), 6.51 (d, $^{3}J_{H-F}$ = 11.0 Hz, 2H, Ar-H), 3.98 (t, $^{3}J_{H-H}$ = 6.4 Hz, 2H, Ar-O$\underline{CH_2}CH_2CH_2(CH_2)_3$-Ar), 2.69 (t, $^{3}J_{H-H}$ = 7.7 Hz, 2H, Ar-$\underline{CH_2}CH_2CH_2(CH_2)_3$O-Ar), 2.61 (t, $^{3}J_{H-H}$ = 7.7 Hz, 2H, Ar-$\underline{CH_2}CH_2(CH_2)_5CH_3$), 1.82 (m, 2H, Ar-$OCH_2\underline{CH_2}CH_2(CH_2)_3$-Ar), 1.71 (p, $^{3}J_{H-H}$ = 7.8 Hz, 2H, Ar-$CH_2\underline{CH_2}CH_2(CH_2)_3$O-Ar), 1.62 (m, 2H, Ar-$CH_2\underline{CH_2}(CH_2)_5CH_3$), 1.48 (m, 4H, Ar-$OCH_2CH_2\underline{CH_2CH_2}CH_2CH_2$-Ar), 1.30 (m, 10H, Ar-$CH_2CH_2\underline{(CH_2)_5}CH_3$), 0.88 (t, $^{3}J_{H-H}$ = 6.7 Hz, 3H, Ar-$CH_2CH_2(CH_2)_5\underline{CH_3}$). $^{13}$C NMR (101 MHz, $CDCl_3$) δ ppm: 163.00 (dd, $^{1}J_{C-F}$ = 256.7 Hz, $^{3}J_{C-F}$ = 9.5 Hz, 2C), 162.40 (t, $^{3}J_{C-F}$ = 14.6 Hz), 150.55 (m), 145.34, 142.43, 141.59, 141.20, 137.80, 137.76, 132.78 (2C), 129.19 (2C), 129.14 (2C), 127.74 (2C), 127.71 (2C), 127.67 (2C), 127.09 (2C), 120.90 (2C), 119.12, 110.99, 107.20 (t, $^{2}J_{C-F}$ = 13.2 Hz), 99.10 (dd, $^{2}J_{C-F}$ = 26.8 Hz, $^{4}J_{C-F}$ = 1.7 Hz, 2C), 68.98, 35.64, 35.60, 32.03, 31.69, 31.39, 29.62, 29.44, 29.41, 28.98, 28.91, 25.91, 22.82, 14.25. $^{19}$F NMR (376 MHz, $CDCl_3$) δ ppm: -111.05 (2F). IR (νmax/cm$^{-1}$): 2923 (sp$^2$ hybridised C-H stretching), 2851 (sp$^2$ hybridised C-H stretching), 2219 (nitrile stretching), 1916 (aromatic overtones), 1624 (imine stretching). HRMS: (TOF ESI$^+$)(m/z): [M+H]$^+$ Calculated for $C_{46}H_{49}N_2OF_2$ = 683.3813; Found 683.3807 (-0.9 ppm).

**CT6O2,6dF.9**

White solid. Yield: 31%.

$^{1}$H NMR (400 MHz, $CDCl_3$) δ ppm: 8.57 (s, 1H, $\underline{CH}$=N), 7.73 (m, appt s, 4H, Ar-H), 7.70 (d, $^{3}J_{H-H}$ = 8.4 Hz, 2H, Ar-H), 7.66 (d, $^{3}J_{H-H}$ = 8.4 Hz, 2H, Ar-H), 7.56 (d, $^{3}J_{H-H}$ = 8.1 Hz, 2H, Ar-H), 7.29 (d, $^{3}J_{H-H}$ = 8.1 Hz, 2H, Ar-H), 7.19 (d, $^{3}J_{H-H}$ = 8.3 Hz, 2H, Ar-H), 7.13 (d, $^{3}J_{H-H}$ = 8.3 Hz, 2H, Ar-H), 6.51 (d, $^{3}J_{H-F}$ = 10.7 Hz, 2H, Ar-H), 3.98 (t, $^{3}J_{H-H}$ = 6.4 Hz, 2H, Ar-O$\underline{CH_2}CH_2CH_2(CH_2)_3$-Ar), 2.69 (t, $^{3}J_{H-H}$ = 7.6 Hz, 2H, Ar-$\underline{CH_2}CH_2CH_2(CH_2)_3$O-Ar), 2.61 (t, $^{3}J_{H-H}$ = 7.6 Hz, 2H, Ar-$\underline{CH_2}CH_2(CH_2)_6CH_3$), 1.82 (m, 2H, Ar-$OCH_2\underline{CH_2}CH_2(CH_2)_3$-Ar), 1.71 (p, $^{3}J_{H-H}$ = 7.7 Hz, 2H, Ar-$CH_2\underline{CH_2}CH_2(CH_2)_3$O-Ar), 1.61 (m, 2H, Ar-$CH_2\underline{CH_2}(CH_2)_6CH_3$), 1.49 (m, 4H, Ar-$OCH_2CH_2\underline{CH_2CH_2}CH_2CH_2$-Ar), 1.30 (m, 12H, Ar-$CH_2CH_2\underline{(CH_2)_6}CH_3$), 0.88 (t, $^{3}J_{H-H}$ = 6.8 Hz, 3H, Ar-$CH_2CH_2(CH_2)_6\underline{CH_3}$). $^{13}$C NMR (101 MHz, $CDCl_3$) δ ppm: 163.19 (dd, $^{1}J_{C-F}$ = 256.1 Hz, $^{3}J_{C-F}$ = 9.7 Hz, 2C), 162.41 (t, $^{3}J_{C-F}$ = 14.5 Hz), 150.54 (m),

145.33, 142.43, 141.58, 141.19, 137.80, 137.75, 132.78 (2C), 129.19 (2C), 129.13 (2C), 127.73 (2C), 127.71 (2C), 127.67 (2C), 127.09 (2C), 120.89 (2C), 119.11, 115.38, 110.98, 107.20 (t, $^{2}J_{C-F}$ = 12.9 Hz), 99.09 (dd, $^{2}J_{C-F}$ = 26.5 Hz, $^{4}J_{C-F}$ = 2.0 Hz, 2C), 68.98, 35.64, 35.60, 32.04, 31.68, 31.39, 29.70, 29.66, 29.47, 29.43, 28.98, 28.91, 25.90, 22.82, 14.26. $^{19}$F NMR (376 MHz, $CDCl_3$) δ ppm: -111.04 (2F). IR (νmax/$cm^{-1}$): 2923 ($sp^2$ hybridised C-H stretching), 2851 ($sp^2$ hybridised C-H stretching), 2219 (nitrile stretching), 1904 (aromatic overtones), 1623 (imine stretching). HRMS: (TOF $ESI^+$)(m/z): $[M+H]^+$ Calculated for $C_{47}H_{51}N_2OF_2$ = 697.3969; Found 697.3973 (+0.6 ppm).

**CT6O2,6dF.10**

White solid. Yield: 10%.

$^{1}$H NMR (400 MHz, $CDCl_3$) δ ppm: 8.57 (s, 1H, <u>CH</u>=N), 7.73 (m, appt s, 4H, Ar-H), 7.70 (d, $^{3}J_{H-H}$ = 8.4 Hz, 2H, Ar-H), 7.66 (d, $^{3}J_{H-H}$ = 8.4 Hz, 2H, Ar-H), 7.56 (d, $^{3}J_{H-H}$ = 8.2 Hz, 2H, Ar-H), 7.29 (d, $^{3}J_{H-H}$ = 8.2 Hz, 2H, Ar-H), 7.19 (d, $^{3}J_{H-H}$ = 8.4 Hz, 2H, Ar-H), 7.13 (d, $^{3}J_{H-H}$ = 8.4 Hz, 2H, Ar-H), 6.51 (d, $^{3}J_{H-F}$ = 10.6 Hz, 2H, Ar-H), 3.98 (t, $^{3}J_{H-H}$ = 6.4 Hz, 2H, Ar-O$\underline{CH_2}CH_2CH_2(CH_2)_3$-Ar), 2.69 (t, $^{3}J_{H-H}$ = 7.7 Hz, 2H, Ar-$\underline{CH_2}CH_2CH_2(CH_2)_3$O-Ar), 2.61 (t, $^{3}J_{H-H}$ = 7.7 Hz, 2H, Ar-$\underline{CH_2}CH_2(CH_2)_7CH_3$), 1.81 (m, 2H, Ar-$OCH_2\underline{CH_2}CH_2(CH_2)_3$-Ar), 1.70 (p, $^{3}J_{H-H}$ = 7.7 Hz, 2H, Ar-$CH_2\underline{CH_2}CH_2(CH_2)_3$O-Ar), 1.60 (m, 2H, Ar-$CH_2\underline{CH_2}(CH_2)_7CH_3$), 1.47 (m, 4H, Ar-$OCH_2CH_2\underline{CH_2}\underline{CH_2}CH_2CH_2$-Ar), 1.29 (m, 14H, Ar-$CH_2CH_2\underline{(CH_2)_7}CH_3$), 0.88 (t, $^{3}J_{H-H}$ = 6.8 Hz, 3H, Ar-$CH_2CH_2(CH_2)_7\underline{CH_3}$). $^{13}$C NMR (101 MHz, $CDCl_3$) δ ppm: 163.18 (dd, $^{1}J_{C-F}$ = 256.3 Hz, $^{3}J_{C-F}$ = 9.7 Hz, 2C), 162.41 (t, $^{3}J_{C-F}$ = 14.3 Hz), 150.54 (m), 145.33, 142.43, 141.57, 141.20, 137.79, 137.74, 132.77 (2C), 129.19 (2C), 129.13 (2C), 127.73 (2C), 127.70 (2C), 127.66 (2C), 127.08 (2C), 120.89 (2C), 119.11, 115.35, 110.96, 107.18 (t, $^{2}J_{C-F}$ = 12.7 Hz), 99.09 (dd, $^{2}J_{C-F}$ = 26.4 Hz, $^{4}J_{C-F}$ = 1.9 Hz. 2C), 68.98, 35.63, 35.59, 32.04, 31.67, 31.38, 29.76, 29.74, 29.65, 29.47, 29.43, 28.98, 28.90, 25.90, 22.83, 14.26. $^{19}$F NMR (376 MHz, $CDCl_3$) δ ppm: -111.03 (2F). IR (νmax/$cm^{-1}$): 2915 ($sp^2$ hybridised C-H stretching), 2845 ($sp^2$ hybridised C-H stretching), 2233 (nitrile stretching), 1905 (aromatic overtones), 1638 (imine stretching). HRMS: (TOF $ESI^+$)(m/z): $[M+H]^+$ Calculated for $C_{48}H_{53}N_2OF_2$ = 711.4126; Found 711.4134 (+1.1 ppm).

**CT6O2,6dF.12**

White solid. Yield: 52%.

$^1$H NMR (400 MHz, $CDCl_3$) δ ppm: 8.57 (s, 1H, $\underline{CH}$=N), 7.73 (m, appt s, 4H, Ar-H), 7.70 (d, $^3J_{H-H}$ = 8.5 Hz, 2H, Ar-H), 7.66 (d, $^3J_{H-H}$ = 8.5 Hz, 2H, Ar-H), 7.56 (d, $^3J_{H-H}$ = 8.2 Hz, 2H, Ar-H), 7.29 (d, $^3J_{H-H}$ = 8.2 Hz, 2H, Ar-H), 7.19 (d, $^3J_{H-H}$ = 8.3 Hz, 2H, Ar-H), 7.13 (d, $^3J_{H-H}$ = 8.3 Hz, 2H, Ar-H), 6.51 (d, $^3J_{H-F}$ = 10.5 Hz, 2H, Ar-H), 3.98 (t, $^3J_{H-H}$ = 6.4 Hz, 2H, Ar-O$\underline{CH_2}$$CH_2CH_2(CH_2)_3$-Ar), 2.69 (t, $^3J_{H-H}$ = 7.7 Hz, 2H, Ar-$\underline{CH_2}$$CH_2CH_2(CH_2)_3$O-Ar), 2.61 (t, $^3J_{H-H}$ = 7.7 Hz, 2H, Ar-$\underline{CH_2}$$CH_2(CH_2)_9CH_3$), 1.82 (m, 2H, Ar-$OCH_2$$\underline{CH_2}$$CH_2(CH_2)_3$-Ar), 1.71 (p, $^3J_{H-H}$ = 7.7 Hz, 2H, Ar-$CH_2$$\underline{CH_2}$$CH_2(CH_2)_3$O-Ar), 1.64 (m, 2H, Ar-$CH_2$$\underline{CH_2}$$(CH_2)_9CH_3$), 1.49 (m, 4H, Ar-$OCH_2CH_2$$\underline{CH_2CH_2}$$CH_2CH_2$-Ar), 1.29 (m, 18H, Ar-$CH_2CH_2$$\underline{(CH_2)_9}$$CH_3$), 0.88 (t, $^3J_{H-H}$ = 6.8 Hz, 3H, Ar-$CH_2CH_2(CH_2)_9$$\underline{CH_3}$). $^{13}$C NMR (101 MHz, $CDCl_3$) δ ppm: 163.18 (dd, $^1J_{C-F}$ = 256.3 Hz, $^3J_{C-F}$ = 9.8 Hz, 2C), 162.41 (t, $^3J_{C-F}$ = 14.4 Hz), 150.54 (m), 145.33, 142.43, 141.58, 141.20, 137.79, 137.75, 132.77 (2C), 129.19 (2C), 129.13 (2C), 127.73 (2C), 127.70 (2C), 127.66 (2C), 127.08 (2C), 120.89 (2C), 119.11, 115.35, 110.97, 107.19 (t, $^2J_{C-F}$ = 12.6 Hz), 99.09 (dd, $^2J_{C-F}$ = 26.6 Hz, $^4J_{C-F}$ = 2.0 Hz, 2C), 68.98, 35.63, 35.59, 32.06, 31.68, 31.38, 29.82, 29.80, 29.78, 29.74, 29.66, 29.50, 29.43, 28.98, 28.90, 25.90, 22.83, 14.26. $^{19}$F NMR (376 MHz, $CDCl_3$) δ ppm: -111.04 (2F). IR (νmax/$cm^{-1}$): 2917 ($sp^2$ hybridised C-H stretching), 2846 ($sp^2$ hybridised C-H stretching), 2231 (nitrile stretching), 1906 (aromatic overtones), 1644 (imine stretching). HRMS: (TOF $ESI^+$)(m/z): $[M+H]^+$ Calculated for $C_{50}H_{57}N_2OF_2$ = 739.4439; Found 739.4420 (-2.6 ppm).

**CT6O2,6dF.14**

White solid. Yield: 34%.

$^1$H NMR (400 MHz, $CDCl_3$) δ ppm: 8.57 (s, 1H, $\underline{CH}$=N), 7.73 (m, appt s, 4H, Ar-H), 7.70 (d, $^3J_{H-H}$ = 8.4 Hz, 2H, Ar-H), 7.66 (d, $^3J_{H-H}$ = 8.4 Hz, 2H, Ar-H), 7.56 (d, $^3J_{H-H}$ = 8.0 Hz, 2H, Ar-H), 7.29 (d, $^3J_{H-H}$ = 8.0 Hz, 2H, Ar-H), 7.19 (d, $^3J_{H-H}$ = 8.3 Hz, 2H, Ar-H), 7.13 (d, $^3J_{H-H}$ = 8.3 Hz, 2H, Ar-H), 6.51 (d, $^3J_{H-F}$ = 10.7 Hz, 2H, Ar-H), 3.98 (t, $^3J_{H-H}$ = 6.4 Hz, 2H, Ar-O$\underline{CH_2}$$CH_2CH_2(CH_2)_3$-Ar), 2.69 (t, $^3J_{H-H}$ = 7.6 Hz, 2H, Ar-$\underline{CH_2}$$CH_2CH_2(CH_2)_3$O-Ar), 2.61 (t, $^3J_{H-H}$ = 7.6 Hz, 2H, Ar-$\underline{CH_2}$$CH_2(CH_2)_{11}CH_3$), 1.82 (m, 2H, Ar-O$\underline{CH_2}$$CH_2(CH_2)_3$-Ar), 1.71 (p, $^3J_{H-H}$ = 7.6 Hz, 2H, Ar-$CH_2$$\underline{CH_2}$$CH_2(CH_2)_3$O-Ar), 1.63 (m, 2H, Ar-$CH_2$$\underline{CH_2}$$(CH_2)_{11}CH_3$), 1.49 (m, 4H, Ar-$OCH_2CH_2$$\underline{CH_2CH_2}$$CH_2CH_2$-Ar), 1.26 (m, 22H, Ar-$CH_2CH_2$$\underline{(CH_2)_{11}}$$CH_3$), 0.88 (t, $^3J_{H-H}$ = 6.7 Hz, 3H, Ar-$CH_2CH_2(CH_2)_{11}$$\underline{CH_3}$). $^{13}$C NMR (101 MHz, $CDCl_3$) δ ppm: 163.19 (dd, $^1J_{C-F}$ = 256.3 Hz, $^3J_{C-F}$ = 9.7 Hz, 2C), 162.40 (t, $^3J_{C-F}$ = 14.9 Hz), 150.54 (m),

145.34, 142.43, 141.58, 141.19, 137.80, 137.75, 132.78 (2C), 129.19 (2C), 129.13 (2C), 127.74 (2C), 127.71 (2C), 127.67 (2C), 127.09 (2C), 120.89 (2C), 119.11, 115.34, 110.99, 107.21 (t, $^{2}J_{C-F}$ = 13.1 Hz), 99.09 (dd, $^{2}J_{C-F}$ = 26.7 Hz, $^{4}J_{C-F}$ = 1.9 Hz, 2C), 68.98, 35.64, 35.60, 32.07, 31.68, 31.39, 29.85, 29.83 (2C), 29.81, 29.80, 29.75, 29.66, 29.51, 29.44, 28.99, 28.91, 25.91, 22.84, 14.27. $^{19}$F NMR (376 MHz, $CDCl_3$) δ ppm: -111.05 (2F). IR (νmax/cm$^{-1}$): 2919 (sp$^2$ hybridised C-H stretching), 2851 (sp$^2$ hybridised C-H stretching), 2220 (nitrile stretching), 1917 (aromatic overtones), 1624 (imine stretching). HRMS: (TOF ESI$^+$)(m/z): $[M+H]^+$ Calculated for $C_{52}H_{61}N_2OF_2$ = 767.4752; Found 767.4719 (-4.3 ppm).

**CT6O2,6dF.16**

White solid. Yield: 52%.

$^{1}$H NMR (400 MHz, $CDCl_3$) δ ppm: 8.57 (s, 1H, $\underline{CH}$=N), 7.73 (m, appt s, 4H, Ar-H), 7.70 (d, $^{3}J_{H-H}$ = 8.4 Hz, 2H, Ar-H), 7.66 (d, $^{3}J_{H-H}$ = 8.4 Hz, 2H, Ar-H), 7.56 (d, $^{3}J_{H-H}$ = 8.0 Hz, 2H, Ar-H), 7.29 (d, $^{3}J_{H-H}$ = 8.0 Hz, 2H, Ar-H), 7.19 (d, $^{3}J_{H-H}$ = 8.3 Hz, 2H, Ar-H), 7.13 (d, $^{3}J_{H-H}$ = 8.3 Hz, 2H, Ar-H), 6.51 (d, $^{3}J_{H-F}$ = 10.7 Hz, 2H, Ar-H), 3.98 (t, $^{3}J_{H-H}$ = 6.4 Hz, 2H, Ar-O$\underline{CH_2}CH_2CH_2(CH_2)_3$-Ar), 2.69 (t, $^{3}J_{H-H}$ = 7.6 Hz, 2H, Ar-$\underline{CH_2}CH_2CH_2(CH_2)_3$O-Ar), 2.61 (t, $^{3}J_{H-H}$ = 7.6 Hz, 2H, Ar-$\underline{CH_2}CH_2(CH_2)_{13}CH_3$), 1.82 (m, 2H, Ar-O$\underline{CH_2}CH_2(CH_2)_3$-Ar), 1.71 (p, $^{3}J_{H-H}$ = 7.6 Hz, 2H, Ar-$CH_2\underline{CH_2}CH_2(CH_2)_3$O-Ar), 1.62 (m, 2H, Ar-$CH_2\underline{CH_2}(CH_2)_{13}CH_3$), 1.48 (m, 4H, Ar-$OCH_2CH_2\underline{CH_2CH_2}CH_2CH_2$-Ar), 1.26 (m, 26H, Ar-$CH_2\underline{(CH_2)_{13}}CH_3$), 0.88 (t, $^{3}J_{H-H}$ = 6.7 Hz, 3H, Ar-$CH_2CH_2(CH_2)_{13}\underline{CH_3}$). $^{13}$C NMR (101 MHz, $CDCl_3$) δ ppm: 163.19 (dd, $^{1}J_{C-F}$ = 256.2 Hz, $^{3}J_{C-F}$ = 9.8 Hz, 2C), 162.40 (t, $^{3}J_{C-F}$ = 14.5 Hz), 150.55 (m), 145.34, 142.44, 141.59, 141.20, 137.80, 137.76, 132.78 (2C), 129.19 (2C), 129.14 (2C), 127.74 (2C), 127.71 (2C), 127.67 (2C), 127.09 (2C), 120.90 (2C), 119.12, 115.34, 110.99, 107.21 (t, $^{2}J_{C-F}$ = 12.9 Hz), 99.10 (dd, $^{2}J_{C-F}$ = 26.7 Hz, $^{4}J_{C-F}$ = 2.2 Hz, 2C), 68.98, 35.64, 35.60, 32.08, 31.69, 31.39, 29.85 (4C), 29.83 (2C), 29.81, 29.76, 29.67, 29.51, 29.45, 28.99, 28.91, 25.91, 22.84, 14.27. $^{19}$F NMR (376 MHz, $CDCl_3$) δ ppm: -111.05 (2C). IR (νmax/cm$^{-1}$): 2919 (sp$^2$ hybridised C-H stretching), 2850 (sp$^2$ hybridised C-H stretching), 2220 (nitrile stretching), 1917 (aromatic overtones), 1627 (imine stretching). HRMS: (TOF ESI$^+$)(m/z): $[M+H]^+$ Calculated for $C_{54}H_{65}N_2OF_2$ = 795.5065; Found 795.5032 (-4.1 ppm).

**CT6O2,6dF.18**

White solid. Yield: 45%.

$^{1}$H NMR (400 MHz, $CDCl_3$) δ ppm: 8.57 (s, 1H, <u>CH</u>=N), 7.73 (m, appt s, 4H, Ar-H), 7.70 (d, $^{3}J_{H-H}$ = 8.5 Hz, 2H, Ar-H), 7.66 (d, $^{3}J_{H-H}$ = 8.5 Hz, 2H, Ar-H), 7.56 (d, $^{3}J_{H-H}$ = 8.2 Hz, 2H, Ar-H), 7.29 (d, $^{3}J_{H-H}$ = 8.2 Hz, 2H, Ar-H), 7.19 (d, $^{3}J_{H-H}$ = 8.4 Hz, 2H, Ar-H), 7.13 (d, $^{3}J_{H-H}$ = 8.4 Hz, 2H, Ar-H), 6.51 (d, $^{3}J_{H-F}$ = 10.8 Hz, 2H, Ar-H), 3.98 (t, $^{3}J_{H-H}$ = 6.4 Hz, 2H, Ar-O$\underline{CH_2}$$CH_2CH_2(CH_2)_3$-Ar), 2.69 (t, $^{3}J_{H-H}$ = 7.6 Hz, 2H, Ar-$\underline{CH_2}$$CH_2CH_2(CH_2)_3$O-Ar), 2.61 (t, $^{3}J_{H-H}$ = 7.7 Hz, 2H, Ar-$\underline{CH_2}$$CH_2(CH_2)_{15}CH_3$), 1.82 (m, 2H, Ar-$OCH_2$$\underline{CH_2}$$CH_2(CH_2)_3$-Ar), 1.71 (p, $^{3}J_{H-H}$ = 7.6 Hz, 2H, Ar-$CH_2$$\underline{CH_2}$$CH_2(CH_2)_3$O-Ar), 1.64 (m, 2H, Ar-$CH_2$$\underline{CH_2}$$(CH_2)_{15}CH_3$), 1.49 (m, 4H, Ar-$OCH_2CH_2$$\underline{CH_2CH_2}$$CH_2CH_2$-Ar), 1.26 (m, 30H, Ar-$CH_2CH_2$$\underline{(CH_2)_{15}}$$CH_3$), 0.88 (t, $^{3}J_{H-H}$ = 6.7 Hz, 3H, Ar-$CH_2CH_2(CH_2)_{15}$$\underline{CH_3}$). $^{13}$C NMR (101 MHz, $CDCl_3$) δ ppm: 163.18 (dd, $^{1}J_{C-F}$ = 256.2 Hz, $^{3}J_{C-F}$ = 9.8 Hz, 2C), 162.41 (t, $^{3}J_{C-F}$ = 14.5 Hz), 150.54 (m), 145.34, 142.43, 141.58, 141.20, 137.79, 137.75, 132.78 (2C), 129.19 (2C), 129.13 (2C), 127.73 (2C), 127.71 (2C), 127.66 (2C), 127.09 (2C), 120.89 (2C), 119.11, 110.97, 107.19 (t, $^{2}J_{C-F}$ = 21.8 Hz), 99.09 (dd, $^{2}J_{C-F}$ = 26.5 Hz, $^{4}J_{C-F}$ 1.9 Hz, 2C), 68.98, 35.64, 35.59, 32.07, 31.68, 31.39, 29.84 (6C), 29.82, 29.80 (2C), 29.75, 29.66, 29.51, 29.44, 28.98, 28.91, 25.90, 22.84, 14.27. $^{19}$F NMR (376 MHz, $CDCl_3$) δ ppm: -111.04 (2F). IR (νmax/$cm^{-1}$): 2918 ($sp^2$ hybridised C-H stretching), 2849 ($sp^2$ hybridised C-H stretching), 2226 (nitrile stretching), 1916 (aromatic overtones), 1629 (imine stretching). HRMS: (TOF $ESI^+$)(m/z): $[M+H]^+$ Calculated for $C_{56}H_{69}N_2OF_2$ = 823.5378; Found 823.5359 (-2.3 ppm).

## 4. Thermal Data

### CT4O2Me.*m* series

**Table S12:** Transition temperatures (in °C) and associated scaled entropies (in parenthesis) for the CT4O2Me.*m* series. Data taken from DSC heating trace unless otherwise noted.

| *m* | Melt | N | $N_{TB}$ | SmA | $SmC_4$ | $SmC_{TB\text{-}DH}$ | $SmC_S$ | $SmC_{TB\text{-}Mod}$ | $SmC_{TB\text{-}SH}$ | $SmC_{TB\text{-}DH}$ | $SmC_{TB\text{-}C}$ |
|---|---|---|---|---|---|---|---|---|---|---|---|
| 1 | 149 (8.91) | 206 (0.15) | 134* | - | - | - | - | - | - | - | - |
| 2 | 132 (8.64) | 193 (0.13) | 123 (0.02)† | - | - | - | - | - | - | - | - |
| 3 | 135 (9.72) | 192 (0.18) | 123 (0.02)† | - | - | - | - | - | - | - | 85 (0.68)† |
| 4 | 130 (11.89) | 181 (0.16) | 117 (0.05)† | - | - | - | - | - | - | - | 97 (1.90)† |
| 5 | 131 (7.54) | 179 (0.11) | 115 (0.02)† | - | - | - | - | - | - | - | 78 (0.59)† |
| 6 | 124 (7.03) | 171 (0.12) | - | 114 | - | - | - | - | 104 (0.03)†‡ | 100‡ | 81 (0.77)† |
| 7 | 123 (7.42) | 169 (0.11) | - | 139 (0.02) | - | - | - | - | 102 (0.01)† | 97 (0.01)† | - |
| 8 | 115 (8.36) | 163 (0.10) | - | 150 (0.05) | - | - | - | - | 89 (0.01)† | 85 (0.01)† | - |
| 9 | 113 (8.92) | 163 (0.46)‡ | - | 157‡ | - | - | - | - | 93 (0.01)† | 88 (0.01)† | - |
| 10 | 112 (8.63) | 160 (0.42)‡ | - | 159‡ | - | - | - | - | 87* | 84* | - |
| 12 | 107 (8.61) | - | - | 163 (0.71) | - | - | - | - | - | 88 (0.01)† | - |
| 14 | 106 (9.66) | - | - | 165 (1.03) | - | - | - | - | - | 93* | - |
| 16 | 105 (10.28) | - | - | 165 (1.22) | - | 100* | 98* | - | - | 95 (0.01) | - |
| 18 | 105 (14.22) | - | - | 166 (1.21) | 145* | 101* | 99* | 96* | - | 96 (~0)† | - |

* denotes temperatures taken from POM observations or measurements of the optical birefringence.
† denotes data taken from the DSC cooling trace.
‡ indicates a combined entropy

**CT6O2Me.S*m* series**

**Table S13**: Transition temperatures (in °C) and associated scaled entropies (in parenthesis) for the CT6O2Me.S*m* series. Data taken from DSC heating trace unless otherwise noted.

| *m* | Melt | N | $N_{TB}$ | SmA | $SmC_{S}$ | $SmC_{4}$ | $SmC_{SRe}$ | SmY | $SmC_{TB\text{-}DH}$ | $SmC_{TB\text{-}C}$ | SmX/B4† |
|---|---|---|---|---|---|---|---|---|---|---|---|
| 1 | 150 (46.86) | 214 (0.25) | 122 (0.01)* | - | - | - | - | - | - | - | - |
| 2 | 134 (39.27) | 196 (0.14) | 119 (0.02)* | - | - | - | - | - | - | 75‡ | - |
| 3 | 122 (47.03) | 175 (0.09) | 95 (0.02)* | - | - | - | - | - | - | 82 (0.79)* | - |
| 4 | 131 (48.17) | 175 (0.12) | 103 (0.02)* | - | - | - | - | - | - | 80 (0.52)* | - |
| 5 | 119 (51.97) | 165 (0.07) | 98 (0.01)* | - | - | - | - | - | - | 84 (0.69)* | - |
| 6 | 117 (60.10) | 167 (0.12) | 100 (0.02)* | - | - | - | - | - | - | 93 (1.09)* | 45 (0.85)* |
| 7 | 112 (66.99) | 161 (0.11) | - | - | - | - | - | - | - | 94 (1.42)* | 51 (0.72)* |
| 8 | 110 (64.19) | 158 (0.06) | - | 113 (~0) | - | - | - | - | 91 (0.01)* | 88 (1.20)* | 54 (0.41)* |
| 9 | 121 (71.62) | 154 (0.09) | - | 131 (0.01) | - | - | - | - | 88‡ | 81‡ | 55‡ |
| 10 | 103 (61.69) | 153 (0.09) | - | 138 (0.02) | - | - | - | - | 82 (0.01)* | 70 (1.54)* | 52 (0.60)* |
| 11 | 105 (51.63) | 149 (0.12) | - | 141 (0.07) | - | - | - | - | 76‡ | 58‡ | - |
| 12 | 89 (56.15) | 148 (0.09) | - | 143 (0.08) | - | - | - | - | 80 (~0)* | - | †71‡ |
| 13 | 87 (29.35) | 146 (0.07) | - | 144 (0.10) | 79‡ | - | 79‡ | - | 76‡ | - | †52‡ |
| 15 | 97 (64.70) | - | - | 143 (0.64) | 117‡ | 107‡ | 81‡ | 80‡ | 77‡ | - | - |
| 16 | 103 (74.88) | - | - | 146 (0.68) | 124 (~0)* | 108‡ | 84‡ | 79‡ | 74‡ | - | - |
| 17 | 113 (76.42) | - | - | 146 (0.70) | 131 (~0)* | 102‡ | 86‡ | 76‡ | 70‡ | - | - |
| 18 | 118 (79.63) | - | - | 145 (0.79) | 134 (~0)* | 97‡ | 88‡ | - | - | - | - |

‡ denotes temperatures taken from measurements of the optical birefringence.

* denotes temperatures taken from the DSC cooling trace

**CT6O2F.*m* series**

**Table S14**: Phase transition temperatures (°C) and associated entropies (in parenthesis, scaled by R) for the CT6O2F.m series. Data taken from the DSC cooling trace unless otherwise noted

| *m* | melt | N | $N_{TB}$ | SmA | $SmC_{TB-SH}$ | $SmC_S$ | $SmC_4$ | $SmC_{TB-DH}$ | $SmC_S$ | $SmC_{TB-DH}$ | $SmC_{TB-C}$ |
|---|---|---|---|---|---|---|---|---|---|---|---|
| 1 | 130 (16.73) | 235 (0.33) | 135 (~0)[*] | - | - | - | - | - | - | - | - |
| 2 | 120 (19.92) | 223 (0.25) | 128 (~0)[*] | - | - | - | - | - | - | - | 116 (1.49)[*] |
| 3 | 114 (8.64) | 221 (0.25) | 134 (~0)[*] | - | - | - | - | - | - | - | 119 (1.52)[*] |
| 4 | 99 (14.66) | 212 (0.22) | 127[‡] | - | - | - | - | - | - | - | 126 (1.79)[*] |
| 5 | 111 (17.73) | 210 (0.24) | 127 (~0)[*] | - | - | - | - | - | - | - | 123 (1.56)[*] |
| 6 | 107 (12.87) | 201 (0.20) | - | - | - | - | - | - | - | - | 125 (1.83)[*] |
| 7 | 113 (19.00) | 199 (0.21) | 120 (~0)[*] | - | - | - | - | - | - | - | 118 (1.59)[*] |
| 8 | 110 (13.20) | 190 (0.15 ) | - | 136[‡] | 114 (~0)[*] | - | - | - | - | - | 111 (1.59)[*] |
| 9 | 109 (15.84) | 187 (0.20) | - | 160 (0.01) | 107 (~0)[*] | - | - | - | - | 96 (0.01)[*] | 89 (1.16)[*] |
| 10 | 111 (33.05) | 183 (0.16) | - | 168 (0.03) | 103[‡] | - | - | - | - | 90[‡] | - |
| 12 | 109 (29.03) | 117 (0.16) | - | 173 (0.19) | 95 (0.01)[*] | - | - | - | - | 90 (0.01)[*] | - |
| 14 | 106 (15.02) | - | - | 175 (1.14) | - | - | - | - | - | 95 (0.01)[*] | - |
| 16 | 104 (12.58) | - | - | 175 (1.45) | - | - | - | - | - | 98 (0.01)[*] | - |
| 18 | 103 (28.23) | - | - | 174 (1.53) | - | 137[‡] | 136[‡] | 100 (~0)[*] | 98 (0.01)[*] | 96[‡] | - |

[‡] denotes data taken from measurements of the optical birefringence

[*] denotes data taken from the DSC cooling trace

**CT6O2,6dF.*m* series**

**Table S15**: Phase transition temperatures (°C) and associated entropies (in parenthesis, scaled by R) for the CT6O2,6dF.m series. Data taken from the DSC cooling trace unless otherwise noted.

| *m* | melt | N | $N_{TB}$ | SmA | $SmC_S$ | $SmC_4$/‡$SmC_{TB-C}$ | $SmC_S$ | $SmC_{TB-DH}$ |
|---|---|---|---|---|---|---|---|---|
| 1 | 143 (18.15)* | 227 (0.33)* | 125† | - | - | - | - | - |
| 2 | 136 (14.93)* | 212 (0.25)* | 110† | - | - | - | - | - |
| 3 | 133 (14.63)* | 213 (0.24)* | 118 (0.01) | - | - | - | - | - |
| 4 | 121 (10.57)* | 202 (0.22)* | 108 (0.01) | - | - | - | - | - |
| 5 | 130 (14.89)* | 201 (0.22)* | 113 (0.01) | - | - | - | - | - |
| 6 | 123 (16.00)* | 193 (0.20)* | 107 (0.01) | - | - | - | - | - |
| 7 | 115 (19.74)* | 190 (0.20)* | 109 (0.01) | - | - | - | - | - |
| 8 | 104 (15.62)* | 184 (0.17)* | - | 101 (0.01) | - | ‡76 (0.23) | - | - |
| 9 | 123 (17.58)* | 181 (0.19)* | - | 136 (~0) | - | - | - | 92 (0.02) |
| 10 | 116 (17.57)* | 176 (0.20)* | - | 153 (0.02) | - | - | - | 92 (~0) |
| 12 | 108 (16.57)* | 170 (0.18)* | - | 159 (0.08) | - | - | 87† | 81† |
| 14 | 104 (17.64)* | 167 (0.12)* | - | 164 (0.21) | - | 111† | 86† | 81 (0.01) |
| 16 | 103 (18.22)* | - | - | 164 (1.18) | 142† | 109 (~0) | 85 (0.01) | 81 (0.02) |
| 18 | 104 (17.18)* | - | - | 162 (1.53) | 137† | 109† | 85 (0.04) | 82 (0.02) |

* denotes data taken from the DSC heating trace.

† denotes temperatures taken from measurements of the optical birefringence.

## References:

[S1] M. Salamończyk, N. Vaupotič, D. Pociecha, R. Walker, J. M. D. Storey, C. T. Imrie, C. Wang, C. Zhu, E. Górecka, Multi-level chirality in liquid crystals formed by achiral molecules, Nat. Commun. 10, 1922 (2019). DOI: 10.1038/s41467-019-09862-y

[S2] A. Pearson, A. Alshammari, G. J. Strachan, M. Majewska, D. Pociecha, J. M. D. Storey, C.T. Imrie, N. Vaupotič, R. Walker, E. Gorecka, Diverse helical structures made of achiral mesogenic dimers, Nat. Commun. **17**, 6090 (2026). DOI:10.1038/s41467-026-72565-8

[S3] V. E. Dmitrienko, Forbidden reflections due to anisotropic X-ray susceptibility of crystals, *Acta Crystallogr. Sect. A* **39**, 29-35 (1983).